\documentclass[prd, aps, showpacs, nofootinbib, amsmath, floats, floatfix, superscriptaddress]{revtex4}

\usepackage{amssymb,amsmath,epsfig}
\usepackage{mathrsfs}
\usepackage[usenames,dvipsnames]{color}
\usepackage[bookmarks]{hyperref}
\usepackage[svgnames]{xcolor}
\usepackage{times}
\usepackage{yhmath}
\usepackage{cancel}
\usepackage{float}
\usepackage{booktabs}
\usepackage{multirow}
\usepackage{ulem}
\usepackage[belowskip=1ex]{subcaption}

\newcommand{\Real}{\mathbb{R}}

\newcommand{\dvol}{\mbox{dvol}}

\newcommand{\ve}[1]{{\bf #1}}

\newcommand{\Torus}{\mathbb{T}}

\begin{document}

\title{Rotating Kinetic Gas Disk Morphology Surrounding a Schwarzschild Black Hole}

\author{Carlos Gabarrete} 
\email{carlos.gabarrete@unah.edu.hn} 
\affiliation{Departamento de Gravitaci\'on, Altas Energ\'ias y Radiaciones, Escuela de F\'isica, Facultad de Ciencias, Universidad Nacional Aut\'onoma de Honduras, Edificio E1, Ciudad Universitaria, Tegucigalpa, Francisco Moraz\'an, Honduras.}

\author{Roger Raudales}
\email{roger.raudales@unah.edu.hn}
\affiliation{Maestría en Física, Escuela de F\'isica, Facultad de Ciencias, Universidad Nacional Aut\'onoma de Honduras, Edificio E1, Ciudad Universitaria, Tegucigalpa, Francisco Moraz\'an, Honduras.}
\affiliation{Departamento de Ciencias Naturales, Facultad de Ciencias Básicas, Universidad Pedagógica Nacional Francisco Morazán, Edificio 3, Col. El Dorado, Tegucigalpa, Francisco Morazán, Honduras}

\begin{abstract}
This paper discusses the behavior of a rotating relativistic kinetic gas surrounding a Schwarzschild black hole. We are interested in the description and analysis of the morphology of the resulting configurations for kinetic gas clouds with and without total angular momentum, and we also compare the macroscopic observables with configurations of finite total mass. Considering models for the one-particle distribution function based on a polytropic ansatz and the inclination angle of the orbits of the particles in the kinetic gas, a collisionless gas in the Schwarzschild spacetime background is analyzed. Profiles of the macroscopic observables of the gas configurations are presented, which are derived from the density current vector field and energy-momentum-stress tensor.
\end{abstract}

\date{\today}


\maketitle

\section{Introduction}
\label{Sec:Introduction}

Relativistic kinetic theory framework can be useful for studying dilute gases in the vicinity of a black hole. Recent studies show that traditional fluid-based models often fail in these scenarios due to the absence of collisions, especially when individual particle dynamics dominate over collective fluid behavior~\cite{rAcGoS2022,oStZ13}. Relativistic kinetic theory is crucial near supermassive black holes like Sgr A* and M87*, where gas dynamics and accretion processes are complex~\cite{pRoS17a,pRoS17b,aCpMaO22,aGetal2021,pMmMoS2025}. Observations from the Event Horizon Telescope require accurate models~\cite{EHTC}. This framework also aids in exploring dark matter interactions and early-universe evolution through the relativistic Boltzmann equation~\cite{pDeJmAeMdN17,pMoA2021b,yZetal2024,gPrBlM2019,Pettinari2016}. 

The main goal of this paper is to establish stationary kinetic gas configurations around black holes, in which we focus on the study of a concrete model that describes the configurations of massive, spinless, and uncharged collisionless kinetic gas particles (with and without total angular momentum) trapped in the gravitational potential of a non-rotating black hole, derived from the results obtained in~\cite{cGoS2022}. We describe the spacetime observables using models parameterized by the inclination angle of particle orbits~\cite{BinneyTremaine-Book} and the relativistic generalization of the polytropic \emph{ansatz}~\cite{eAhAaL16,eAhAaL2019,cGoS2023b}.

In contrast to previous studies such as \cite{cGoS2022,cGoS2023b}, which characterized relativistic gas but did not systematically account for gas rotation via orbital inclination, nor provide analytic spatial bounds for finite kinetic configurations, we present explicit analytical expressions for the particle current density and the energy–momentum–stress tensor in stationary relativistic kinetic gas models that incorporate a polytropic ansatz together with a parametrized inclination law for particle orbits. From these expressions, we compute macroscopic observables, such as particle density, energy density, and principal pressures, and derive analytical bounds for the inner and outer radii of finite, energy–cutoff configurations. This allows for a qualitative comparison with finite fluid configurations, such as the polish doughnuts~\cite{Rezzolla-Book} outlined in Section~\ref{Sec:Results}. By explicitly including the azimuthal contribution to the current density and the corresponding off-diagonal components of the energy–momentum–stress tensor, we identify the morphology of rotating kinetic gas configurations.

In this study we model the gas under a set of simplifying but well-motivated assumptions. We describe it as collisionless, so that individual particle interactions can be neglected, an approximation valid for dilute astrophysical systems such as dark matter or stellar distributions. The particles are assumed to be identical, massive, and uncharged, which rules out electromagnetic effects but remains appropriate in contexts where gravity dominates. The self-gravity of the gas is neglected as well, since the central black hole provides the dominant contribution to the gravitational field; in view of the uniqueness theorems, we may regard it as belonging to the Kerr family, although here we further restrict to the non-rotating case, so the exterior geometry is Schwarzschild. Our analysis is confined to stationary configurations in which particles follow bound geodesics of the exterior Schwarzschild spacetime, and for simplicity we focus on axisymmetric states where the distribution function depends only on the energy and the azimuthal angular momentum. Finally, the construction is based on generalized polytropic ansatz for the distribution function, previously mentioned, which allows us to derive explicit single-integral expressions for the current density and the energy–momentum–stress tensor.

A key limitation of the present study is the restriction to a Schwarzschild background, however further analyses performed in Kerr spacetime should reduce, in the limit of vanishing black-hole spin, to the corresponding Schwarzschild results for bounded orbits as the ones showed in this study. In Kerr spacetime, we expect, by analogy with the Schwarzschild case, the existence of a minimum radius for bounded trajectories; however, a systematic investigation of the corresponding bounds in Kerr has not yet been undertaken.

Regarding the neglecting of the self-gravity, this can always be achieved by choosing the amplitude parameter in the distribution function sufficiently small. In fact, the total mass of the gas scales linearly with this amplitude, such that one can always guarantee that it becomes negligible compared to the mass of the central black hole, which validates the test-particle approximation and confirms that neglecting the self-gravity of the gas is fully consistent within the parameter space considered here.

In the development of this study, we find solutions that describe a collisionless kinetic gas around a non-rotating black hole, beginning in Section~\ref{Sec:SchwarzschildSpacetime} with a summary of the principal properties of the Schwarzschild exterior spacetime and some relevant implications for this article. Then, we proceed with a brief review of the principal ideas of the collisionless Boltzmann equation in Section~\ref{Sec:CollisionlessBoltzmann} for the construction of the spacetime observables based on the inclination angle models and the polytropic \emph{ansatz} for the one-particle distribution function (DF). In Section~\ref{Sec:Models}, we introduce a one-particle DF based on the polytropic ansatz and inclination angle of the orbits for configurations of non-rotating and rotating kinetic gas. Additionally, in this section, we construct the spacetime observables for both models.  In Section~\ref{Sec:TotalMass}, we calculate the total particle number, energy and angular momentum by establishing a comparison between configurations, which allows us to explore graphically the resulting configurations as normalized profiles of particle density, energy density and principal pressures. In section~\ref{Sec:Results} we present the results and a comparison of the configurations through the behavior of the macroscopic quantities explored. We compare the resulting behavior of spacetime observables between models with and without total angular momentum, and develop gas configurations with finite extension and compare them with configurations that extend asymptotically to infinity. Finally, in Section~\ref{Sec:Conclusions} we summarize the conclusions drawn from this study.

In other work in development~\cite{dMcG2025}, we present other results associated to the principal pressures and the entropy density to bring completeness to this work. We use the signature convention $(-, +, +, +)$ for the spacetime metric and geometrized units in which Newton’s constant and the speed of light are equal to one, i.e. $G_N = c = 1$.


\section{Summary of the main properties of Schwarzschild exterior spacetime}
\label{Sec:SchwarzschildSpacetime}

In order to describe a relativistic collisionless kinetic gas propagating in the exterior of a spherically-symmetric black hole spacetime, the formalism necessary for this paper is briefly revisited here. Neglecting the self-gravity of the gas particles and assuming the Schwarzschild spacetime background of fixed mass $M_{\mathrm{BH}} := M > 0$, we describe the individual gas particles with rest mass $m$ which follow future-directed spatially bounded timelike geodesics. The spacetime metric described by standard Schwarzschild coordinates $(x^\mu) = (t,r,\vartheta,\varphi)$ is expressed as,
\begin{equation}
g := -N(r) dt^2 + \frac{dr^2}{N(r)} + r^2 \left( d\vartheta^2 + \sin^2\vartheta d\varphi^2 \right), \qquad N(r) := 1 - \frac{2M}{r} > 0.
\end{equation}
The geodesic motion possesses the following integrals of motion: the energy $E$ and the total angular momentum $L^2 := L_x^2 + L_y^2 + L_z^2$, due to the timelike Killing vector fields $\hat{k}=\partial_t$, and $\hat{\ell} = \partial_\varphi$ respectively. Also, the  particle's rest mass $m > 0$ is conserved, which is related to the free-particle Hamiltonian $\mathcal{H}$. For this spacetime manifold $(\mathcal{M},g)$ with associated adapted local coordinates $(x^\mu, p_\mu) = (t, r, \vartheta, \varphi, p_t, p_r, p_\vartheta, p_\varphi)$ on the cotangent bundle $T^*\mathcal{M}$ the free-particle Hamiltonian is defined by
\begin{equation}
\label{Eq:FreeParticleHamiltonian}
\mathcal{H}(x, p) := \frac{1}{2} g_x^{-1}(p,p) 
 = \frac{1}{2}\left( -\frac{p_t^2}{N(r)} + N(r)p_r^2 + \frac{p_\vartheta^2}{r^2} + \frac{p_\varphi^2}{r^2\sin^2\vartheta} \right),
\end{equation}
and it possesses the following conserved quantities associated with this spacetime manifold:
\begin{equation}
\label{Eq:ConservedQuantities}
m = \sqrt{-2\mathcal{H}}, \quad
E = -p_t, \quad 
L_z = p_\varphi, \quad  
L^2 = p^2_\vartheta + \frac{L_z^2}{\sin^2\vartheta},
\end{equation}
where $L:=|\ve{L}|$. It is convenient to introduce the following orthonormal basis of vector fields
$\displaystyle e^{\hat{\mu}} = \left(-N(r)^{-1/2}, N(r)^{1/2}, r^{-1}, (r\sin\vartheta)^{-1} \right)$, in which, if we expand the covector field $p$ in terms of this orthonormal basis $p = p_{\hat{\mu}} e^{\hat{\mu}}$, one obtains
\begin{equation}
\label{Eq:Orthonormalbasis}
p_{\hat{\mu}} = \left(-\frac{E}{\sqrt{N(r)}}, \frac{\epsilon_r}{\sqrt{N(r)}} \sqrt{E^2 - V_{m,L}(r)}, \frac{\epsilon_\vartheta}{r}\sqrt{L^2 - \frac{L_z^2}{\sin^2\vartheta}}, \frac{L_z}{r\sin\vartheta} \right).
\end{equation}
The signs $\epsilon_r = \epsilon_\vartheta =\pm 1$ determine the corresponding signs of $p_{\hat{1}}$ and $p_{\hat{2}}$. The radial effective potential $V_{m,L}(r)$ for the geodesic motion in the Schwarzschild spacetime is defined by
\begin{equation}
\label{Eq:SchEffectivePotential}
V_{m,L}(r) := N(r) \left(m^2 + \frac{L^2}{r^2} \right).
\end{equation}
\begin{figure}[b]
\centering
\includegraphics[scale=0.4]{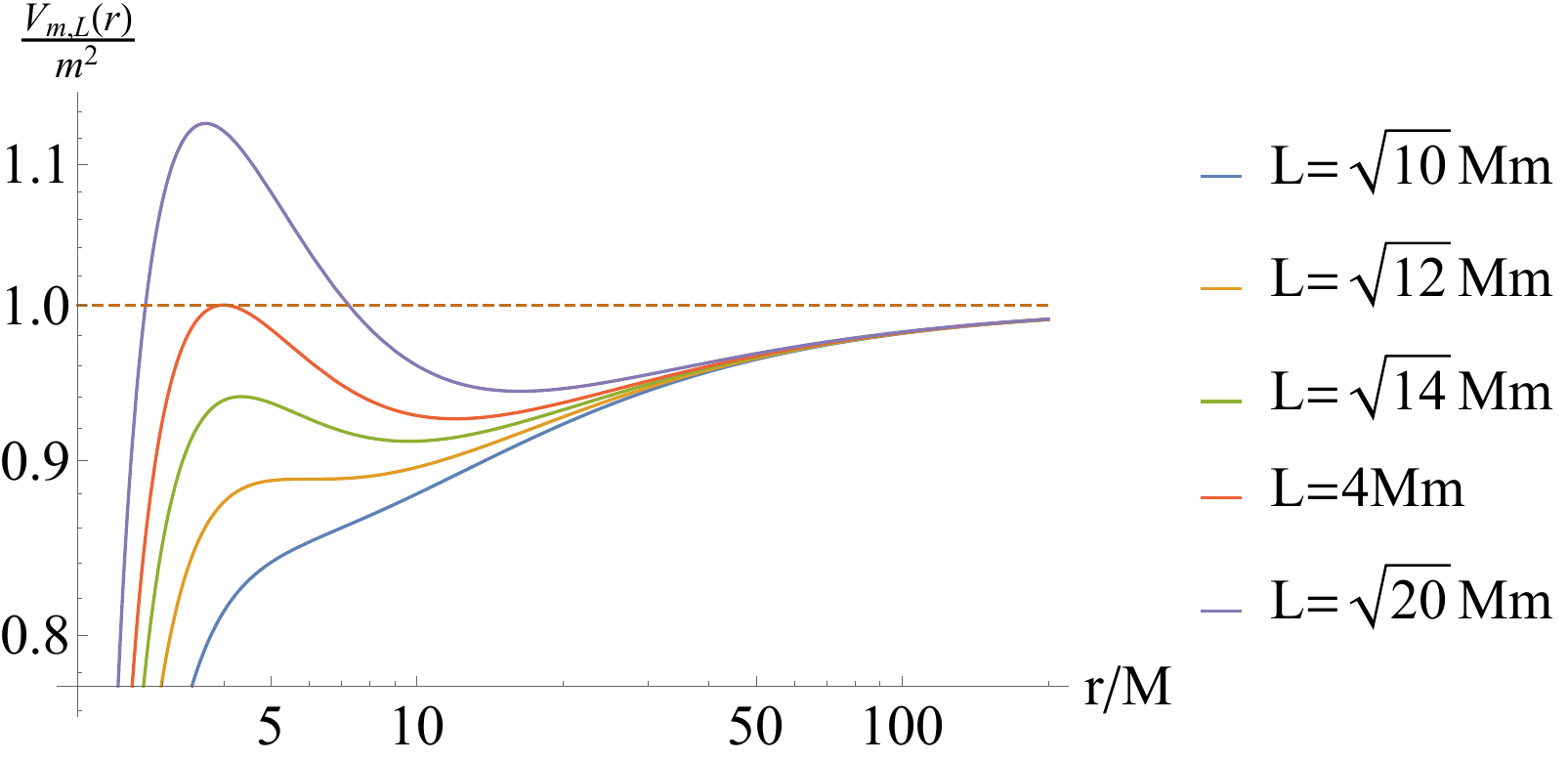}
\caption[Behavior of the effective potential $V_{m,L}$ as a function of the radial coordinate $r$ for different values of $L$]{Behavior of the effective potential $V_{m,L}$ as a function of the radial coordinate $r$ for different values of $L$.}
\label{potentials}
\end{figure}
Depending on the angular momentum of the particles, the effective potential exhibits different behaviors, as shown in figure~\ref{potentials}. In this case, we have:
\begin{enumerate}
\item[(i)] If $L<\sqrt{12}Mm$ (red curve in figure \ref{potentials}), the potential has no critical points.
\item[(ii)] If $L=\sqrt{12}Mm$ (blue curve in figure \ref{potentials}), the potential has an inflection point at $r=6M$.
\item[(iii)] If $L>\sqrt{12}Mm$ (green, purple and orange curves in the figure \ref{potentials}), the potential has two critical points:
\begin{equation}
r_{\text{max}}=\dfrac{6M}{1-\sqrt{1-12 M^2 m^2/L^2}},\ \ \ \ \ \
r_{\text{min}}=\dfrac{6M}{1+\sqrt{1-12 M^2 m^2/L^2}}
\end{equation}
which represent the radial coordinates $r$ where the effective potential is maximum and minimum, respectively. For more details, see~\cite{Misner73,Carroll-Book,Straumann-Book}, and Appendix~A in~\cite{pRoS17a} or~\cite{aCpM2022,cGoS2023b}, and references therein.
\end{enumerate}
Furthermore, the behavior of the geodesics followed by particles under the effective potential also depends on the selection of a given angular momentum and energy (see figure \ref{trajectory}). Then, the trajectories followed by particles can be classified as:
\begin{enumerate}
\item[(i)] Absorbed: These trajectories correspond to particles that come from infinity and fall into the black hole. This occurs for particles whose energy $E$ is greater than the maximum of the effective potential $V_{m,L}(r_{\text{max}})$.
\item[(ii)] Scattered: These trajectories correspond to incoming particles that ``bounce'' off the potential barrier and then ``bounce'' back, returning to the asymptotic region. These particles have an energy between $V_{m,L}(r_{\text{max}})$ and the asymptotic limit of energy $E=m$.
\item[(iii)] Bounded: These are the trajectories corresponding to particles trapped in the potential well, whose energy oscillates between the asymptotic limit $E=m$ and the minimum of the effective potential $V_{m,L}(r_{\text{min}})$.
\end{enumerate}
\begin{figure}[h]
\centering
\includegraphics[scale=0.4]{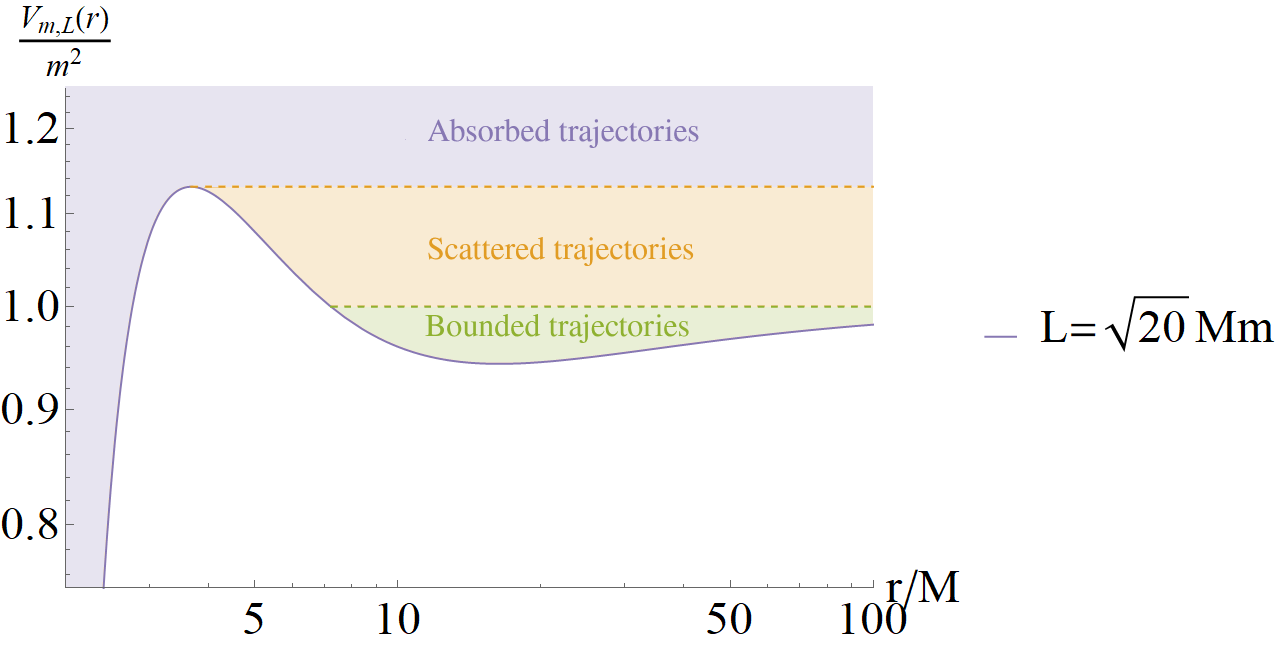}
\caption[Behavior of the trajectories followed by particles from an effective potential associated with a given value of $L$]{Effective potential $V_{m,L}(r)$ for $L=\sqrt{20}Mm$. The purple shaded area represents the trajectories of particles absorbed by the black hole. The yellow shaded area corresponds to the trajectories of particles that are scattered by the potential barrier. Finally, the green shaded area corresponds to particles whose trajectories are bounded.}
\label{trajectory}
\end{figure}
The results determining the three regions in terms of a parameter space $(E,L)$~\cite{pRoS17a,aCpM2022} are summarized below. For a given energy level $E$, the range of the total angular momentum $L$ and its corresponding trajectory characterizations are:
\begin{equation}
\sqrt{\dfrac{8}{9}}m<E<m: \left\{ \begin{array}{lcl}
L<L_{\textrm{c}}(E) & \rightarrow & \text{absorbed trajectory} \\
& & \\
L_c(E)<L<L_{\textrm{ub}} (E)& \rightarrow & \text{bounded trajectory}
\end{array}
\right.
\label{Eq:EnergyLimits}
\end{equation}
and
\begin{equation}
E>m: \left\{ \begin{array}{lcl}
L<L_{\textrm{c}}(E) & \rightarrow & \text{absorbed trajectory} \\
& & \\
L>L_{\textrm{c}}(E)& \rightarrow & \text{scattered trajectory}
\end{array}
\right.
\end{equation}
where $L_{\textrm{c}}(E)$ is the critical value of the total angular momentum, corresponding to the scenario where the maximum in the effective potential is equal to $E^2$. The critical angular momentum is given by:
\begin{equation}
\label{Eq:AngularMomentumC}
L_{\textrm{c}}(E) = \frac{4\sqrt{2}Mm^3}{\sqrt{36m^2 E^2 - 8m^4 - 27E^4 + E\left( 9E^2 - 8m^2 \right)^{3/2} }}.
\end{equation}
$L_{\textrm{ub}}(E)$ is the critical angular momentum corresponding to the scenario where the minimum in the effective potential is equal to $E^2$, given by:
\begin{equation}
\label{Eq:AngularMomentumUB}
L_{\textrm{ub}}(E) = \frac{4\sqrt{2}Mm^3}{\sqrt{36m^2 E^2 - 8m^4 - 27E^4 - E\left( 9E^2 - 8m^2 \right)^{3/2} }}.
\end{equation}

For this particular research, the angular momentum $L$ has to be large enough in magnitude, such that $L^2>L_{\textrm{ms}}^2$, with $L_{\textrm{ms}}=\sqrt{12}Mm$ being the angular momentum corresponding to the marginally stable circular orbit, and the energy $E$ has to be in a certain interval $E_{\text{min}} (L) <E<E_{\text{max}} (L)$, with $E_{\text{min}} (L)$ being the energy of the stable circular orbit with angular momentum $L$ and $E_{\text{max}} (L)\leq m$, the maximum of the potential well, since only bounded orbits are considered. The range of parameters $(E,L,L_z)$ corresponding to the bounded orbits is given by the set~\cite{pRoS2020}:
\begin{equation}
\Omega:=\{(E,L,L_z)\in\mathbb{R}^3: E_{\textrm{c}}(r) < E < m, L_{\textrm{c}}(E) \leq L \leq L_{\textrm{max}}(E,r),\ |L_z| <L\} 
\label{Eq:Limits}
\end{equation}
where $E_{\textrm{c}}(r)$ corresponds to the minimum energy at radius $r$, defined as~\cite{cGoS2023b,aGetal2021}:
\begin{equation}
E_{\textrm{c}}(r) = \left\{
\begin{array}{lcl}
\displaystyle m\frac{1-2M/r}{\sqrt{1-3M/r}}, && 4M \leq r \leq 6M, \\ 
& & \\
\displaystyle m\frac{1 + 2M/r}{\sqrt{1 + 6M/r}}, && r \geq 6M.
\end{array}
\right.
\label{Eq:MinimumEnergy2}
\end{equation}
and $L_{\textrm{max}}(E,r)$ is the maximum allowed total angular momentum for an energy $E$ at an arbitrary radius $r$:
\begin{equation}
L_{\textrm{max}}(E,r) = r\sqrt{\frac{E^2}{N(r)} - m^2}
\label{Eq:MaximumAngularMomentum}
\end{equation}
It is worth noting that the range~(\ref{Eq:Limits}) is zero if $r < 4M$, since the minimum radius required for bounded orbits due to the influence of a Schwarzschild black hole is $r = 4M$. 
In the following chapter, we will describe the macroscopic observables of interest with their respective integration ranges that allow establishing stationary configurations of kinetic gas, considering the region of trajectories trapped by the effective potential.


\section{Collisionless Boltzmann equation and the spacetime observables}
\label{Sec:CollisionlessBoltzmann}

The relativistic Boltzmann equation for a kinetic gas is given by $\displaystyle L[f] = C_W[f,f]$~\cite{rAcGoS2022}, where $L$ is the Liouville operator and $C_W$ is the bilinear collision operator. When the collisions are neglected in a stationary system, the Boltzmann equation reduces to the \emph{collisionless} Boltzmann equation~\footnote{In the literature this equation can be called as Vlasov or Liouville equation.}, which on the curved spacetime manifold $(\mathcal{M}, g)$ is simplified as
\begin{equation}
L[f] = 0,
\label{Eq:Vlasov}
\end{equation}
where $f: T^*\mathcal{M}\to \Real$ is a function denoted as the one-particle DF on the future mass shell, given by
\begin{equation}
\label{Eq:FutureMassShell}
\Gamma_m^+ := \{ (x,p) : x\in \mathcal{M}, p\in P_x^+(m) \}.
\end{equation}
The domain is restricted to the subset $\Gamma_{\textrm{bound}}$ of $T^*\mathcal{M}$, since we are working in the regime of particles moving in bounded orbits. In this subset, one can transform from the adapted local coordinates to generalized action-angle variables, that is, $(x^\mu,p_\mu)\mapsto (\mathcal{Q}^\mu, \mathcal{J}_\mu)$, which is a symplectic transformation. The solution for the DF in terms of these variables is given in~\cite{pRoS2020}. For the following, the gas configurations consisting of identical particles of positive rest mass were restricted, and the momentum is confined to the future mass hyperboloid,
\begin{equation}
P_x^+(m) := \left\{p\in T_x^* \mathcal{M} \: : \: g_x^{-1}(p,p) = -m^2, \hbox{ the vector dual to $p$ is future directed} \right\}.
\end{equation}
It is expected that any gas configuration consisting of purely bounded particles (with a DF $f$ supported in $\Gamma_{\textrm{bound}}$) relaxes in time to a stationary configuration which can be described by a one-particle DF $f$ depending only on the integrals of motion, due to phase space mixing~\cite{pRoS18,pRoS2020,pRoS2024}. This means that, the DF $f$ depends only on $(E,L_x,L_y,L_z)$. Here we assume that $f$ depends on energy $E$, the total angular momentum $L$ and the azimuthal component of the angular momentum $L_z$, due to the gas configuration being stationary and axisymmetric, such that
\begin{equation}
\label{Eq:OneParticleDistributionFunction}
f(x,p) = F(E,L,L_z),
\end{equation}
for some suitable function $F$. If the amplitude of the DF is sufficiently small and $f$ vanishes outside $\Gamma_{\textrm{bound}}$, it has been showed in~\cite{fJ2021,fJ2022} that these configurations can be constructed with the gas’s self-gravity included, thereby providing a natural starting point for future studies that couple the dynamics to the Poisson equation. 

From the one-particle DF $f$, one can construct the physical observables $C^\infty$-smooth tensor fields, which are the most important ones on the spacetime manifold $(\mathcal{M}, g)$. These tensor fields are the particle current density vector field $J\in\mathcal{X}(\mathcal{M})$, the energy-momentum-stress tensor field $T\in\mathcal{T}^{2}{}_{0}(\mathcal{M})$, and the entropy flux vector field $S\in\mathcal{X}(\mathcal{M})$, obtained by suitable fibre integrals over the momenta. However, in this study, we analyze the macroscopic observables derived from the fields $J$ and $T$, leaving the resulting configuration and implications of the field $S$ to the work in progress~\cite{dMcG2025}. 

The particle current density $J$ is defined as
\begin{equation}
\label{Eq:CurrentDensity}
J_\mu (x) := \int\limits_{P_x^+(m)} f(x,p) p_\mu \dvol_x(p),
\end{equation}
and the energy-momentum-stress tensor $T$ is given by
\begin{equation}
\label{Eq:EnergyMomentumStress}
T_{\mu \nu} (x) := \int\limits_{P_x^+(m)} f(x,p) p_\mu p_\nu \dvol_x(p), 
\end{equation}
where $\dvol_x(p)$ is the Lorentz-invariant volume form (see, for instance, Refs.~\cite{CercignaniKremer-Book,rAcGoS2022}). This volume form in terms of the orthonormal basis~(\ref{Eq:Orthonormalbasis}) yields
\begin{equation}
\label{Eq:VolumeForm}
\dvol_x(p) := \frac{dp_{\hat{1}}\wedge dp_{\hat{2}}\wedge dp_{\hat{3}} }{\sqrt{m^2 + p^2_{\hat{1}} + p^2_{\hat{2}} + p^2_{\hat{3}}}}
 = \frac{dp_{\hat{1}}\wedge dp_{\hat{2}}\wedge dp_{\hat{3}} }{|p_{\hat{0}}|}.
\end{equation}
These fields, $J$ and $T$ are divergence-free
\begin{equation}
\nabla^\mu J_\mu = 0, \qquad \nabla^\mu T_{\mu\nu} = 0,
\end{equation}
satisfying the collisionless Boltzmann equation~(\ref{Eq:Vlasov}) for any positive DF $f$. One can rewrite $J$ and $T$ in terms of integrals over the conserved quantities $(E,L,L_z)$. The explicit representation of the volume form~(\ref{Eq:VolumeForm}) in terms of the conserved quantities $(E,L,L_z)$ is given by
\begin{equation}
\label{Eq:VolumeFormELLz}
\dvol_x(p) = \frac{1}{r^2\sin\vartheta } \frac{dE \wedge LdL \wedge dL_z}{\sqrt{E^2 - V_{m,L}(r)} \sqrt{L^2 - L_z^2 /\sin^2\vartheta}}.
\end{equation}
From now on, it is convenient to introduce the angle $\chi$ defined as $\displaystyle (p_{\hat{2}}, p_{\hat{3}}) = (L/r)\left( \cos\chi, \sin\chi\right)$, which implies that
\begin{equation}
\cos (i) := \frac{L_z}{L} = \sin\vartheta \sin\chi,
\end{equation}
where the angle $i$ is the inclination angle of the particle orbits~\cite{BinneyTremaine-Book,cGoS2022}. The purpose of the introduction of the $\chi$-angle is to re-express the volume form~(\ref{Eq:VolumeFormELLz}) in a compact way as
\begin{equation}
\label{Eq:VolumeFormELchi}
\dvol_x(p) = \frac{1}{r^2} \frac{dE \wedge LdL \wedge d\chi}{\sqrt{E^2 - V_{m,L}(r)}}.
\end{equation}
For this, we establish the domain of parameter space $(E,L,\chi)$ for bounded orbits (see~\cite{pRoS17a,cGoS2023b,cGoS2022}):
\begin{equation}
\label{Eq:Limits2}
E_{\textrm{c}}(r) < E < m, \quad L_{\textrm{c}}(E) \leq L \leq L_{\textrm{max}}(E,r), \quad \hbox{and} \quad 0 \leq \chi \leq 2\pi,
\end{equation}
where $E_{\textrm{c}}$ is the minimum energy at radius $r$, $L_{\textrm{c}}$ is the critical value for the total angular momentum, and $L_{\textrm{max}}$ is the maximum angular momentum allowed at an energy $E$ and radius $r$ defined in equations~(\ref{Eq:MinimumEnergy2}),~(\ref{Eq:AngularMomentumC}), and~(\ref{Eq:MaximumAngularMomentum}) respectively.

Finally, in this section, the explicit form for the particle current density~(\ref{Eq:CurrentDensity}) and the energy-momentum-stress tensor~(\ref{Eq:EnergyMomentumStress}) in terms of conserved quantities for the domain of bounded orbits~(\ref{Eq:Limits}) using the volume form~(\ref{Eq:VolumeFormELchi}) are:
\begin{eqnarray}
\label{Eq:Jhatmu}
J_{\hat{\mu}}(x) &=& \frac{1}{r^2} \sum_{\epsilon_r, \epsilon_\vartheta = \pm 1} \int\limits_{E_{\textrm{c}}(r)}^m \int\limits_{L_{\textrm{c}}(E)}^{L_{\textrm{max}}(E,r)} \int\limits_0^{2\pi} f(x,p) p_{\hat{\mu}}(\epsilon_r,\epsilon_\vartheta) \frac{dE \wedge LdL \wedge d\chi}{\sqrt{E^2 - V_{m,L}(r)}}, \\
\label{Eq:Thatmunu}
T_{\hat{\mu}\hat{\nu}}(x) &=& \frac{1}{r^2} \sum_{\epsilon_r, \epsilon_\vartheta = \pm 1} \int\limits_{E_{\textrm{c}}(r)}^m \int\limits_{L_{\textrm{c}}(E)}^{L_{\textrm{max}}(E,r)} \int\limits_0^{2\pi} f(x,p) p_{\hat{\mu}}(\epsilon_r,\epsilon_\vartheta) p_{\hat{\nu}}(\epsilon_r,\epsilon_\vartheta) \frac{dE \wedge LdL \wedge d\chi}{\sqrt{E^2 - V_{m,L}(r)}},
\end{eqnarray}
where the functions $p_{\hat{\mu}}(\epsilon_r,\epsilon_\vartheta)$ are given by Eq.~(\ref{Eq:Orthonormalbasis}) and the effective potential $V_{m,L}$ is defined in Eq.~(\ref{Eq:SchEffectivePotential}).

From the particle current density, one can determine the invariant particle density and four-velocity of the gas:
\begin{equation}
n := \sqrt{-J^{\hat{\mu}} J_{\hat{\mu}}}, \qquad 
u^{\hat{\mu}} = \frac{1}{n} J^{\hat{\mu}}.
\label{Eq:n}
\end{equation}
From the components of the energy-momentum-stress tensor one can construct the energy density $\varepsilon$ and principal pressures $P_{\hat{r}}$, $P_{\hat{\vartheta}}$, $P_{\hat{\varphi}}$, which can be determined by diagonalizing $T^{\hat{\mu}}{}_{\hat{\nu}}$ \cite{Synge2-Book,oStZ13}. More specifically, $-\varepsilon$ is the eigenvalue of $T^{\hat{\mu}}{}_{\hat{\nu}}$ corresponding to the timelike eigenvector, and $P_{\hat{r}}$, $P_{\hat{\vartheta}}$, $P_{\hat{\varphi}}$ are the eigenvalues belonging to the spacelike ones~\cite{cGoS2023b}. Then, for the analytic form of these fields one needs to assume a model for the one-particle DF in terms of the conserved quantities, which is explained in the following section.


\section{The energy and inclination angle models for the one-particle DF}
\label{Sec:Models}

As mentioned in the previous section, we assume the one-particle DF depending only on the integral of motion~(\ref{Eq:OneParticleDistributionFunction}) for a collisionless relativistic gas consisting of identical massive particles trapped in the effective potential. For the following, we focus on the particular \emph{ansatz} in which the one-particle DF is a product of two functions, one depending only on the energy of the rest-mass particles and the other depending on the inclination angle of the orbits. The energy-dependent part of the distribution function is motivated by the relativistic polytropic ansatz employed in earlier studies~\cite{eAhAaL16,eAhAaL2019,cGoS2023b}, which provides power-law control of the energy weighting. For the inclination dependence, we adopt a profile inspired by~\cite{cGoS2022,BinneyTremaine-Book}. We also examine configurations in which the gas rotates with non-zero angular momentum.

The one-particle distribution function is given by
\begin{equation}
\label{Eq:OneParticleDistrFunct}
F(E,L,L_z) := F_0(E) \times G(i),
\end{equation}
with
\begin{equation}
\label{Eq:PolyDF}
F_0(E) = \alpha \left(1 - \frac{E}{E_0} \right)_+^{k-\frac{3}{2}},
\end{equation}
where $k > 1/2$, $\alpha > 0$, and $E_0 > 0$ are constants. The meaning of the parameter $E_0 \leq m$, introduced in~\cite{eAhAaL16,eAhAaL2019,cGoS2023b}, is a cut-off in energy which provide an upper bound for that quantity, while for $E_0 < m$ the configurations have finite extension, torus-like, as it will be seen in Section~\ref{Sec:Results}. On the other hand, the notation $\mathcal{F}_+$ refers to the positive part of the quantity $\mathcal{F}$, that is $\mathcal{F}_+ = \mathcal{F}$ if $\mathcal{F} > 0$ and $\mathcal{F}_+ = 0$ otherwise. 

For the function, $G(i)$ we will assume the models:
\begin{equation}
\label{Eq:GiModels}
\displaystyle
G(i) := \left\{
\begin{matrix}
G_i^{\textrm{(even)}}\left(\vartheta,\chi\right), \\
& \\
G_{i/2}^{\textrm{(rot)}}\left(\vartheta,\chi\right),
\end{matrix}
\right.\nonumber
\end{equation}
where the superscript even refers to a nonrotating gas configuration while the superscript rot is meaning for a rotating gas cloud with non-zero total angular momentum. These functions are defined as:
\begin{eqnarray}
\label{Eq:GEven}
G_i^{\textrm{(even)}}\left(\vartheta,\chi\right) &:=& \cos^{2s} (i) = \left(\frac{L_z}{L}\right)^{2s} = (\sin\vartheta \sin\chi)^{2s}, \\
\label{Eq:GRot}
G_{i/2}^{\textrm{(rot)}}\left(\vartheta,\chi\right) &:=& \cos^{2s} (i/2) = \frac{1+s}{1+2s}\frac{1}{2^s} \left(1+\frac{L_z}{L}\right)^s = \frac{1+s}{1+2s} \frac{1}{2^s} \left(1+\sin\vartheta\sin\chi\right)^s,
\end{eqnarray}
where $s\geq 0$ is a parameter that controls the concentration of the orbits near the equatorial plane. An illustration of the effect of the parameter $s$ can be seen in figure 1 in~\cite{cGoS2022}.

An important requirement for obtaining finite total-mass configurations is a constraint linking the parameters $k$ and $s$, which respectively govern the energy-dependent and angular-momentum-dependent factors of the distribution function. Relativistic kinetic gas configurations have finite mass \((E_0 \leq m)\) provided that \(2k > s + 7\). Consequently, the values of \(k\) and \(s\) adopted for the profiles that will be shown in chapter~\ref{Sec:Results} are not arbitrary. A more detailed analysis of the interplay between these parameters is given in~\cite{cGoS2023a}. For completeness, Table~\ref{tableI} summarizes some of the parameter choices (values of $k$ and $s$) used in the profiles discussed in Sec.~\ref{Sec:Results}.

\begin{table}[ht]
\centering
\setlength{\tabcolsep}{12pt}
\renewcommand{\arraystretch}{1.15}
\begin{tabular}{@{}c c@{}}
\toprule
$k$ & Allowed values of $s$ \\ \hline
\midrule
5 & 1, 2 \\
\textbf{6} & \textbf{1, 2, 3}, 4 \\
\textbf{7} & \textbf{1, 2, 3}, 4, 5, 6 \\
\textbf{8} & \textbf{1, 2, 3}, 4, 5, 6, 7, 8 \\ \hline
\bottomrule
\end{tabular}
\caption{Some allowed values of parameters $k$, $s$. Values of $k$ and $s$ used to compute the profiles of the spacetime observables in Sec.~\ref{Sec:Results} are highlighted in boldface.}
\label{tableI}
\end{table}

The particle current density covector field~(\ref{Eq:Jhatmu}) and the energy-momentum-stress tensor~(\ref{Eq:Thatmunu}) take the following forms
\begin{eqnarray}
\label{Eq:J2}
J_{\hat{\mu}}(x) &=& \frac{1}{r^2} \sum\limits_{\epsilon_r,\epsilon_\vartheta = \pm 1} \int\limits_{E_{\textrm{c}}(r)}^m F_0(E) dE \int\limits_{L_{\textrm{c}}(E)}^{L_{\textrm{max}}(E,r)} \frac{L}{\sqrt{E^2 - V_{m,L}(r)}} dL \int\limits_0^{2\pi} p_{\hat{\mu}}(\epsilon_r,\epsilon_\vartheta) G(i) d\chi, \\
\label{Eq:T2}
T_{\hat{\mu}\hat{\nu}}(x) &=& \frac{1}{r^2} \sum\limits_{\epsilon_r,\epsilon_\vartheta = \pm 1} \int\limits_{E_{\textrm{c}}(r)}^m F_0(E) dE \int\limits_{L_{\textrm{c}}(E)}^{L_{\textrm{max}}(E,r)} \frac{L}{\sqrt{E^2 - V_{m,L}(r)}} dL \int\limits_0^{2\pi} p_{\hat{\mu}}(\epsilon_r,\epsilon_\vartheta) p_{\hat{\nu}}(\epsilon_r,\epsilon_\vartheta) G(i) d\chi.
\end{eqnarray}

For the \emph{ansatz} of the non-rotating model~(\ref{Eq:GEven}) and DF~(\ref{Eq:OneParticleDistrFunct}) the resulting non-zero components of~(\ref{Eq:J2}) yield
\begin{equation}
J^{\textrm{(even)}}_{\hat{0}}(x) = -\frac{2}{r N(r)} \mathcal{I}_i(\vartheta) \int\limits_{E_{\textrm{c}}(r)}^m dE F_0(E) E  L_{\textrm{max}}(E,r) \sqrt{1 - b(E,r)^2},
\end{equation}
and the non-vanishing components of~(\ref{Eq:T2}) are
\begin{eqnarray}
    T^{\textrm{(even)}}_{\hat{0}\hat{0}}(x) &=& \frac{2}{r N(r)^{3/2}} \mathcal{I}_i(\vartheta) \int\limits_{E_{\textrm{c}}(r)}^m dE F_0(E) E^2 L_{\textrm{max}}(E,r) \sqrt{1 - b(E,r)^2}, \\
    T^{\textrm{(even)}}_{\hat{1}\hat{1}}(x) &=& \frac{2}{3 r^3 \sqrt{N(r)}} \mathcal{I}_i(\vartheta) \int\limits_{E_{\textrm{c}}(r)}^m dE F_0(E) L_{\textrm{max}}(E,r)^3 \left( 1 - b(E,r)^2\right)^{3/2},\\
    T^{\textrm{(even)}}_{\hat{2}\hat{2}}(x) &=& \frac{2}{3 r^3 \sqrt{N(r)}} \frac{1}{s+1}\mathcal{I}_i(\vartheta) \int\limits_{E_{\textrm{c}}(r)}^m dE F_0(E) L_{\textrm{max}}(E,r)^3 \left( 1 + \frac{b(E,r)^2}{2}\right)\sqrt{1 - b(E,r)^2}, \\
    T^{\textrm{(even)}}_{\hat{3}\hat{3}}(x) &=& (2s+1)T^{\textrm{(even)}}_{\hat{2}\hat{2}}(x),
\end{eqnarray}
according to~\cite{cGoS2022}.

Now, for the rotating model~(\ref{Eq:GRot}) the non-vanishing components of~(\ref{Eq:J2}) are
\begin{eqnarray}
    \label{Eq:J0rot}
    J^{\textrm{(rot)}}_{\hat{0}}(x) &=& -\frac{2}{rN(r)} \mathcal{I}_{i/2}(\vartheta) \int\limits_{E_{\textrm{c}}(r)}^m dE F_0(E) E L_{\textrm{max}}(E,r) \sqrt{1-b(E,r)^2}, \\
    \label{Eq:J3rot}
    J^{\textrm{(rot)}}_{\hat{3}}(x) &=& \frac{1}{r^2 \sqrt{N(r)}} \mathcal{I}^\ddag_{i/2}(\vartheta) \int\limits_{E_{\textrm{c}}(r)}^m dE F_0(E) L_{\textrm{max}}(E,r)^2 \left( \frac{\pi}{2} + b(E,r)\sqrt{1-b(E,r)^2} - \arctan\frac{b(E,r)}{\sqrt{1-b(E,r)^2}} \right),
\end{eqnarray}
according to~\cite{cGoS2023b} in which the rotating models have a contribution in the azimuthal component. For this formulation we used the expressions in the appendices~\ref{Appx:A} and~\ref{Appx:B}.

The non-vanishing components of~(\ref{Eq:T2}) in the rotating model are
\begin{eqnarray}
    T^{\textrm{(rot)}}_{\hat{0}\hat{0}}(x) &=& \frac{2}{r N(r)^{3/2}} \mathcal{I}_{i/2}(\vartheta) \int\limits_{E_{\textrm{c}}(r)}^m dE E^2 F_0(E) L_{\textrm{max}}(E,r) \sqrt{1-b(E,r)^2}, \\
    T^{\textrm{(rot)}}_{\hat{1}\hat{1}}(x) &=& \frac{2}{3r^3 \sqrt{N(r)}} \mathcal{I}_{i/2}(\vartheta) \int\limits_{E_{\textrm{c}}(r)}^m dE F_0(E) L_{\textrm{max}}(E,r)^3 \left(1 - b(E,r)^2 \right)^{3/2} , \\
    T^{\textrm{(rot)}}_{\hat{2}\hat{2}}(x) &=& \frac{4}{3 r^3 \sqrt{N(r)}} \mathcal{I}^*_{i/2}(\vartheta) \int\limits_{E_{\textrm{c}}(r)}^m dE F_0(E) L_{\textrm{max}}(E,r)^3 \left( 1 + \frac{b(E,r)^2}{2}\right)\sqrt{1 - b(E,r)^2},
\end{eqnarray}    
\begin{eqnarray}
    T^{\textrm{(rot)}}_{\hat{3}\hat{3}}(x) &=& \frac{4}{3 r^3 \sqrt{N(r)}} \mathcal{I}^\dag_{i/2}(\vartheta) \int\limits_{E_{\textrm{c}}(r)}^m dE F_0(E) L_{\textrm{max}}(E,r)^3 \left( 1 + \frac{b(E,r)^2}{2}\right)\sqrt{1 - b(E,r)^2} \nonumber\\
    &=& \frac{\mathcal{I}^\dag_{i/2}(\vartheta)}{\mathcal{I}^*_{i/2}(\vartheta)} T^{\textrm{(rot)}}_{\hat{2}\hat{2}}(x), \\
    T^{\textrm{(rot)}}_{\hat{0}\hat{3}}(x) &=& \frac{1}{r^2 N(r)} \mathcal{I}^\ddag_{i/2}(\vartheta) \int\limits_{E_{\textrm{c}}(r)}^m dE E F_0(E) L_{\textrm{max}}(E,r)^2 \left( \frac{\pi}{2} + b(E,r)\sqrt{1-b(E,r)^2} - \arctan\frac{b(E,r)}{\sqrt{1-b(E,r)^2}} \right),
\end{eqnarray}
according also with~\cite{cGoS2023b} in which the energy-momentum-stress tensor has a non-zero component in the direction $(\hat{0}\hat{3})$. All integrals over $\chi$-variable which appear in the components of the vector $J$ and tensor $T$ fields are resumed in Appendix~\ref{Appx:A}, however, the notation used in the $\mathcal{I}$-integral is
\begin{equation}
    \mathcal{I}_{\textrm{i-model}}, \nonumber
\end{equation}
with the subscript $(\textrm{i-model})$ corresponds to even ($i$) or rotating model ($i/2$). Also, we are using the function $b$ given by
\begin{equation}
    \label{Eq:b}
    b(E,r) := \frac{L_{\textrm{c}}(E)}{L_{\textrm{max}}(E,r)}.
\end{equation}

For the even model~(\ref{Eq:GEven}) one finds 
\begin{eqnarray}
    \label{Eq:ObservablesEven}
    n^{(\textrm{even})} &=& J^{\hat{0}\textrm{(even)}}, \\
    \label{Eq:ObservablesEvene}
    -\varepsilon^{(\textrm{even})} &=& T^{\hat{0}}{}_{\hat{0}}^{(\textrm{even})}, \\ 
    P^{(\textrm{even})}_{\hat{r}} &=& T^{\hat{1}}{}_{\hat{1}}^{(\textrm{even})}, \\
    P^{(\textrm{even})}_{\hat{\vartheta}} &=& T^{\hat{2}}{}_{\hat{2}}^{(\textrm{even})}, \\
    \label{Eq:ObservablesEvenPphi}
    P^{(\textrm{even})}_{\hat{\varphi}} &=& T^{\hat{3}}{}_{\hat{3}}^{(\textrm{even})}.
\end{eqnarray}
while for the rotating models~(\ref{Eq:GRot}) the existence of the non-diagonal components $(\hat{\mu}\hat{\nu}) = (\hat{0}\hat{3})$ in the energy-momentum-stress tensor allows to construct the energy density $\varepsilon$ and principal pressures $P_{\hat{r}}$, $P_{\hat{\vartheta}}$, $P_{\hat{\varphi}}$, via diagonalization of tensor $T^{\hat{\mu}}{}_{\hat{\nu}}$ as shown in~\cite{Synge2-Book,oStZ13}. Explicitly, in general this yields
\begin{eqnarray}
\label{Eq:ObservablesRot}
n^{(\textrm{rot})} &=& \sqrt{ \left( J^{(\textrm{rot})}_{\hat{0}} \right)^2 - \left( J^{(\textrm{rot})}_{\hat{3}} \right)^2}, \\
\label{Eq:ObservablesRote}
\varepsilon^{(\textrm{rot})} &=& \frac{1}{2}\left[T^{(\textrm{rot})}_{\hat{0}\hat{0}} - T^{(\textrm{rot})}_{\hat{3}\hat{3}} + \sqrt{ \left(T^{(\textrm{rot})}_{\hat{3}\hat{3}} + T^{(\textrm{rot})}_{\hat{0}\hat{0}}\right)^2 - 4 \left(T^{(\textrm{rot})}_{\hat{0}\hat{3}}\right)^2 }\right], \\ 
P^{(\textrm{rot})}_{\hat{r}} &=& T^{(\textrm{rot})}_{\hat{1}\hat{1}}, \\ 
P^{(\textrm{rot})}_{\hat{\vartheta}} &=& T^{(\textrm{rot})}_{\hat{2}\hat{2}}, \\
P^{(\textrm{rot})}_{\hat{\varphi}} &=& \frac{1}{2}\left[ -T^{(\textrm{rot})}_{\hat{0}\hat{0}} + T^{(\textrm{rot})}_{\hat{3}\hat{3}} + \sqrt{ \left(T^{(\textrm{rot})}_{\hat{3}\hat{3}} + T^{(\textrm{rot})}_{\hat{0}\hat{0}}\right)^2 - 4 \left(T^{(\textrm{rot})}_{\hat{0}\hat{3}}\right)^2 }\right].
\label{Eq:ObservablesRotPphi}
\end{eqnarray}
In these results for the observables, when the $(\hat{0}\hat{3})$-component is vanishing, the results are different between the non-rotating and the rotating models due to the integrals $\mathcal{I}_i$ and $\mathcal{I}_{i/2}$ being not equal. See Appendix~\ref{Appx:A} for more details. From the observables computed in this Section, we will review the behavior of these ones graphically in Section~\ref{Sec:Results}. In order to accomplish this, the parameter of amplitude $\alpha$ from which depends the DF must be fixed. In the next Section, we calculate the total mass of the configuration of gas, which allows us to fix $\alpha$, and also, it let us compare between finite mass configurations.


\section{Total particle number, energy and angular momentum}
\label{Sec:TotalMass}

In order to eliminate the dependence of the observables discussed in the last Section from the amplitude parameter $\alpha$ which appears in the energy function~(\ref{Eq:OneParticleDistrFunct}), it is useful to obtain the total mass of the gas cloud, which depends on the same parameter. For this (and the following calculations), it is convenient to introduce the following dimensionless quantities:
\begin{equation}
\label{Eq:Scale}
\xi := \frac{r}{M}, \quad 
\lambda := \frac{L}{m M}, \quad 
\varepsilon := \frac{E}{m}, \quad
\varepsilon_0 := \frac{E_0}{m}, \quad \hbox{and} \quad 
U_\lambda(\xi) := \frac{V_{m,L}(r)}{m^2}.     
\end{equation}

The total particle number of the gas configuration is a conserved quantity associated with the divergence-free current $J^\mu$. Denoting by $\mathcal{S}$ a spacelike Cauchy surface and by $\hat{n}$ the associated future-directed unit normal, the total particle number is given by
\begin{equation}
\mathcal{N}_{\textrm{gas}} = -\int\limits_\mathcal{S} J^\mu \hat{n}_\mu \eta_\mathcal{S},
\label{Eq:Ngas}
\end{equation}
with $\eta_\mathcal{S}$ is the induced volume form on $\mathcal{S}$. This total particle number, in terms of a hypersurface of constant time $t$ and spatial coordinates $(x^1,x^2,x^3)$, yields~\cite{rAcGoS2022}
\begin{equation}
\mathcal{N}_{\textrm{gas}} = \int\limits_{\Real^6} f(x,p_\mu dx^\mu) dx^1 dx^2 dx^3 dp_1 dp_2 dp_3,
\label{Eq:NumberOfParticles2}
\end{equation}
where $x$ is the manifold point with local coordinates $(t,x^1,x^2,x^3)$ and $p_\mu dx^\mu\in P_x^+(m)$ is the momentum covector on the future mass shell with spatial coordinates $(p_1,p_2,p_3)$. On the other hand, is very useful to transform the spatial phase space coordinates $(x^i,p_i)$ to action-angle variables $(\mathcal{Q}^i,\mathcal{J}_i)$ (see~\cite{pRoS2020}). Since the transformation $(x^i, p_i)\mapsto (\mathcal{Q}^i,\mathcal{J}_i)$ is symplectic, the number of particles~(\ref{Eq:NumberOfParticles2}) transforms as
\begin{equation}
\mathcal{N}_{\textrm{gas}} = \int\limits_{\Omega_J}\int\limits_{\Torus^3} f(x,p) 
d\mathcal{Q}^1 d\mathcal{Q}^2 d\mathcal{Q}^3
d\mathcal{J}_1 d\mathcal{J}_2 d\mathcal{J}_3,
\label{Eq:NumberOfParticles3}
\end{equation}
with $\Omega_J\subset\Real^3$ the range of the action variables $\mathcal{J}_i$. In this study the DF $f$ only depends on the integrals of motion and hence only on $\mathcal{J}_i$. Therefore, the integral over the angle variables $\mathcal{Q}^i$ yields $(2\pi)^3$. However, the action variables can be expressed in terms of the conserved quantities $(E,L,L_z)$ in which their transformation is 
\begin{equation}
d\mathcal{J}_1 d\mathcal{J}_2 d\mathcal{J}_3 = \frac{1}{2\pi} T_r(E,L) dE dL dL_z,
\end{equation}
where $T_r(E,L)$ is the period function for the radial motion (see e.g.~\cite{pRoS2020,cGoS2023b} for more details). With this, the total particle number then yields
\begin{equation}
\mathcal{N}_{\textrm{gas}} = 4\pi^2 \int\limits_{\Omega} F(E,L,L_z) T_r(E,L) dE dL dL_z,
\label{Eq:NumberOfParticles4}
\end{equation}
where $\Omega$ is the range for $(E,L,L_z)$ corresponding to $\Omega_J$. As it showed in~\cite{pRoS18} the period function can be expressed in terms of Legendre's elliptic integrals and the roots $r_0 < r_1 < r_2$ of the cubic equation $r^3(E^2 - V_{m,L}(r)) = 0$, with $r_1$ and $r_2$ being the turning points. The period function has the form (cf. Appendix in~\cite{pRoS18} and Section 3.3 in~\cite{cGoS2023b})
\begin{equation}
T_r(\varepsilon,\lambda) = 2 M \varepsilon \left[ \mathbb{H}_2 - \mathbb{H}_0 \right],
\label{Eq:PeriodFunction2}
\end{equation}
where 
\begin{eqnarray}
\mathbb{H}_0 &:=& -\sqrt{\frac{\xi_{012} }{2\xi_1(\xi_2 -\xi_0)}} \left[ (\xi_0 \xi_{012} - \xi_1 \xi_2) \mathbb{F}(\kappa) + \xi_1(\xi_2 - \xi_0 )\mathbb{E}(\kappa) + \xi_{012} (\xi_1 - \xi_0)\Pi\left( b^2, \kappa \right) \right], \\
\mathbb{H}_2 &:=& \sqrt{\frac{8\xi_{012} }{\xi_1(\xi_2 -\xi_0)}} \left[ \frac{\xi_0^2}{\xi_0-2} \mathbb{F}(\kappa) + (\xi_1 -\xi_0)\Pi\left(b^2, \kappa \right) -\frac{4(\xi_1 -\xi_0)}{(\xi_1-2)(\xi_0-2)}\Pi\left(\beta^2, \kappa\right) \right].
\end{eqnarray}
Here, $\mathbb{F}(\kappa)$, $\mathbb{E}(\kappa)$, and $\Pi(b^2,\kappa)$ are Legendre's complete elliptic integrals of the first, second and third kind, respectively, as defined in~\cite{DLMF}. Further, we have abbreviated $\xi_{012} := \xi_0 + \xi_1 + \xi_2$ and have defined
\begin{equation}
b := \sqrt{\frac{\xi_2 -\xi_1}{\xi_2 -\xi_0}},\quad 
\kappa := \sqrt{\frac{\xi_0}{\xi_1}}b, \quad
\beta := \sqrt{\frac{\xi_0 - 2}{\xi_1 - 2}} b.
\label{Eq:bkappabeta}
\end{equation}

Integrating directly on $L_z$ for the models~(\ref{Eq:OneParticleDistrFunct}-\ref{Eq:GRot}) the total particle number~(\ref{Eq:NumberOfParticles4}) is given by
\begin{equation}
\label{Eq:TotalNumberParticles}
\frac{\mathcal{N}_{\textrm{gas}}}{M^3 m^3 \alpha} =  \frac{16 \pi^2}{2s+1} \int\limits_{\varepsilon_{\text{min}}}^1 d\varepsilon \; \varepsilon\left(1-\frac{\varepsilon}{\varepsilon_0} \right)^{k-\frac{3}{2}}_+ \int\limits_{\lambda_{\textrm{c}}(\varepsilon)}^{\lambda_{\textrm{ub}}(\varepsilon)} d\lambda\; \lambda (\mathbb{H}_2-\mathbb{H}_0),    
\end{equation}

for both models. The performing of the integral above is challenging, due to the dependency of the complicated form of the integrand and its limits of integration; however, is very useful to take a change of variables from $(\varepsilon,\lambda)$ to $(p,e)$ which generalizes the semi-latus rectum and eccentricity to the Schwarzschild case (see~\cite{wS02},~\cite{jBmGtH15},~\cite{pRoS18},~\cite{cGoS2022}). These new variables are related to the turning points through
\begin{equation}
\xi_1 = \frac{p}{1+e}, \quad \xi_2 = \frac{p}{1-e}.
\end{equation}
The main advantage of this transformation is the fact that it maps the region of integration to the simpler region $0 < e < 1$ and $p > 6 + 2e$. Furthermore, the third root $\xi_0$ and the dimensionless energy and angular momentum can be expressed explicitly in terms of $(p,e)$ as
\begin{equation}
\xi_0 = \frac{2p}{p-4},\quad
\varepsilon^2 = \frac{(p-2)^2 - 4e^2 }{p\left( p - e^2 - 3\right)},\quad
\lambda^2 = \frac{p^2}{p - e^2 - 3}.
\end{equation}
The resulting integral is then calculated numerically using Wolfram Mathematica~\cite{Mathematica} as follows
\begin{equation}
\label{Eq:TotalNumberParticlesep}
\frac{\mathcal{N}_{\textrm{gas}}}{M^3 m^3 \alpha} = \frac{8\pi^2}{2s+1} \int\limits_{0}^1 de \int\limits_{6+2e}^{\infty} \left(1-\frac{\varepsilon(e,p)}{\varepsilon_0} \right)^{k-\frac{3}{2}}_+ \left(\mathbb{H}_2(e,p)-\mathbb{H}_0(e,p)\right) \dfrac{e\left((p-6)^2-4e^2\right)}{(p-e^2-3)^3}dp.
\end{equation}
Completing the expressions for the total energy and the total angular momentum for the gas cloud, we follow the same way as in the total particle number, and analogously these quantities are defined by:
\begin{equation}
    \label{Eq:EgasJgas}
    \mathcal{E}_{\text{gas}} = \int\limits_S T^{\mu}{}_{\nu} \hat{k}^{\nu} \hat{n}_{\mu}\ \eta_S, \qquad \hbox{and} \qquad
    \mathcal{L}_{\text{gas}} = -\int\limits_S T^{\mu}{}_{\nu} \hat{\ell}^{\nu}  \hat{n}_{\mu}\ \eta_S,
\end{equation}
where $S$ is a Cauchy hypersurface with its respective induced volume form $\eta_S$, $\hat{n}$ its respective future-directed unit vector field, and recalling $\hat{k}$ and $\hat{\ell}$ are the killing time vector and the azimuthal Killing vector of the Schwarzschild metric, respectively. Integrating first in the azimuthal angular momentum variable and using the Keplerian variables, we have that the total energy and angular momentum of the configuration~(\ref{Eq:EgasJgas}) are expressed respectively as:
\begin{eqnarray}
    \label{Eq:Egasep}
    \frac{\mathcal{E}_{\textrm{gas}}}{M^3 m^4 \alpha} &=& \frac{8\pi^2}{2s+1} \int\limits_{0}^1 e\ de \int\limits_{6+2e}^{\infty} \left(1-\frac{\varepsilon}{\varepsilon_0} \right)^{k-\frac{3}{2}}_+ \left(\mathbb{H}_2 -\mathbb{H}_0 \right)\left((p-6)^2-4e^2\right)\left(\dfrac{(p-2)^2-4e^2}{p(p-e^2-3)^7}\right)^{1/2}
    dp, \\
    \label{Eq:Jgasep}
    \frac{\mathcal{L}_{\textrm{gas}}}{M^4 m^4 \alpha} &=& \frac{8\pi^2 s}{(2s+1)(s+2)} \int\limits_{0}^1 de\int\limits_{6+2e}^{\infty} \left(1-\frac{\varepsilon}{\varepsilon_0} \right)^{k-\frac{3}{2}}_+ \left(\mathbb{H}_2 -\mathbb{H}_0 \right) \dfrac{e\left((p-6)^2-4e^2\right)p}{(p-e^2-3)^{7/2}} dp.
\end{eqnarray}
The above expression for the total angular momentum is only non-zero for the rotating case, since the integral over the azimuthal angular momentum $L_z$ is an odd function for any $s$, and this is defined on a symmetric interval, so in the even angle model it is such that the total angular momentum of the system is zero.


\section{Results and comparison}
\label{Sec:Results}

In this section, we provide the results obtained for the profiles of the macroscopic observables for the kinetic gas models explored in this study. We will focus on the particle density obtained from the particle current density vector field and the profiles associated with the macroscopic observables derived from the energy-momentum-stress tensor~(\ref{Eq:ObservablesEven}-\ref{Eq:ObservablesRotPphi}). The resulting configurations derived from the entropy vector field will be explored in~\cite{dMcG2025}. 

The behavior of the observables for different values of the parameters $(k,s,\varepsilon_0)$ are shown in the equatorial plane ($\vartheta=\pi/2$) and in the $xz$-plane. For values of $\varepsilon_0 =1$ the configurations are infinitely extended, while for values of $8/9< \varepsilon_0^2 < 1$ the resulting configurations are finite extended. This range of values of $\varepsilon_0$ is due to the restriction in the values of energy given by (\ref{Eq:EnergyLimits}). The analytical determination of the radii around which finite configurations of kinetic gas correspond to the lower bound on the extension of the gas can be determined by equating the expression for $\varepsilon_c(\xi)$ for $\xi \geq 4$ in~(\ref{Eq:MinimumEnergy2}) with $\varepsilon_0$, while the upper bound is obtained by equating the corresponding expression for $\varepsilon_c(\xi)$ for $ \xi \geq 6$ with $\varepsilon_0$. This guarantees that when calculating the macroscopic observables, the configuration vanishes in the coordinates described above, given by:
\begin{eqnarray}
    \label{Eq:rmin}
    \xi_{-} &=& \frac{8}{-3\varepsilon_0^2 + 4 + \varepsilon_0\sqrt{9\varepsilon_0^2 - 8}}, \\
    \label{Eq:rmax}
    \xi_{+} &=& \frac{4}{3\varepsilon_0^2 - 2 - \varepsilon_0\sqrt{9\varepsilon_0^2 - 8}}.
\end{eqnarray}
By evaluating (\ref{Eq:rmin}) and (\ref{Eq:rmax}) for the values of $\varepsilon_0$ given before, we can observe that the greater the energy cutoff imposed by $\varepsilon_0$, the larger the spatial extent of the gas configuration within a finite region around the black hole.

The profiles associated with the macroscopic observables will be developed both for relativistic kinetic gas configurations that extend asymptotically to infinity and for configurations with a given cutoff energy whose profiles fall within a finite radius dependent on said energy. Since the expressions of the macroscopic spacetime observables explicitly depend on the amplitude $\alpha$ associated with the energy-dependent part in the proposed distribution function model~(\ref{Eq:PolyDF}), in this article we adopt a normalized macroscopic observable were their expressions are divided by the total number of particles $\mathcal{N}_{\textrm{gas}}$ given by the equation~(\ref{Eq:TotalNumberParticlesep}). In this way, a complete description of the profiles can be obtained without needing to specify the value of $\alpha$. With this, the normalized profiles of particle density, energy density, and principal gas pressures for each proposed model respectively are given by:
\begin{equation}
\bar{n} = \frac{M^3}{\mathcal{N}_{\text{gas}}}n, \qquad \bar{\varepsilon} = \frac{M^3}{m\mathcal{N}_{\text{gas}}} \varepsilon, \qquad  \bar{P}_i=\frac{M^3}{m\mathcal{N}_{\text{gas}}}P_i, \qquad \hbox{for} \qquad i=r,\vartheta,\varphi.
\end{equation}

\subsection{Normalized profiles of particle density}
\label{SubSec:Results01}

\begin{figure}[t]
\centerline{
\includegraphics[scale=0.35]{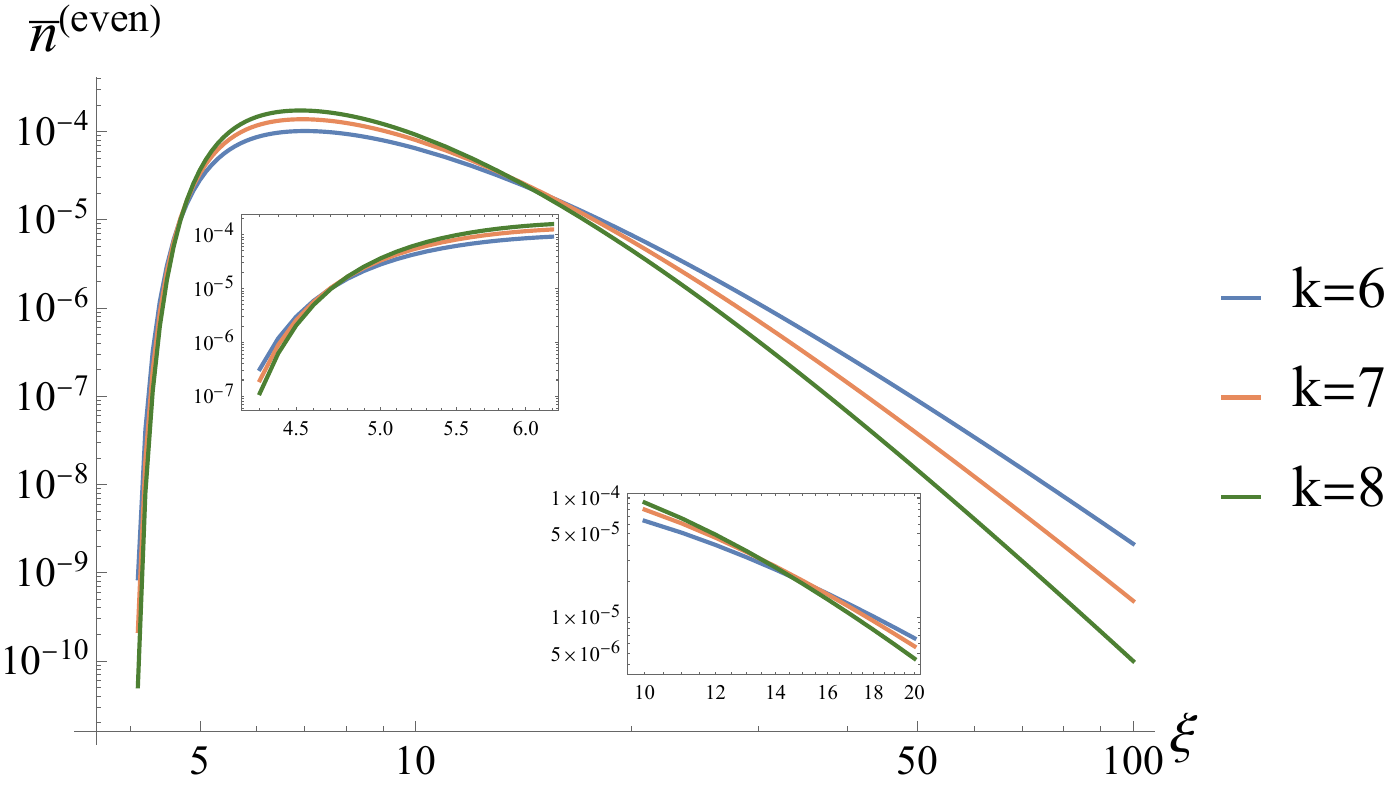}
\includegraphics[scale=0.35]{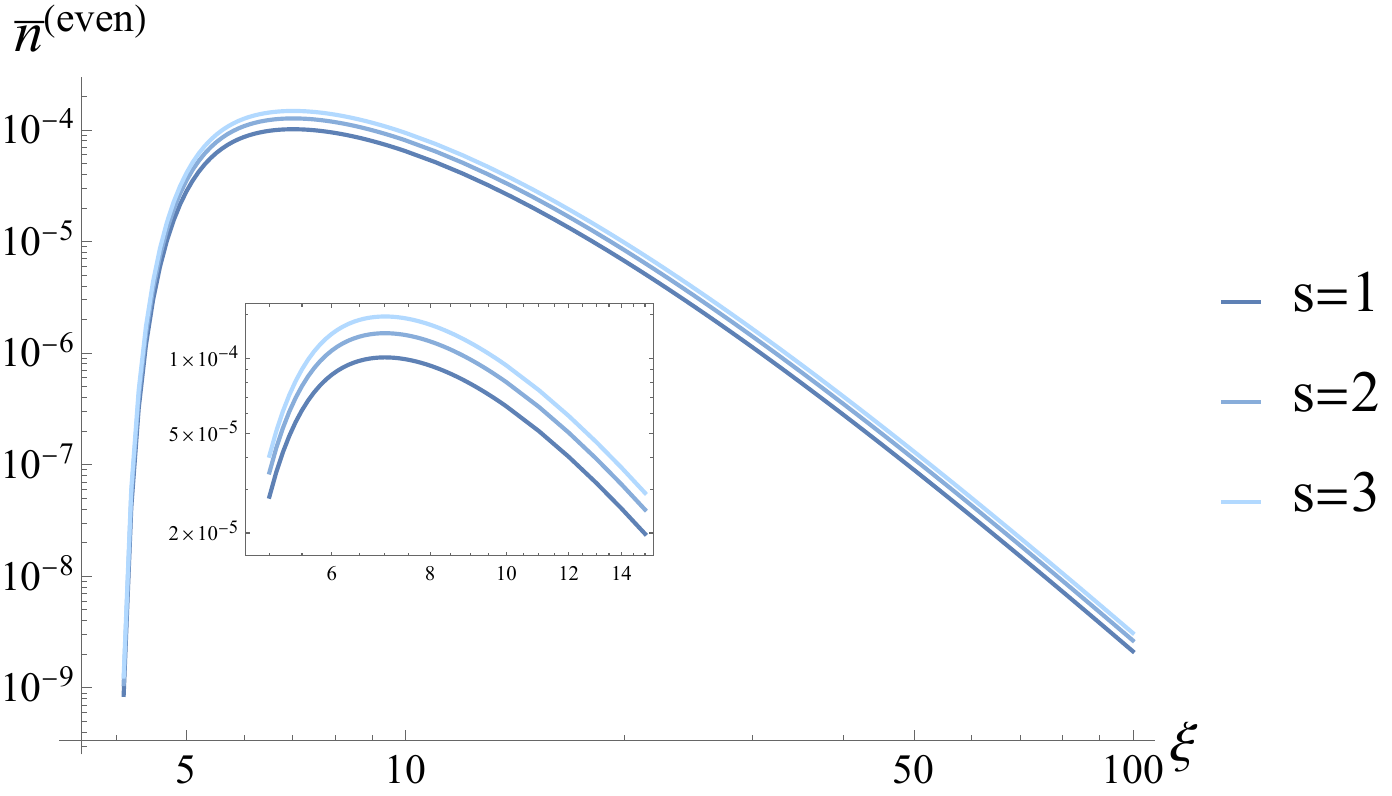}}
\caption{Log-log plot showing the behavior of the  particle density normalized by $\mathcal{N}_{\textrm{gas}}$ as a function of the dimensionless areal radius $\xi$ in the equatorial plane for an asymptotically infinite extended non-rotating configuration of gas. Left panel: plot for different parameter values of $k=6,7,8$ and $(s,\varepsilon_0)=(1,1)$. Right panel: plot for different parameter values of $s=1,2,3$ and $(k,\varepsilon_0)=(6,1)$.}
\label{Fig:neven01}
\end{figure}

\begin{figure}[b]
\centerline{
\includegraphics[scale=0.35]{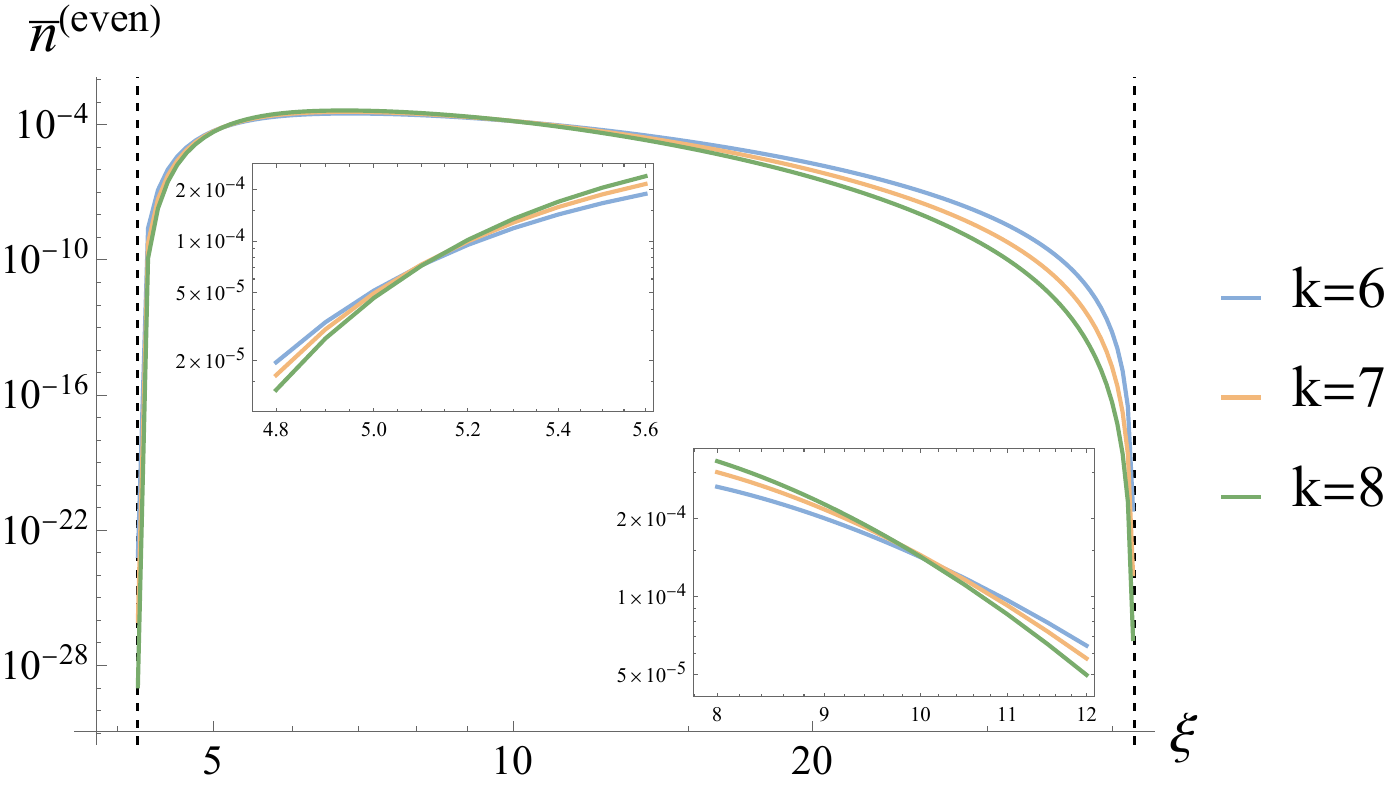}
\includegraphics[scale=0.35]{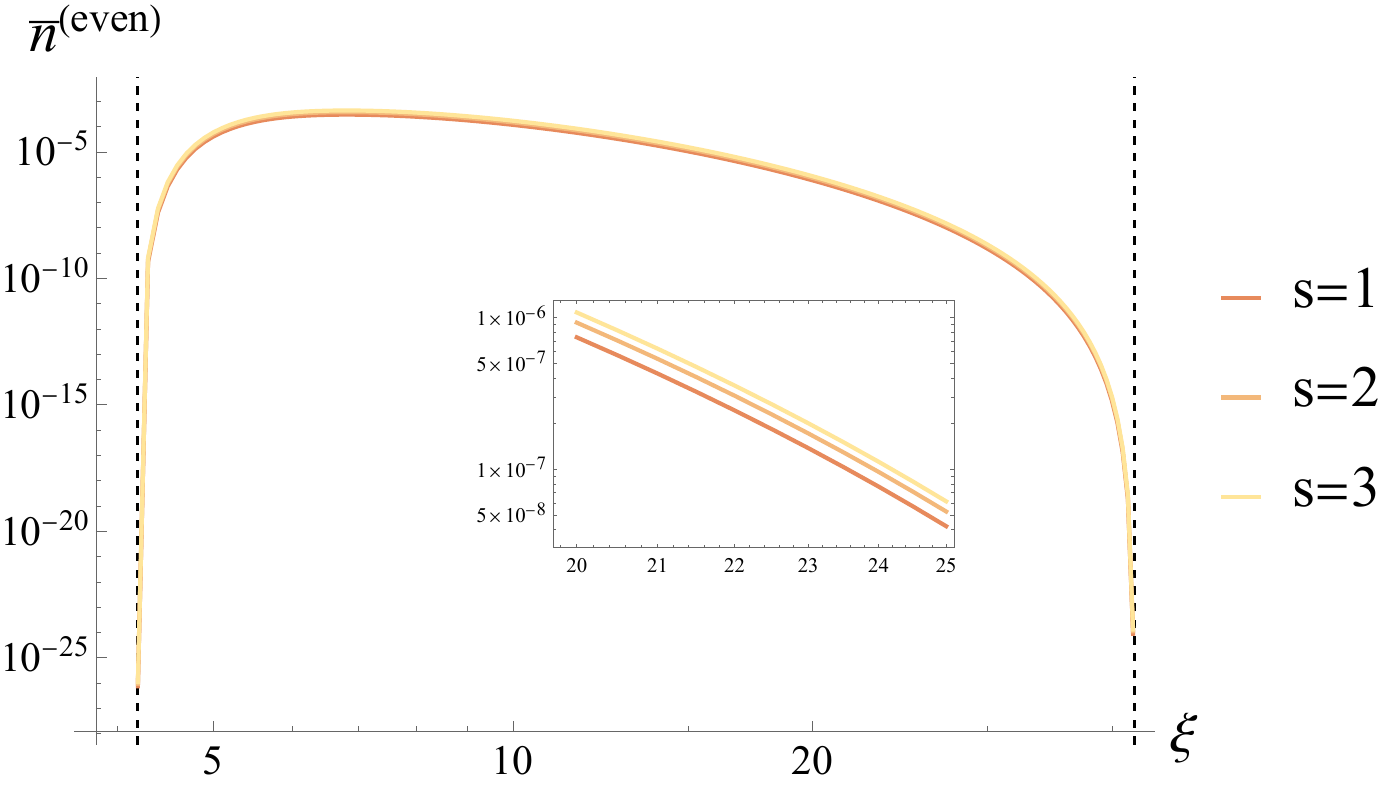}}
\caption{Log-log plot showing the behavior of the particle density normalized by $\mathcal{N}_{\textrm{gas}}$ as a function of the dimensionless areal radius $\xi$ in the equatorial plane for a finite extended non-rotating configuration of gas. Left panel: plot for different parameter values of $k=6,7,8$ and $(s,\varepsilon_0)=(1,0.98)$. Right panel: plot for different parameter values of $s=1,2,3$ and $(k,\varepsilon_0)=(6,0.98)$. The dotted vertical lines represent the radial coordinates where the configuration drops to zero.}
\label{Fig:neven02}
\end{figure}

In this subsection, we explore the behavior of the normalized particle density in both models (even and rot) for different choices of the parameter space shown in table~\ref{tableI}. The figures~\ref{Fig:neven01} show the normalized particle density for different values of the parameters $k$ and $s$ in a gas configuration that tends asymptotically to infinity.
From these graphs, we can see that the larger the value of $s$, the more concentrated the particle density will be in a given region. This is consistent with the proposed model: by increasing or decreasing $s$, the concentration of particles around the black hole's equatorial plane increases or decreases (since the inclination angle of their orbits varies). As $s$ increases, the configuration described above tends to become a disk-like configuration. In the case of varying parameter $k$, it can be noted that the configuration reaches maximum levels of particle density, since the configurations favored by the polytropic \emph{ansatz} are those with minimum energy, but in turn, the particle density drops significantly as the gas configuration moves away from its maximum value as $k$ becomes larger, that is, the configuration decays at long radii, as expected. This occurs regardless of whether the configuration extends asymptotically to infinity, or whether it is bounded due to the cutoff energy $\varepsilon_0$ with finite extension between~(\ref{Eq:rmin}-\ref{Eq:rmax}). The figures~\ref{Fig:neven02} show the normalized particle density for different values of the parameters $k$ and $s$ in a finite gas configuration.
The aforementioned particle density behaviors are similarly exhibited for other values of $k$ and $s$. \\

\begin{figure}[t]
\centerline{
\includegraphics[scale=0.24]{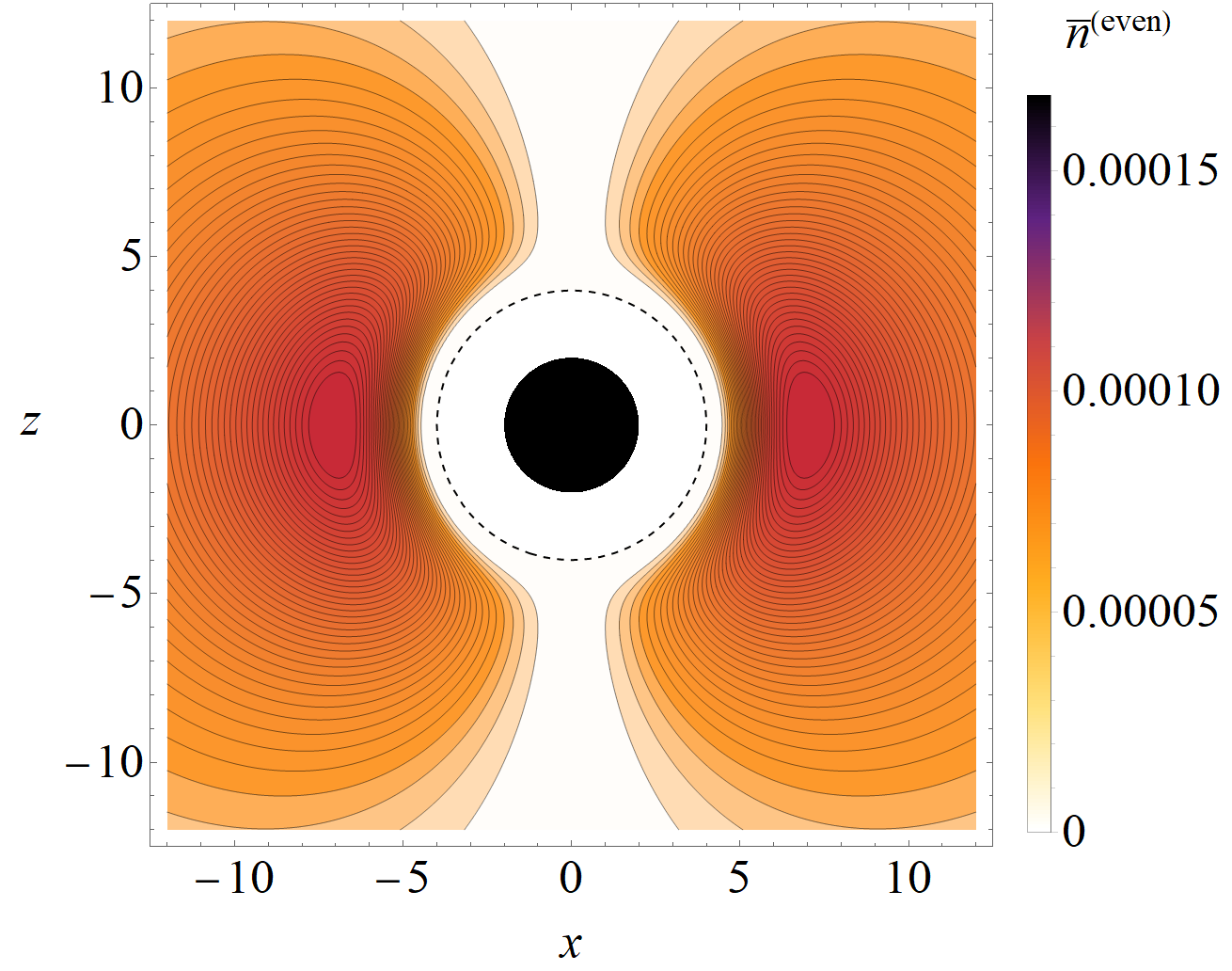}
\includegraphics[scale=0.24]{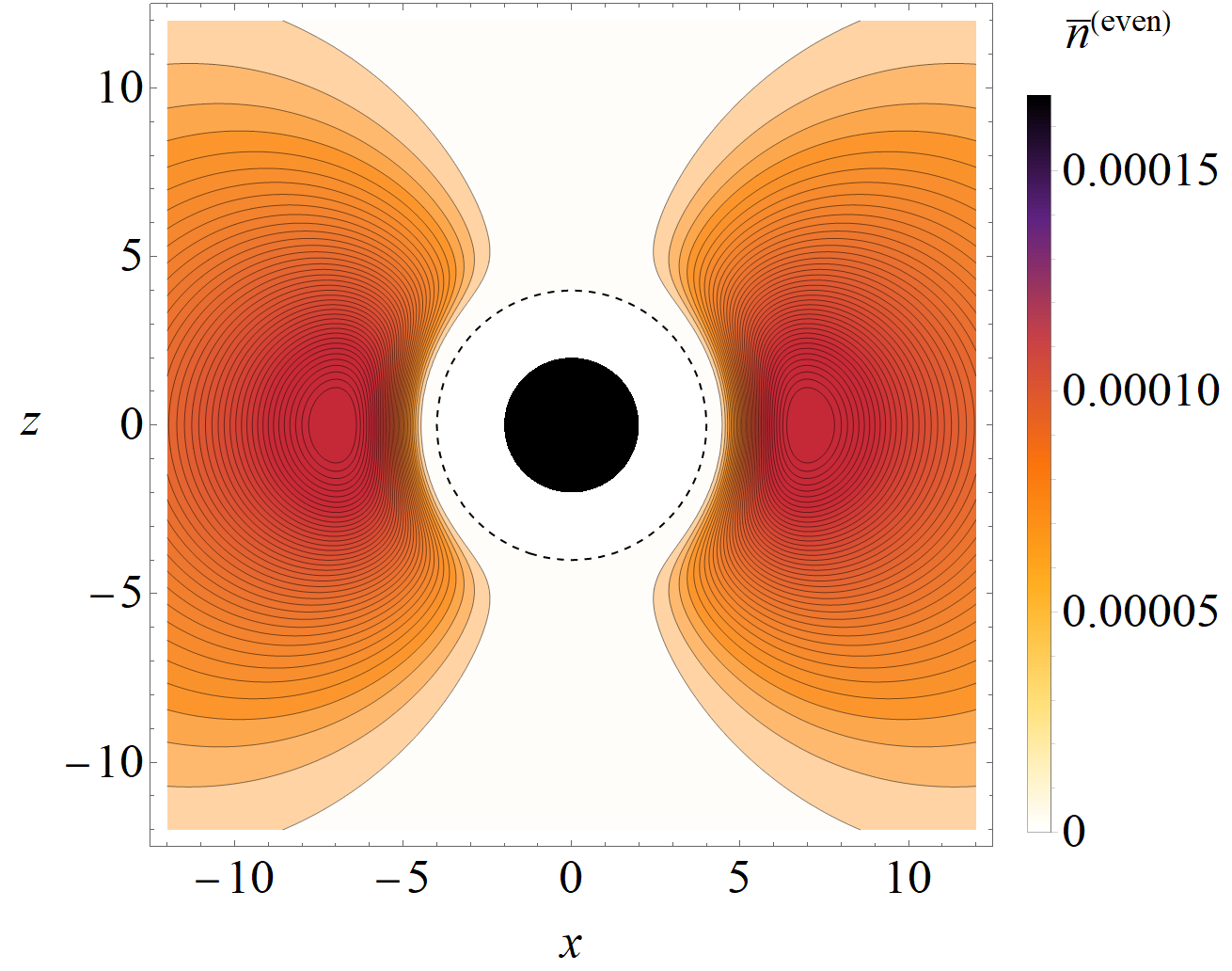}}
\caption{Contour plots of the particle density normalized by $\mathcal{N}_{\textrm{gas}}$ for a fixed value of $k$ and different values of $s$ in a non-rotating kinetic gas configuration of asymptotically infinite extension. Left panel: plot for different parameter values of $(k,s,\varepsilon_0)=(6,1,1)$. Right panel: plot for different parameter values of $(k,s,\varepsilon_0)=(6,2,1)$.}
\label{Fig:neven03}
\end{figure}

\begin{figure}[b]
\centerline{
\includegraphics[scale=0.19]{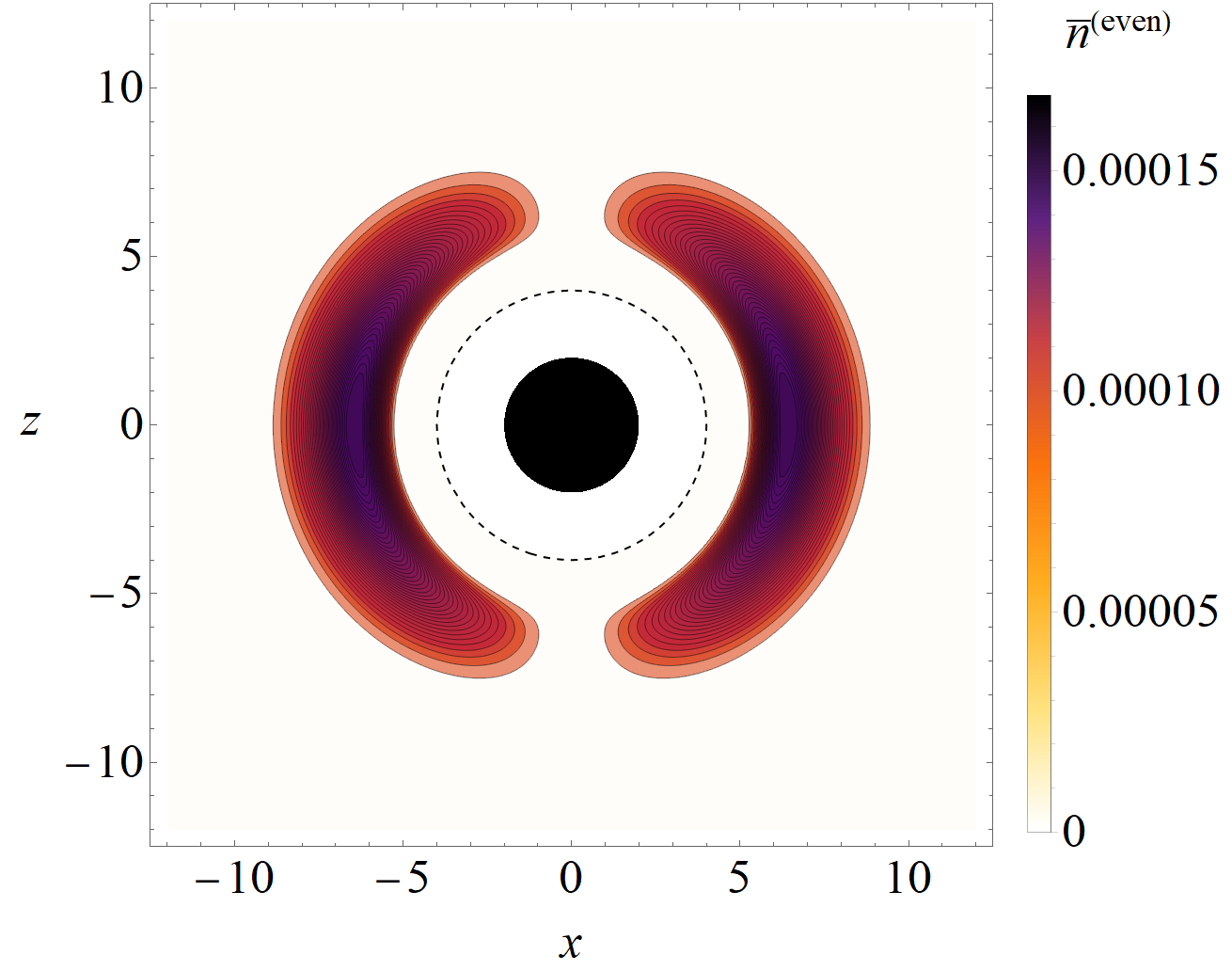}
\includegraphics[scale=0.19]{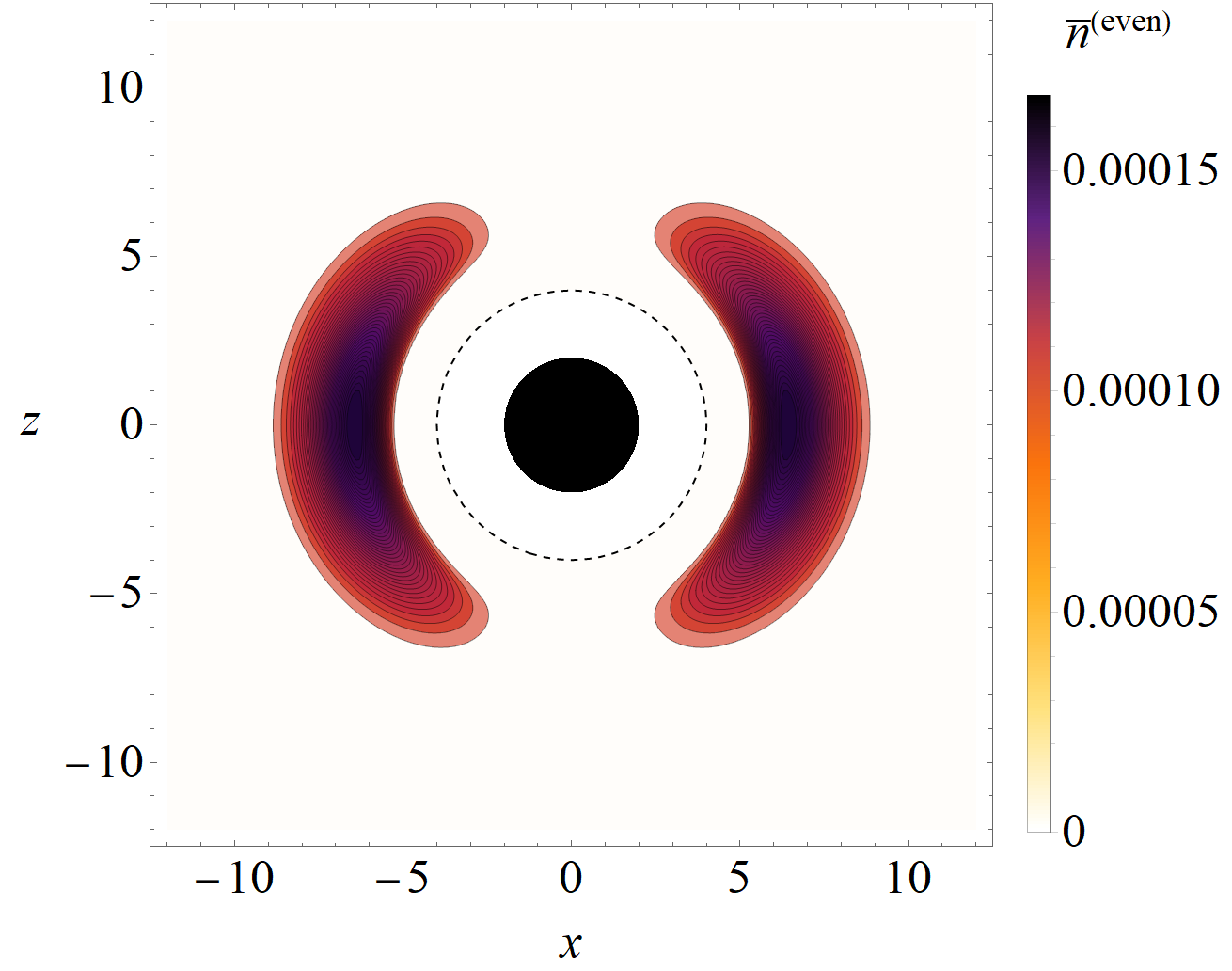}
\includegraphics[scale=0.19]{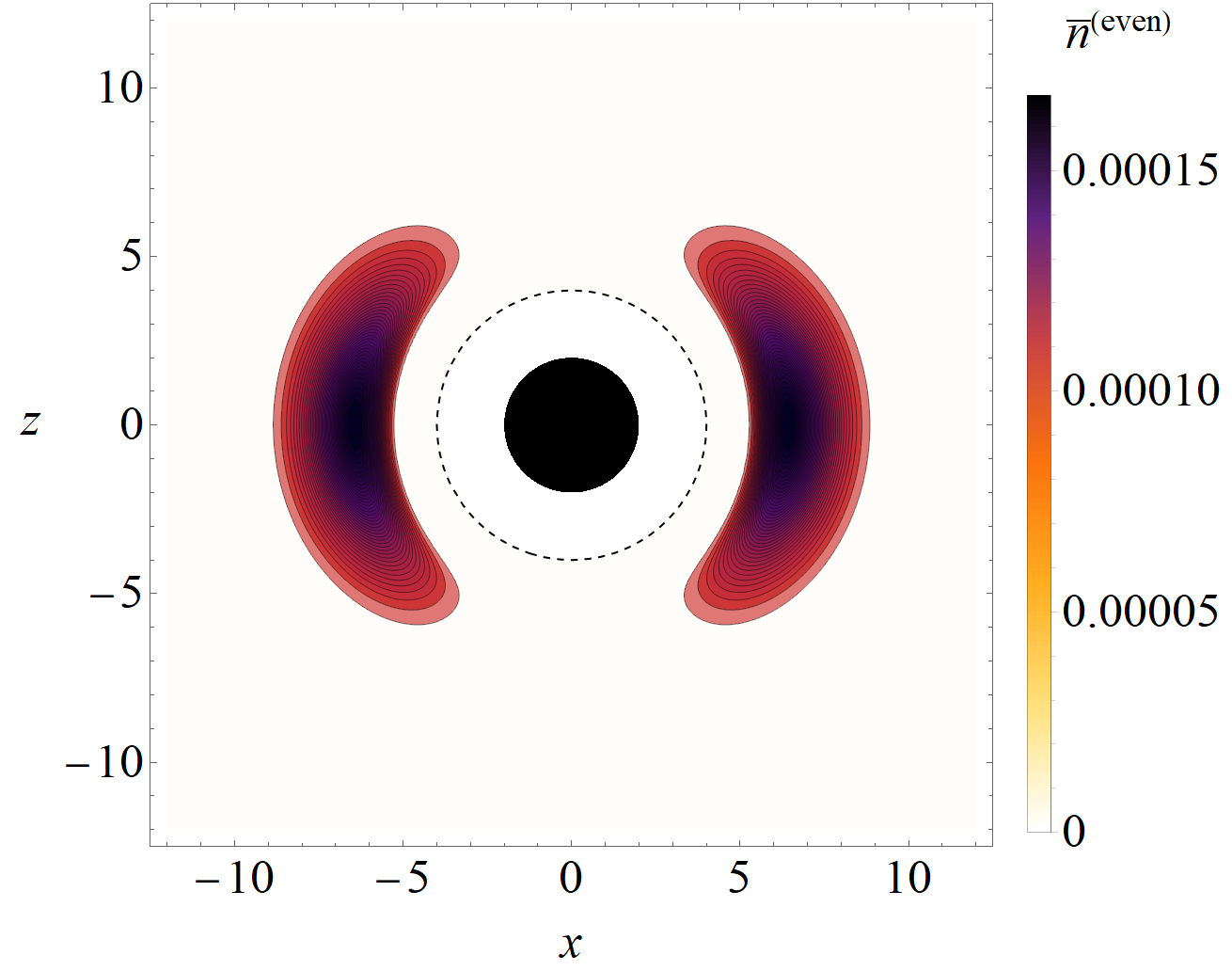}}
\caption{Contour plots of the particle density normalized by $\mathcal{N}_{\textrm{gas}}$ for fixed values of $k$ and $\varepsilon_0$, and different values of $s$ in a non-rotating kinetic gas configuration of finite extension. Left panel: $(k,s,\varepsilon_0)=(6,1,0.95)$. Middle panel: $(k,s,\varepsilon_0)=(6,2,0.95)$. Right panel: $(k,s,\varepsilon_0)=(6,3,0.95)$.}
\label{Fig:neven04}
\end{figure}

To close the discussion on the results for a non-rotating gas configuration, figures~\ref{Fig:neven03} and~\ref{Fig:neven04} show contour plots of the particle density for all $\vartheta$ in an asymptotically infinite (or finite) non-rotating kinetic gas configuration for different values of $s$. The effect of varying the parameter $s$ can be noted as this: the larger the value of $s$, the higher the concentration of particles around the equatorial plane. In all figures in the $xz$-plane the black hole is represented in the center by a black circle and the black dotted circle line represents the minimum region permitted by the bounded trajectories as~\cite{cGoS2023b}. This behavior also occurs analogously for finite extension configurations (though with higher concentrations of particles around the equatorial plane), as shown in figure \ref{Fig:neven04}.

As shown in the previous plots, the configurations like torus or doughnuts are established for an even model in configurations in which the gas is described by relativistic collisionless kinetic gas in a non-rotating model of inclination angle of the particle's orbit. 

\begin{figure}[t]
\centerline{
\includegraphics[scale=0.35]{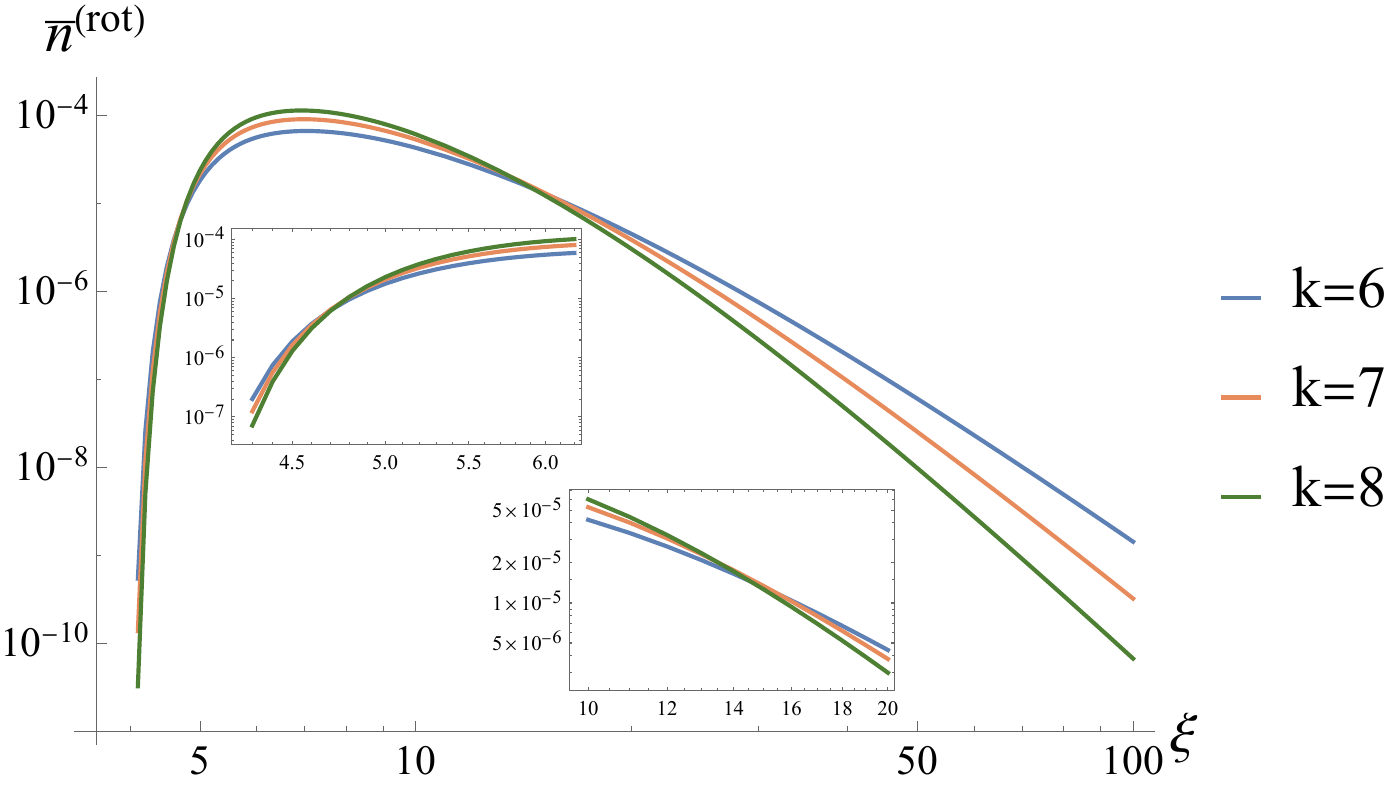}
\includegraphics[scale=0.35]{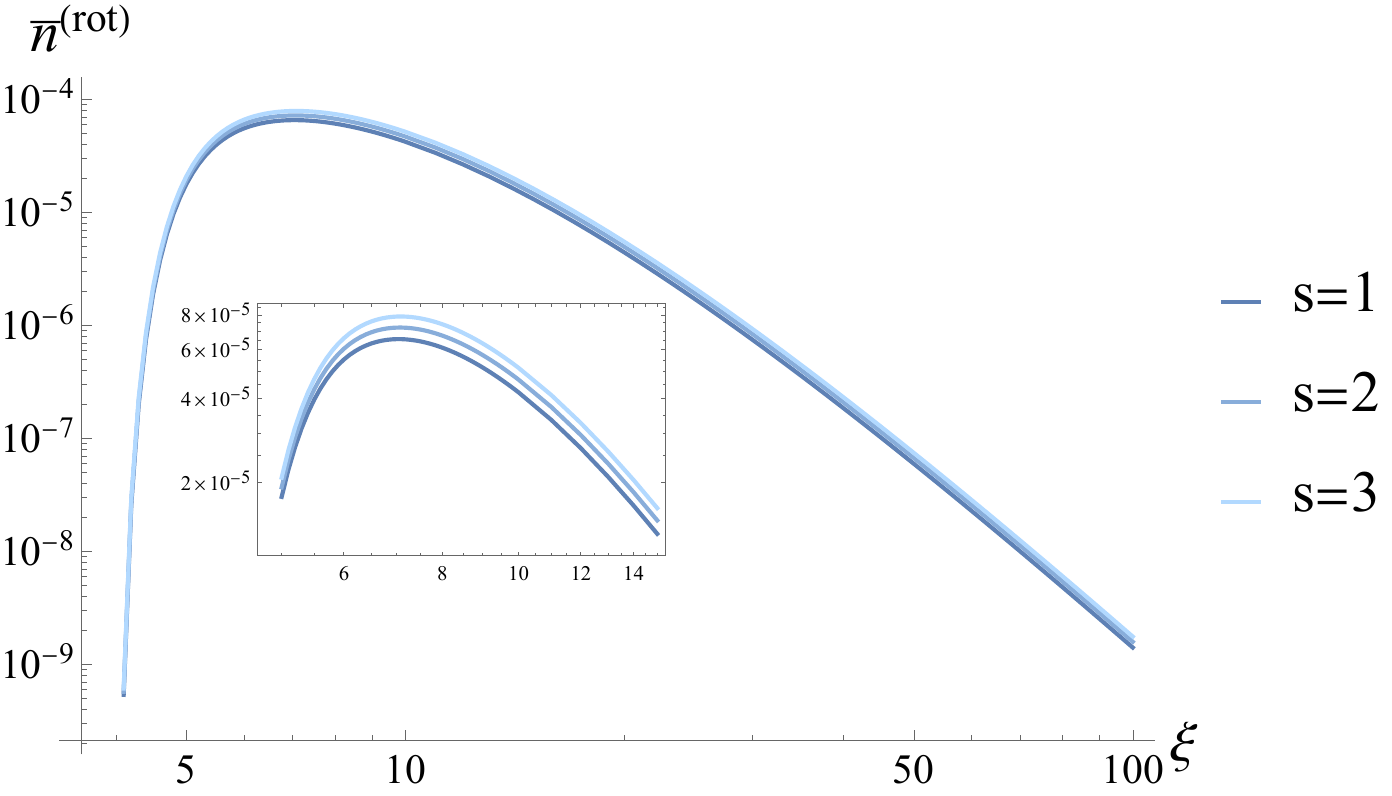}}
\caption{Log-log plot showing the behavior of the  particle density normalized by $\mathcal{N}_{\textrm{gas}}$ as a function of the dimensionless areal radius $\xi$ in the equatorial plane for an asymptotically infinite extended rotating configuration of gas. Left panel: plot for different parameter values of $k=6,7,8$ and $(s,\varepsilon_0)=(1,1)$. Right panel: plot for different parameter values of $s=1,2,3$ and $(k,\varepsilon_0)=(6,1)$.}
\label{Fig:nrot01}
\end{figure}

Now, we show the behavior of the rotating counterpart and explain some differences in the resulting configurations. For this, the figures~\ref{Fig:nrot01} show the normalized particle density for different values of the parameters $k$ and $s$ in a gas configuration that tends asymptotically to infinity.

\begin{figure}[H]
\centerline{
\includegraphics[scale=0.35]{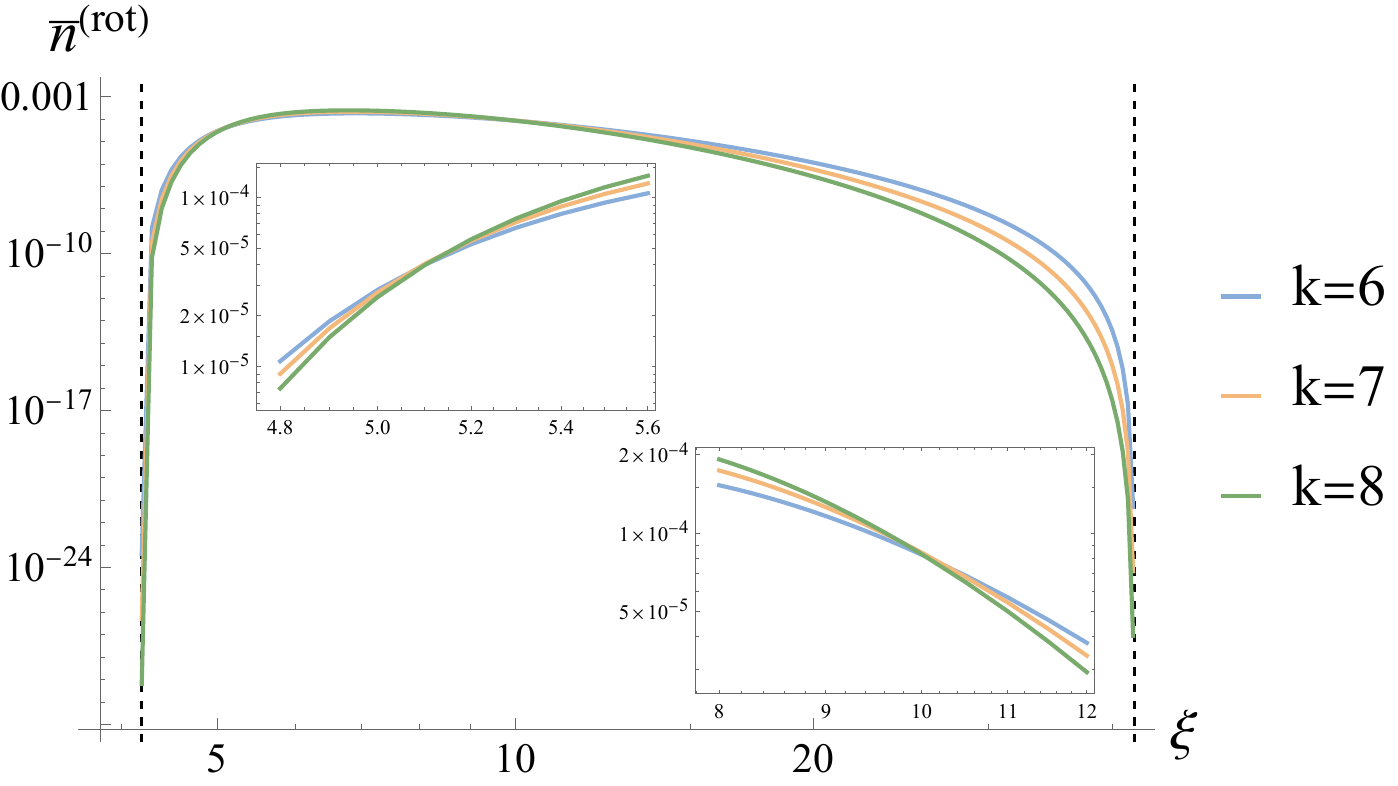}
\includegraphics[scale=0.35]{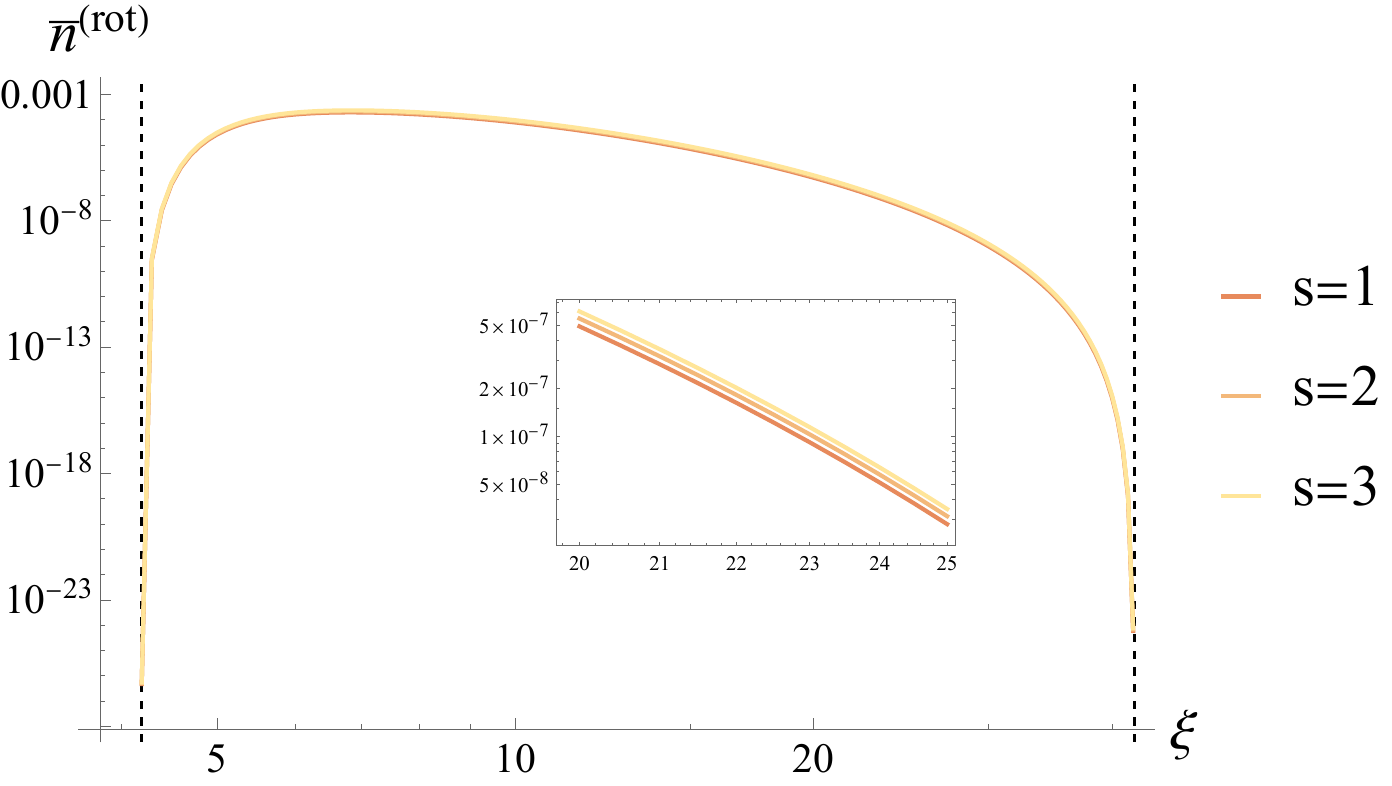}}
\caption{Log-log plot showing the behavior of the particle density normalized by $\mathcal{N}_{\textrm{gas}}$ as a function of the dimensionless areal radius $\xi$ in the equatorial plane for a finite extended rotating configuration of gas. Left panel: plot for different parameter values of $k=6,7,8$ and $(s,\varepsilon_0)=(1,0.98)$. Right panel: plot for different parameter values of $s=1,2,3$ and $(k,\varepsilon_0)=(6,0.98)$. The dotted vertical lines represent the radial coordinates where the configuration drops to zero.}
\label{Fig:nrot02}
\end{figure}

For the rotating gas, the behavior is the same as in the non-rotating counterpart: the greater the values of $s$, the more concentrated the particle density is. The figures~\ref{Fig:nrot02} show the normalized particle density for different values of the parameters $k$ and $s$ in a finite gas configuration.

.

The behavior of the particle density is similar for other values of $k$ and $s$. \\

\begin{figure}[b]
\centerline{
\includegraphics[scale=0.24]{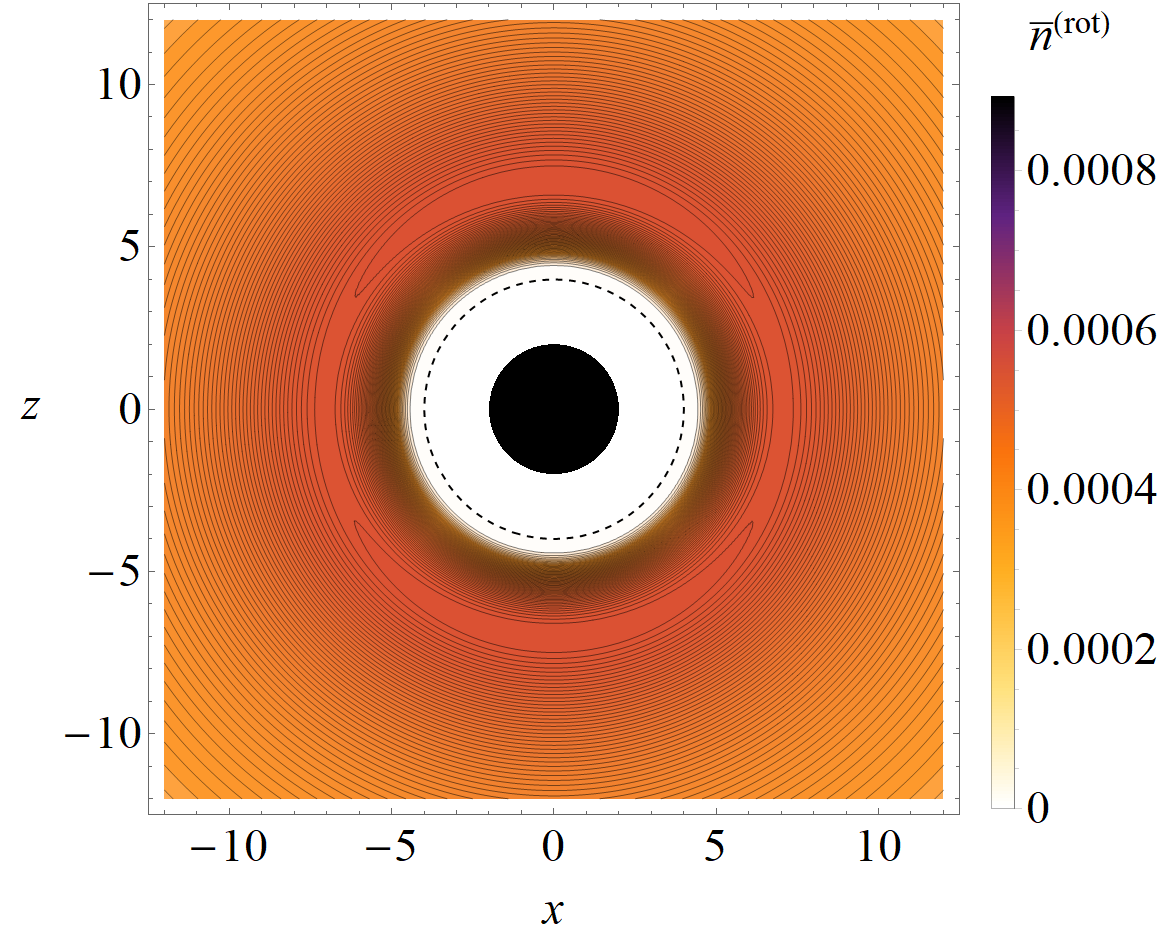}
\includegraphics[scale=0.24]{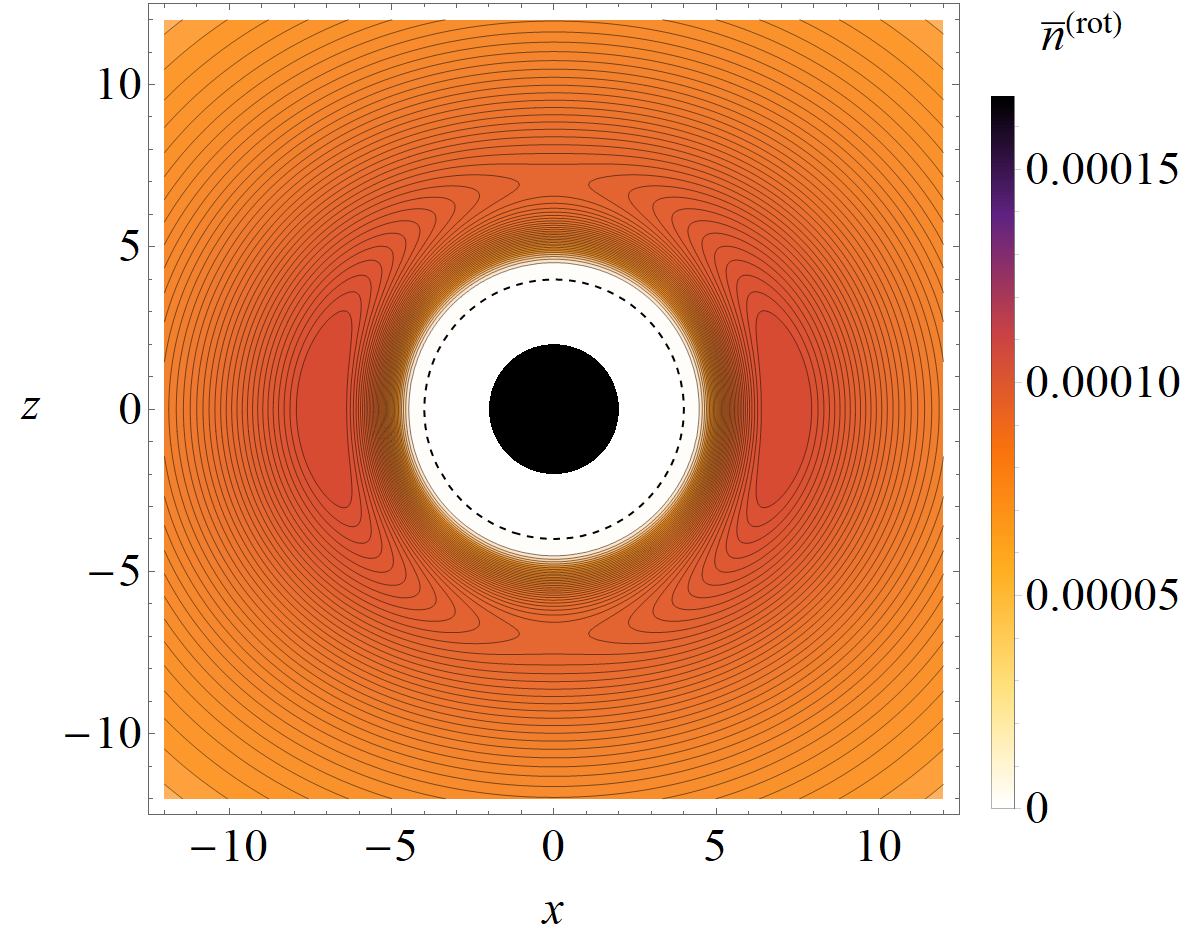}}
\caption{Contour plots of the particle density normalized by $\mathcal{N}_{\textrm{gas}}$ for a fixed value of $k$ and different values of $s$ in a rotating kinetic gas configuration of asymptotically infinite extension. Left panel: plot for different parameter values of $(k,s,\varepsilon_0)=(6,1,1)$. Right panel: plot for different parameter values of $(k,s,\varepsilon_0)=(6,2,1)$.}
\label{Fig:nrot03}
\end{figure}

\begin{figure}[h!]
\centerline{
\includegraphics[scale=0.19]{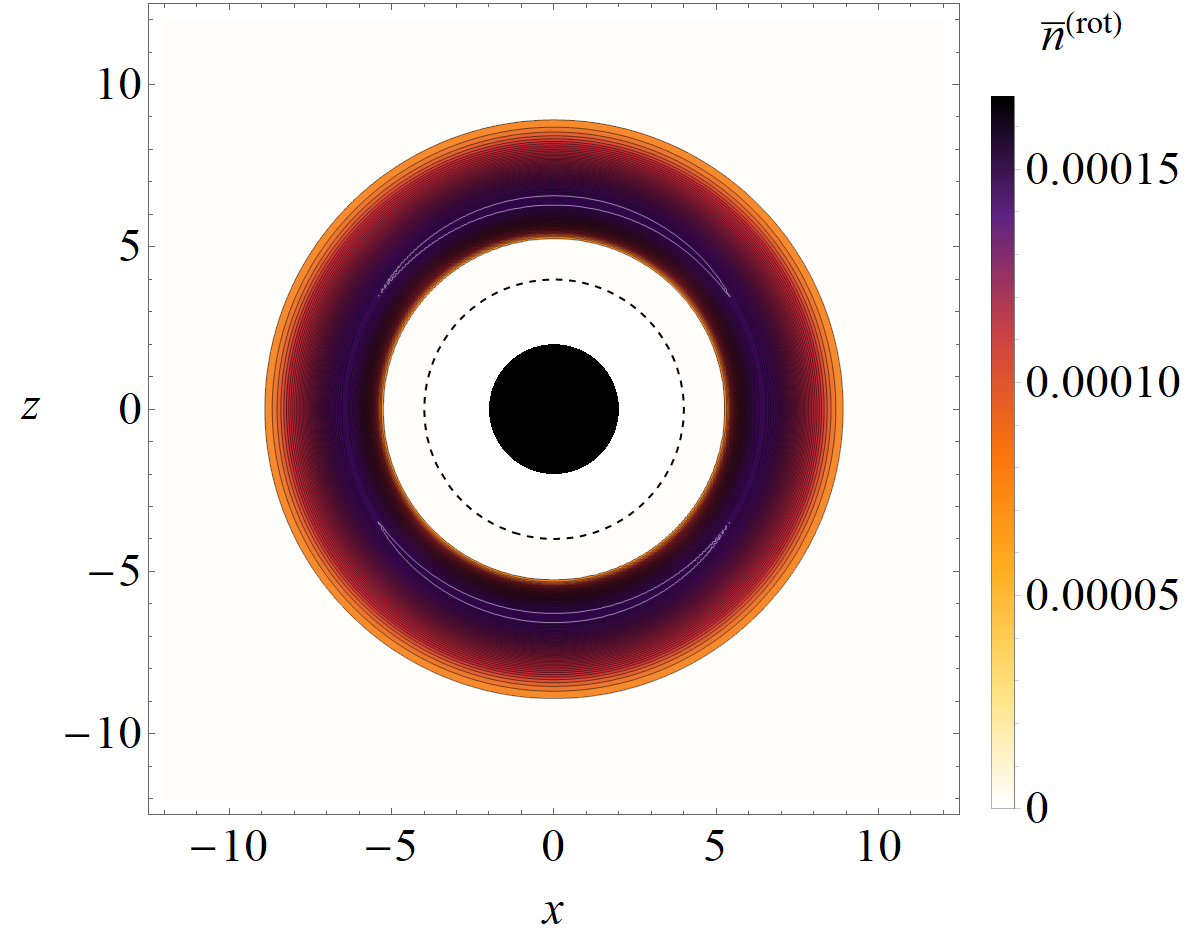}
\includegraphics[scale=0.19]{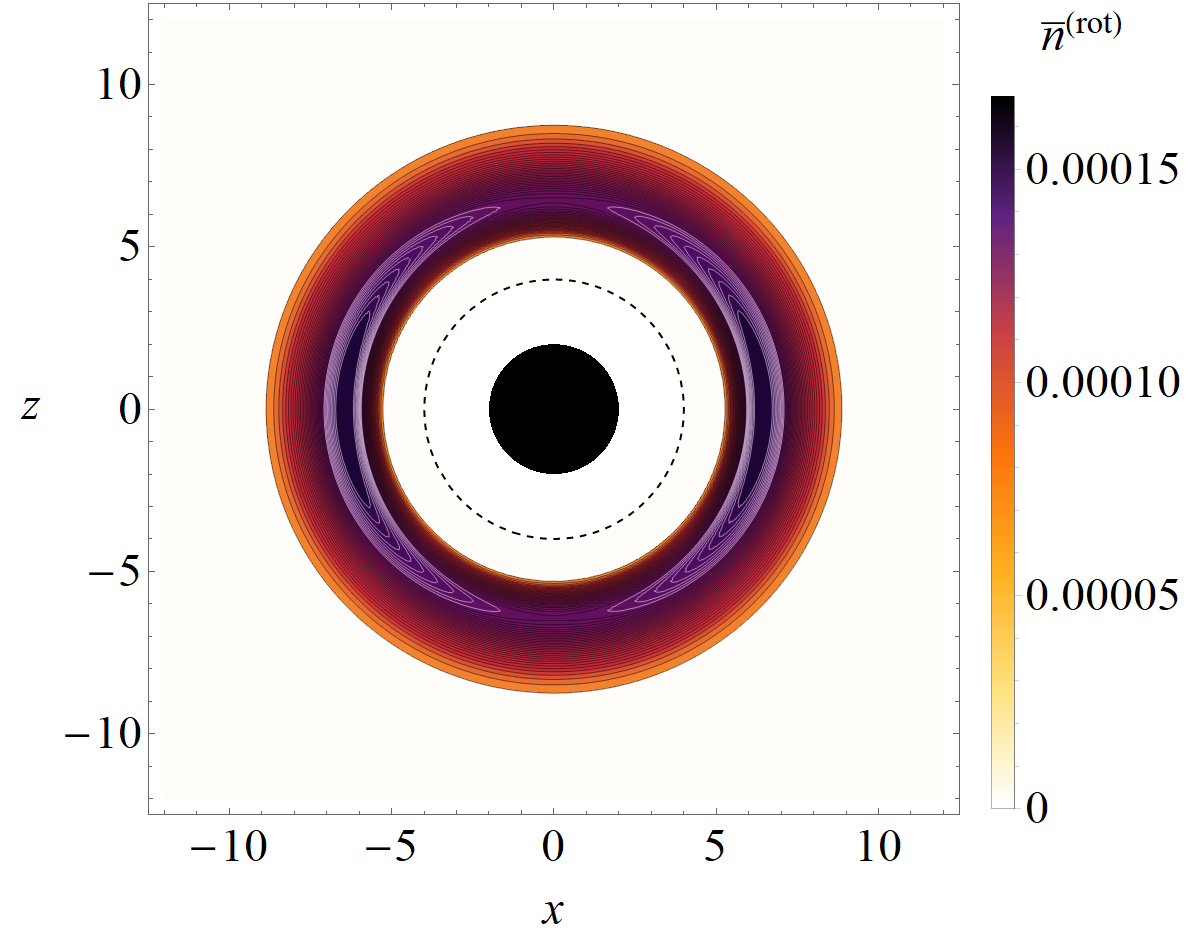}
\includegraphics[scale=0.19]{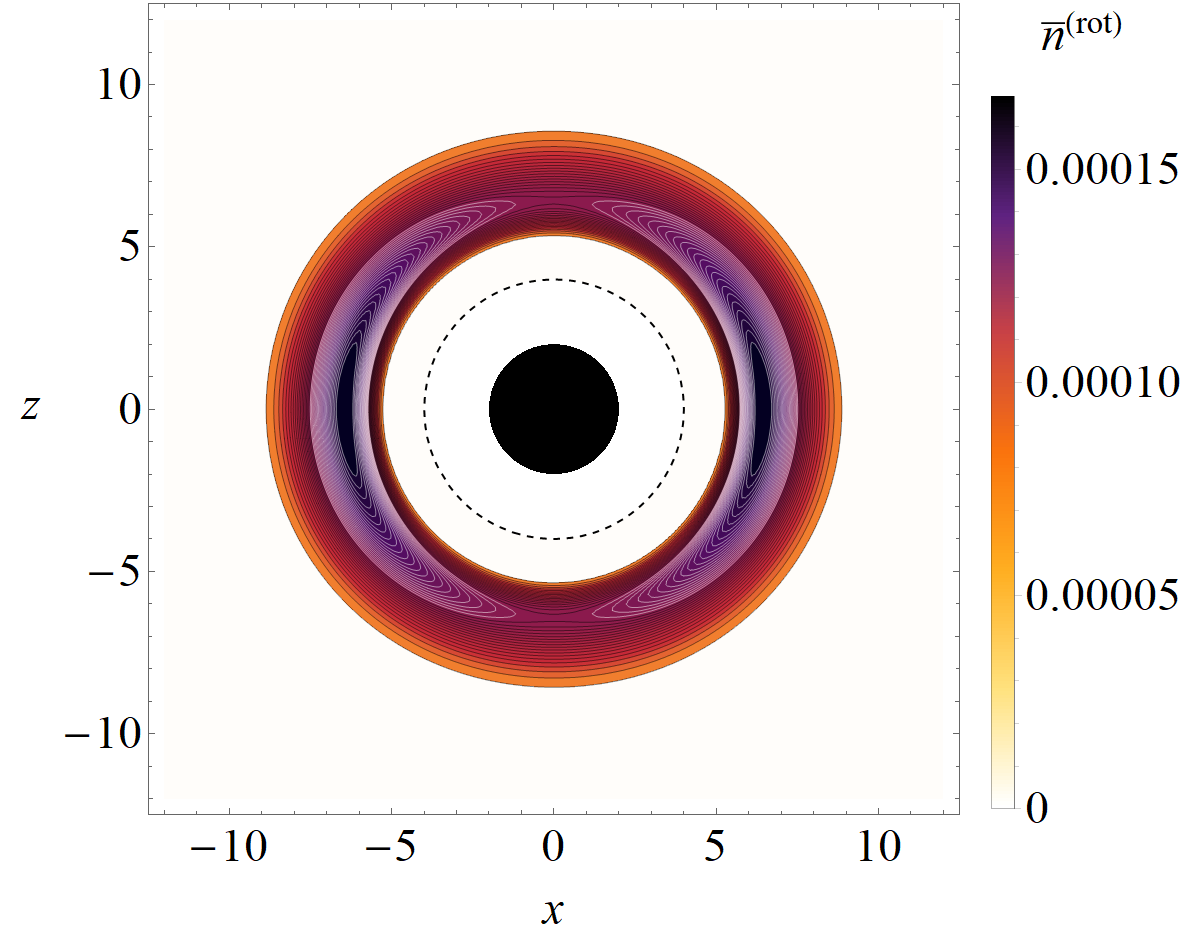}}
\caption{Contour plots of the particle density normalized by $\mathcal{N}_{\textrm{gas}}$ for fixed values of $k$ and $\varepsilon_0$, and different values of $s$ in a rotating kinetic gas configuration of finite extension. Left panel: $(k,s,\varepsilon_0)=(6,1,0.95)$. Middle panel: $(k,s,\varepsilon_0)=(6,2,0.95)$. Right panel: $(k,s,\varepsilon_0)=(6,3,0.95)$.}
\label{Fig:nrot04}
\end{figure}

\begin{figure}[b]
\centerline{
\includegraphics[scale=0.25]{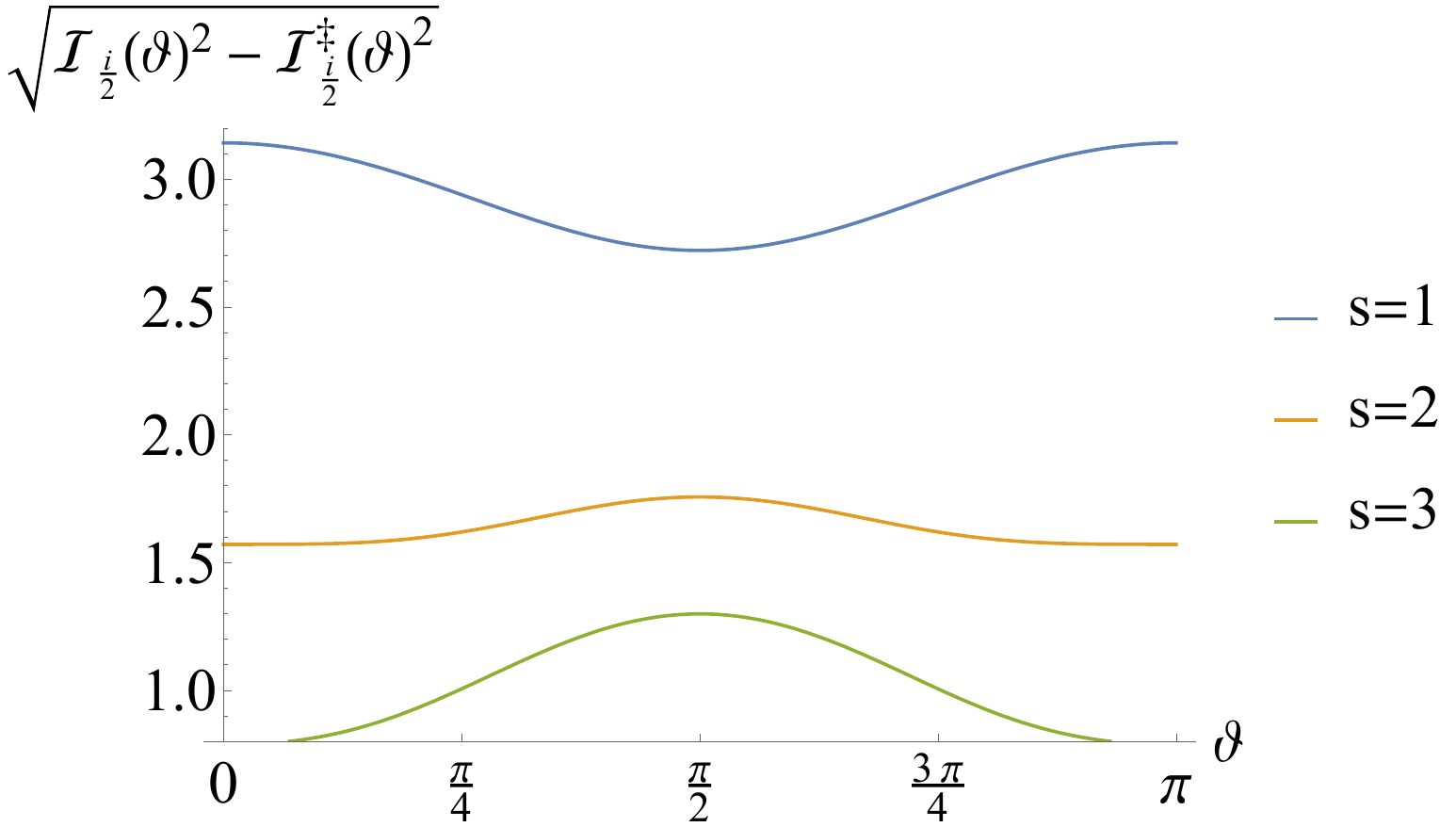}
\includegraphics[scale=0.25]{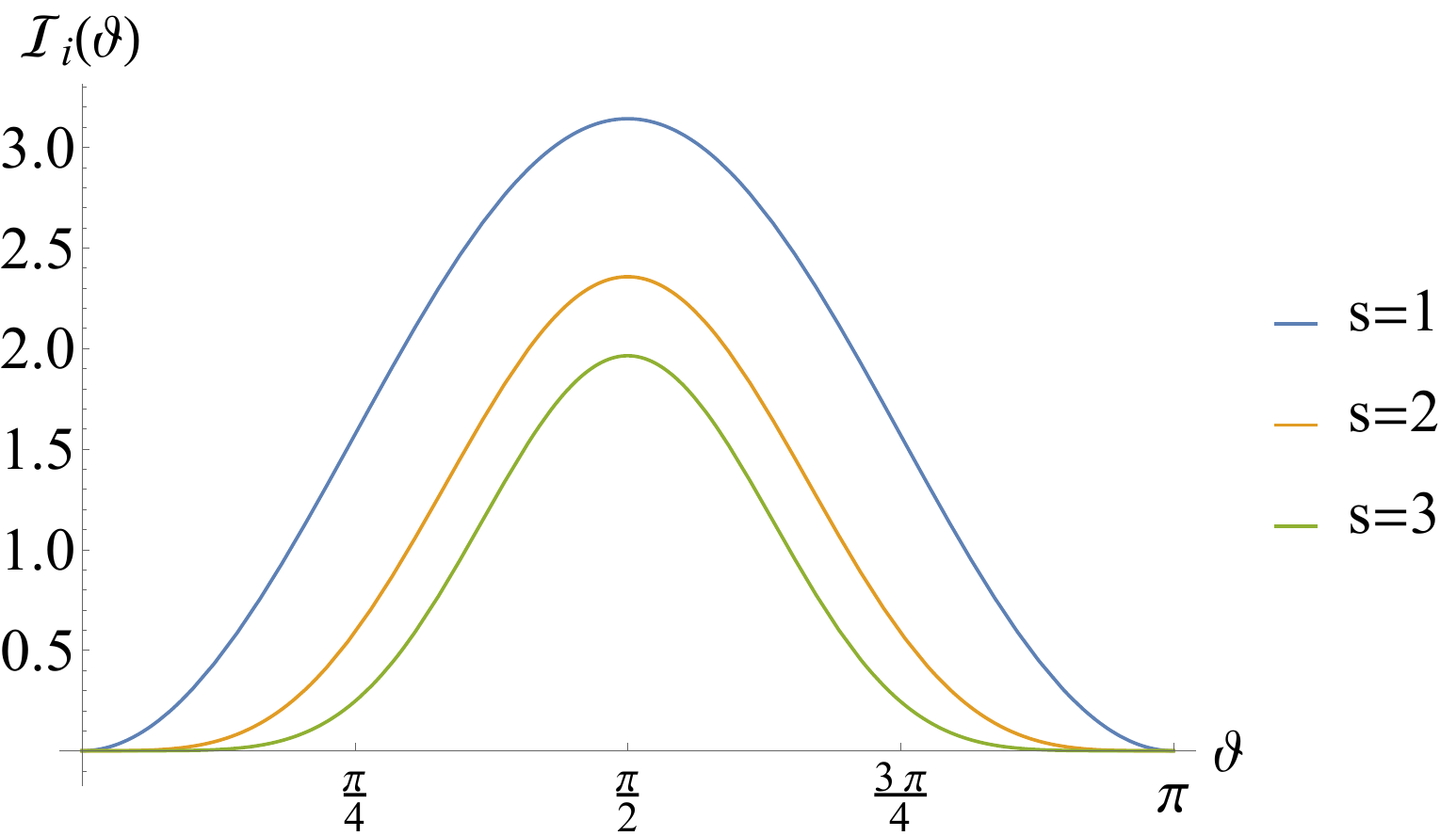}}
\caption{Plots of the angular part of the particle density for both kinetic gas models. Left panel: plot of the function~(\ref{Eq:WFunction}) (setting $W_0 = W_3=1$) for the particle density in the rotating model. Right panel: plot of the function~(\ref{Eq:I01}) for the particle density in the non-rotating model.}
\label{Fig:ifigure01}
\end{figure}

The figures~\ref{Fig:nrot03} and~\ref{Fig:nrot04} show contour plots of the particle density for all $\vartheta$ in an asymptotically infinite (or finite) rotating kinetic gas configuration for different values of $s$.
As showed in the previous plots, the configurations like torus or doughnuts are established for a rotating model in configurations in which the gas is described by relativistic collisionless kinetic gas and these are very similar to their non-rotating counterparts, however some differences can be appreciated between models (see left panel of figure~\ref{Fig:neven04} and left panel of figure~\ref{Fig:nrot04}). In these figures, we notice that the rotating configuration is more compact at the polar axis $\vartheta=0,\pi$, in contrast to the non-rotating configuration, which is always more compact at the equatorial plane $\vartheta=\pi/2$. This difference is an effect of the model proposed for the angular part of the distribution function~(\ref{Eq:OneParticleDistrFunct}). By replacing~(\ref{Eq:J0rot}) and~(\ref{Eq:J3rot}) into the particle density $n^{\textrm{(rot)}}$, it exhibits the form:
\begin{equation} 
\sqrt{W_0(r,\varphi)[\mathcal{I}_{i/2}(\vartheta)]^2-W_3(r,\varphi)[\mathcal{I}^{\ddag}_{i/2}(\vartheta)]^2} 
\label{Eq:WFunction}
\end{equation}
where $W_0$ and $W_3$ are weighting functions accompanying the angular functions described above. These functions $\mathcal{I}_{i/2}$ and $\mathcal{I}^{\ddag}_{i/2}$ are defined in the Appendix~\ref{Appx:A}, equations~(\ref{Eq:App01}) and~(\ref{Eq:App02}) respectively. The figure~\ref{Fig:ifigure01} shows graphs of~(\ref{Eq:WFunction}) for different values of $s$, setting $W_0(r,\varphi) = W_3(r,\varphi)=1$ for simplicity. The behavior of the function corresponding to $s=1$ is opposite to that of the functions corresponding to $s=2$ and $s=3$, which justifies the $\pi/2$ radian rotation in the particle concentration seen in the left panel of the figure~\ref{Fig:nrot04}, relative to the concentrations in the middle and right panels in the same figure. In contrast, in the angular part of the particle density for the non-rotating model we have that for all values of $s$, the $\vartheta$-dependent functions are in phase, therefore all particle concentrations in the right panel of the figure~\ref{Fig:nrot04} are around the equatorial plane of the black hole.

\begin{figure}[t]
\centerline{
\includegraphics[scale=0.25]{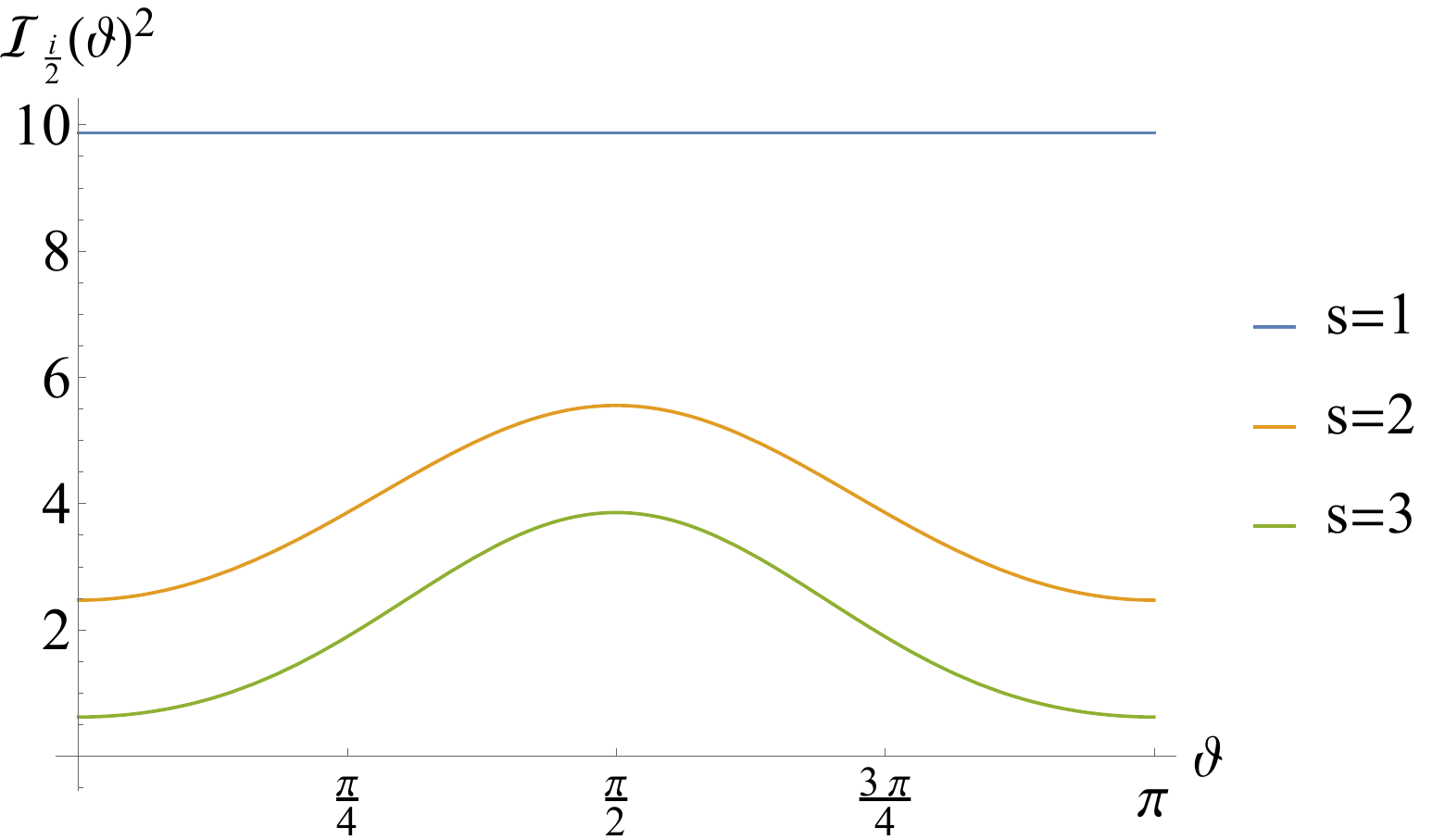}}
\caption{Graphs of the angular part of $J_{\hat{0}}^{\textrm{(rot)}}$ for different values of $s$.}
\label{Fig:ifigure02}
\end{figure}

In an effort to explain the anomalous behavior of the particle density for the rotating model, we can analyze figure~\ref{Fig:ifigure02}, where it can be noted that the angular function for $s=1$ of $J_{\hat{0}}^{\textrm{(rot)}}$ is a constant, unlike the other angular functions for other values of $s$. Since the particle density $n^{\textrm{(rot)}}$ is the subtraction of two quadratic terms within a square root, the term $J_{\hat{0}}^{\textrm{(rot)}}$ must predominate over $J_{\hat{3}}^{\textrm{(rot)}}$. Evaluating (\ref{Eq:ObservablesRot}) for $s=1$ yields 
\begin{equation}
    \left. n^{\textrm{(rot)}}\right|_{s=1}=\sqrt{ \left[ J^{(\textrm{rot})}_{\hat{0}}(r) \right]^2 - \left[ J^{(\textrm{rot})}_{\hat{3}}(r,\vartheta) \right]^2},
\end{equation}
One observes that $J_{\hat{0}}$ depends only on the radial coordinate $r$, whereas $J_{\hat{3}}$ depends on both $r$ and the polar angle $\vartheta$. Because the $\vartheta$-dependence enters through $\sin\vartheta$ and appears quadratically, $n^{\textrm{(rot)}}$ is even in $\vartheta$.
This results in the phase shift of (\ref{Eq:WFunction}) for that specific value of $s$ and, consequently, the accumulation of particles at the black hole's poles instead of in its equatorial plane. It is inferred that such behavior is an eminently relativistic effect: the particle density is measured by an observer comoving with the particle flux given by the particle current density $J$, as defined in equation~(\ref{Eq:n}). For $s=1$, a Lorentz contraction effect is recorded, in which the effect of the particle flux associated with the azimuthal motion of these (described by $J_{\hat{3}}^{\textrm{(rot)}}$) is minimal in relation to the effect of the flux described by $J_{\hat{0}}^{\textrm{(rot)}}$. Therefore, for that particular choice of the parameter $s$, the tendency of the particles is to concentrate at the poles of the black hole.

For values of $s$ other than $s=1$, when evaluating (\ref{Eq:ObservablesRot}), it yields 
 \begin{equation}
     \left. n^{\textrm{(rot)}}\right|_{s \geq 2}=\sqrt{ \left[ J^{(\textrm{rot})}_{\hat{0}}(r,\vartheta) \right]^2 - \left[ J^{(\textrm{rot})}_{\hat{3}}(r,\vartheta) \right]^2}
 \end{equation}
In this regime, $J_{\hat{0}}^{\textrm{(rot)}}$ acquires a polar dependence with a power of $\sin\vartheta$ distinct from that in $J_{\hat{3}}^{\textrm{(rot)}}$, which renders $n^{\textrm{(rot)}}$ odd in $\vartheta$, in contrast with the even behavior found at $s=1$. Consequently, for configurations with $s\geq 2$, the particle density is concentrated about the equatorial plane $\vartheta=\pi/2$.

To provide an example of mapping between the normalized density profiles and physical masses, we consider the supermassive black hole M87*, whose mass is $M_{\rm BH} \approx 6.6 \times 10^{9}\,M_\odot$~\cite{Akiyama_2019} and observations estimate a molecular gas mass of $M_{\rm gas}^{\rm obs} \approx 1.08 \times 10^{7}\,M_\odot$~\cite{Ray_2024}. This corresponds to a mass ratio: $M_{\rm gas}^{\rm obs}/M_{\rm BH} \approx 1.6\times 10^{-3}$, i.e., well below the percent level. In our kinetic models, the total gas mass scales linearly with the amplitude parameter $\alpha$, since the distribution function satisfies $F_0\propto \alpha$, implying $M_{\rm gas}\propto \alpha$. One can fix $\alpha$ such that $M_{\rm gas}$ lies to $M_{\rm gas}^{\rm obs}$, in which even adopting the maximal observational estimate, the resulting mass fraction remains $M_{\rm gas}/M_{\rm BH} \sim 10^{-3}$, ensuring that the gas self-gravity is dynamically subdominant compared to the black hole gravitational field. This explicit scaling demonstrates that the normalized density profiles employed in the present work can reproduce astrophysically realistic mass values while justifying the test-particle approximation.

\subsection{Normalized profiles of energy density and principal pressures}
\label{SubSec:Results02}

\begin{figure}[H]
\centerline{
\includegraphics[scale=0.325]{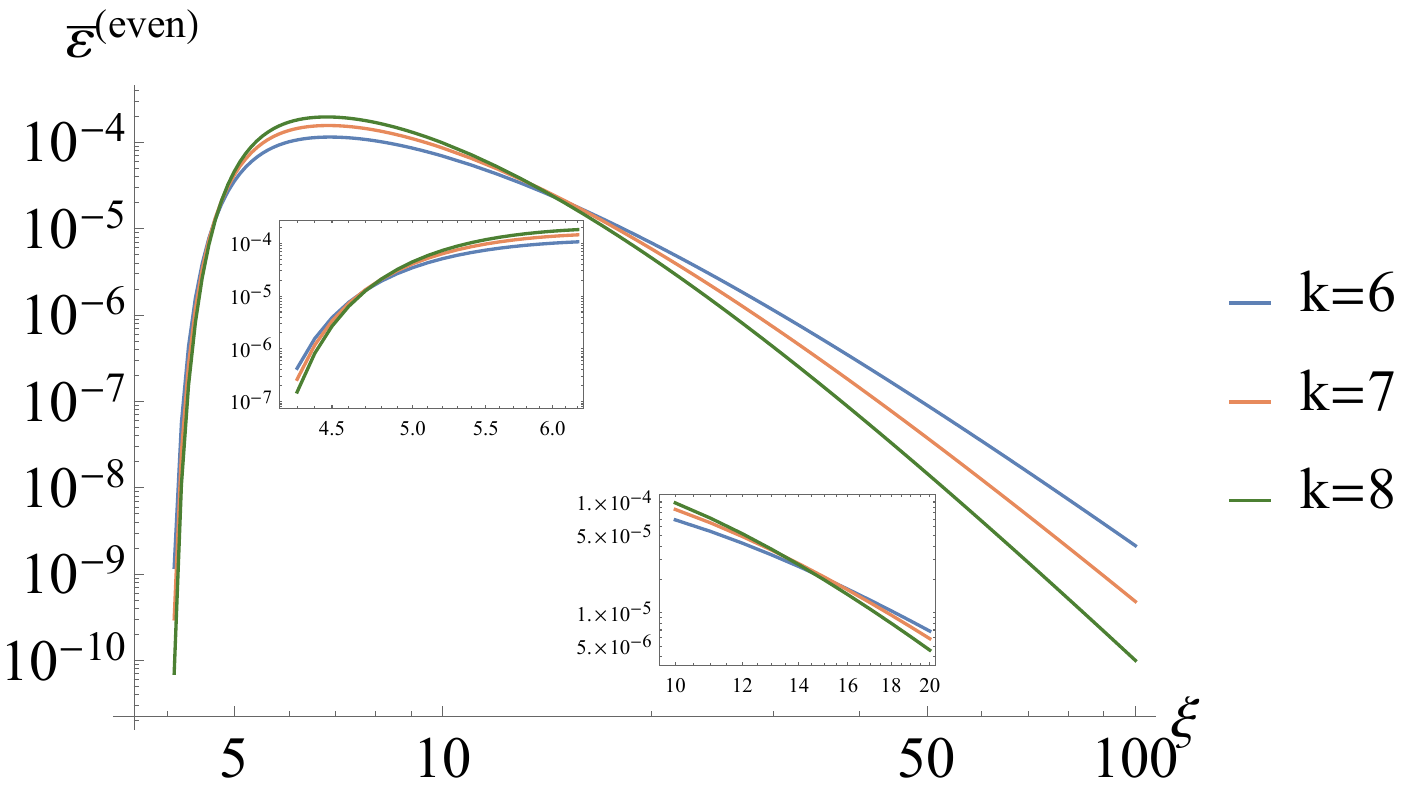}
\includegraphics[scale=0.325]{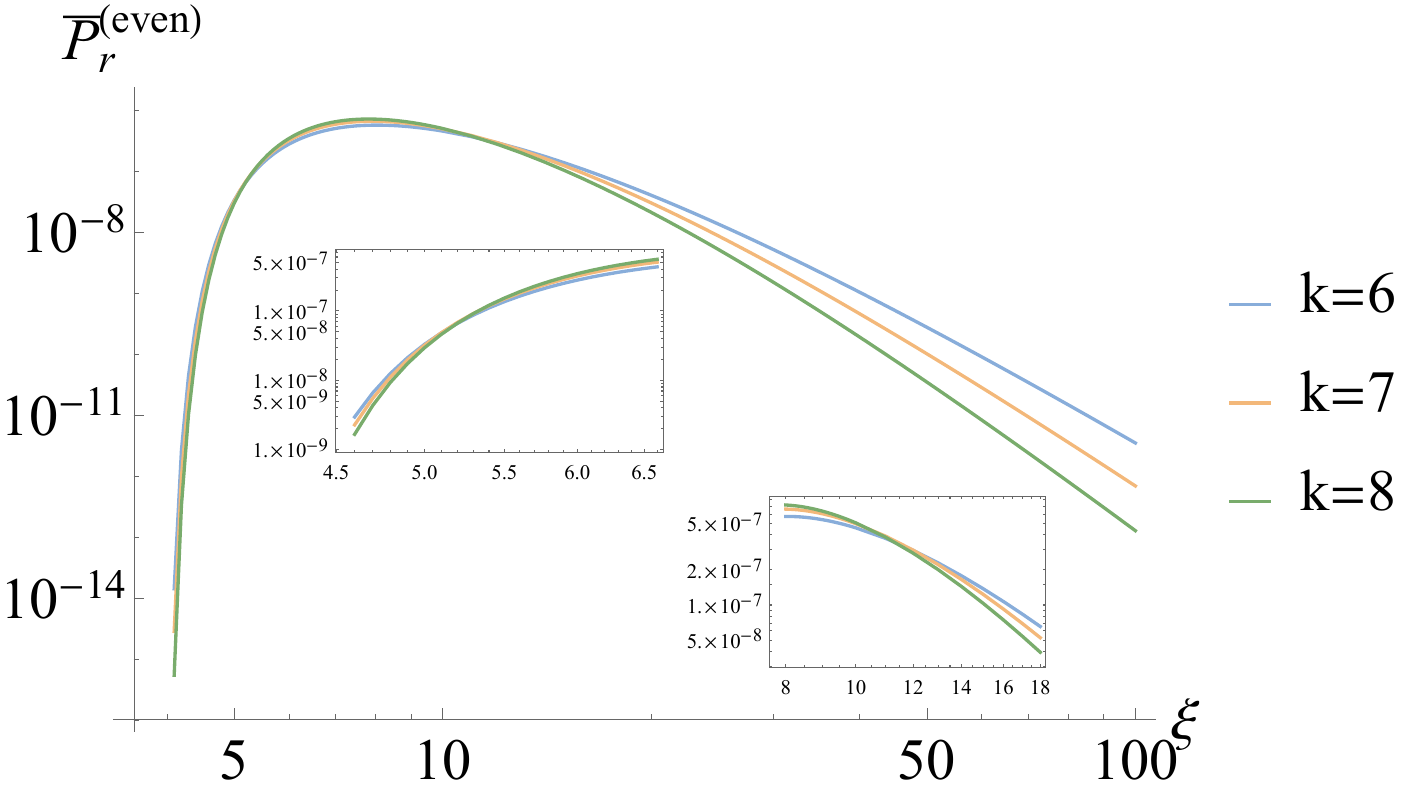}}
\centerline{
\includegraphics[scale=0.325]{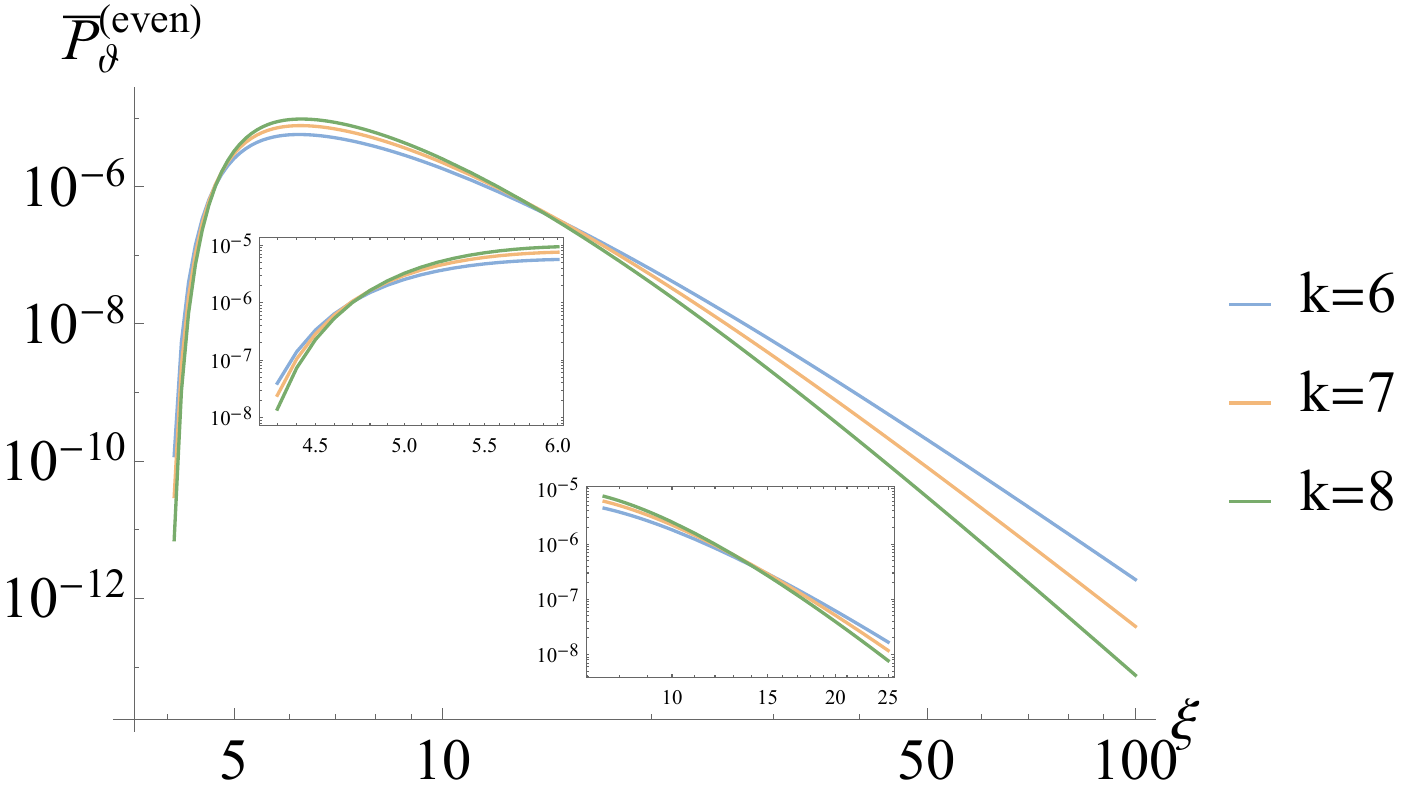}
\includegraphics[scale=0.325]{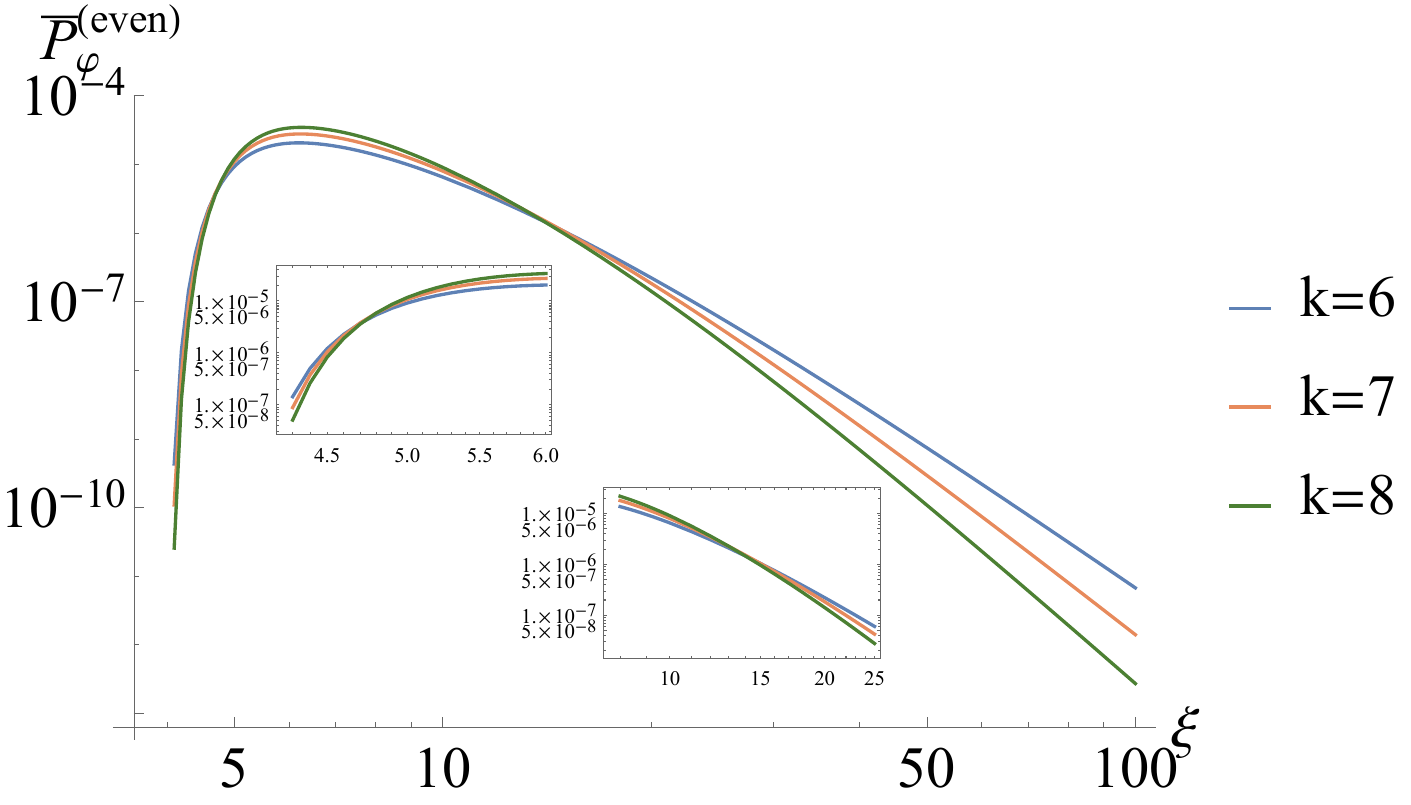}}
\caption{Profiles of the macroscopic observables derived from the energy-momentum-stress tensor: the energy density, the radial, polar and azimuthal pressures, for fixed values of $k=6,7,8$, $\varepsilon_0=1$ and some values of $s$, all of them normalized by $\mathcal{N}_{\textrm{gas}}$. All profiles are normalized by $\mathcal{N}_{\textrm{gas}}$, and describe asymptotically infinite gas configurations with zero total angular momentum. Left up panel: plot of the energy density for $s=1$. Right up panel: plot of the radial pressure for $s=2$. Left down panel: plot of the polar pressure for $s=2$. Right down panel: plot of the azimuthal pressure for $s=1$.}
\label{Fig:ePrPtPp01}
\end{figure}

In this subsection we describe the behavior of the normalized energy density and the principal pressures in a graphical way to complete the results of the macroscopic observables, which were derived in~(\ref{Eq:ObservablesEvene}-\ref{Eq:ObservablesEvenPphi}) and~(\ref{Eq:ObservablesRote}-\ref{Eq:ObservablesRotPphi}) for both models. Figures~\ref{Fig:ePrPtPp01} shows the particle density and the principal pressures for fixed values of parameters $(s,\varepsilon_0)$ and varying the $k$-parameter. One should observe that, when $k$ is larger (keeping $s$ at a fixed value) the maxima of the macroscopic observables increases too, as expected, and the internal sub-plots show the intersection between different graphs.

\begin{figure}[H]
\centerline{
\includegraphics[scale=0.325]{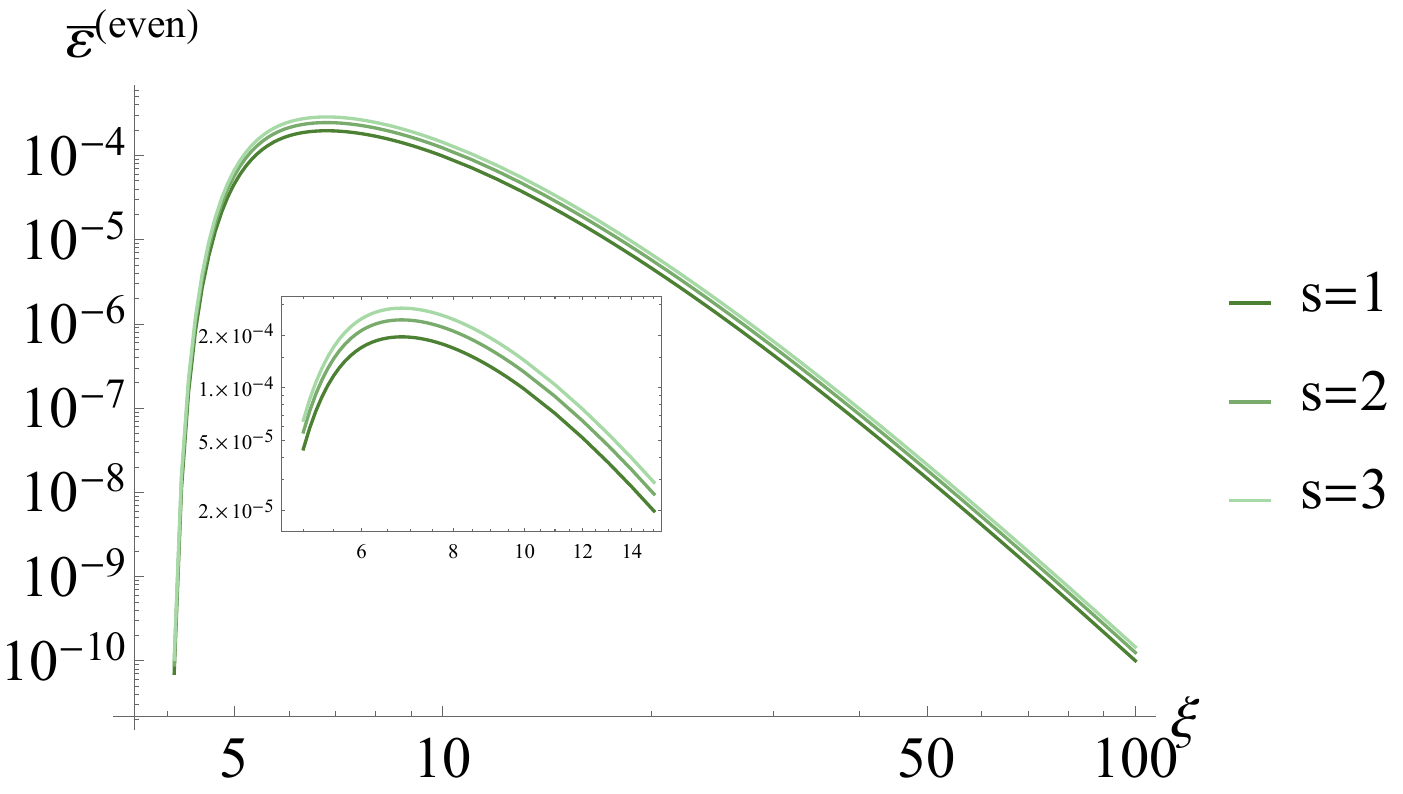}
\includegraphics[scale=0.325]{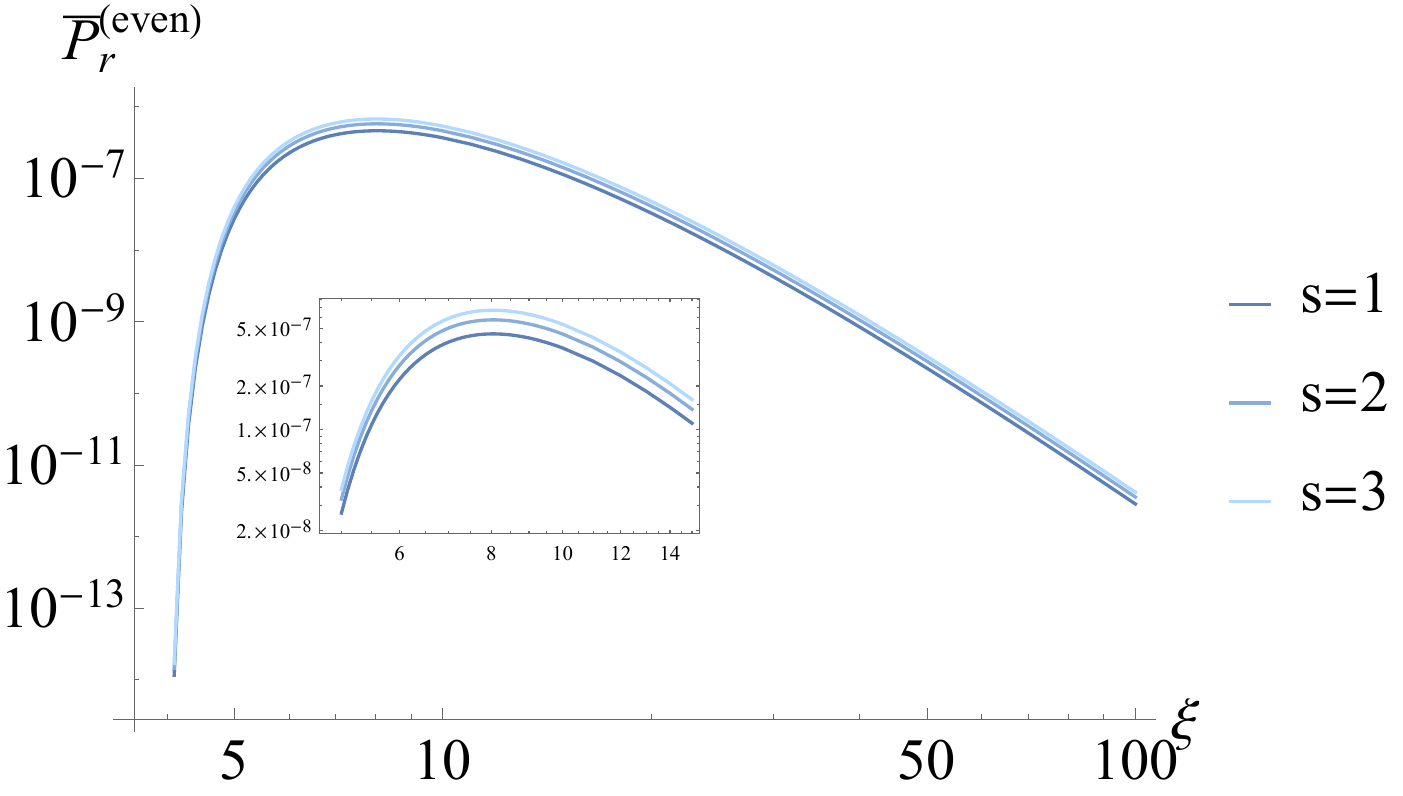}}
\centerline{
\includegraphics[scale=0.325]{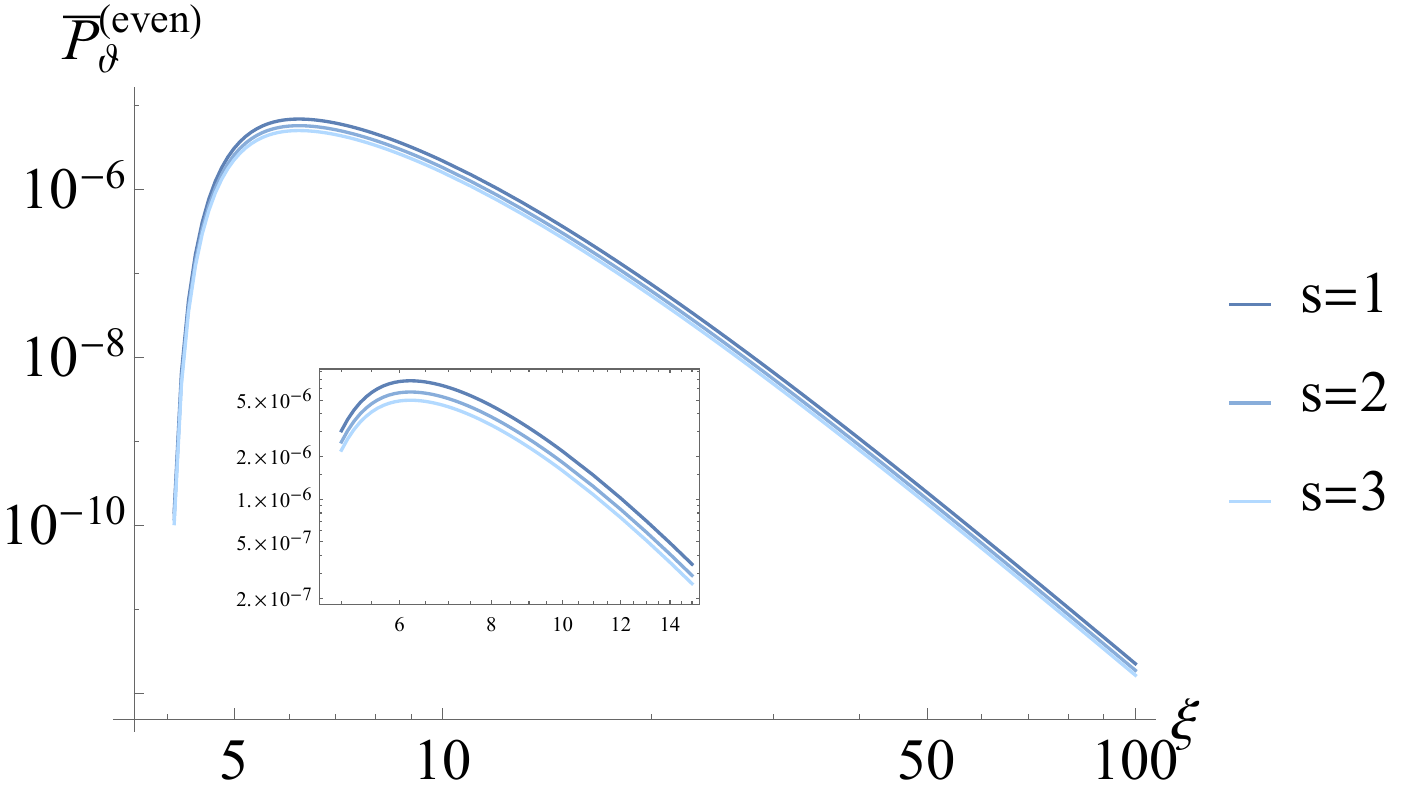}
\includegraphics[scale=0.325]{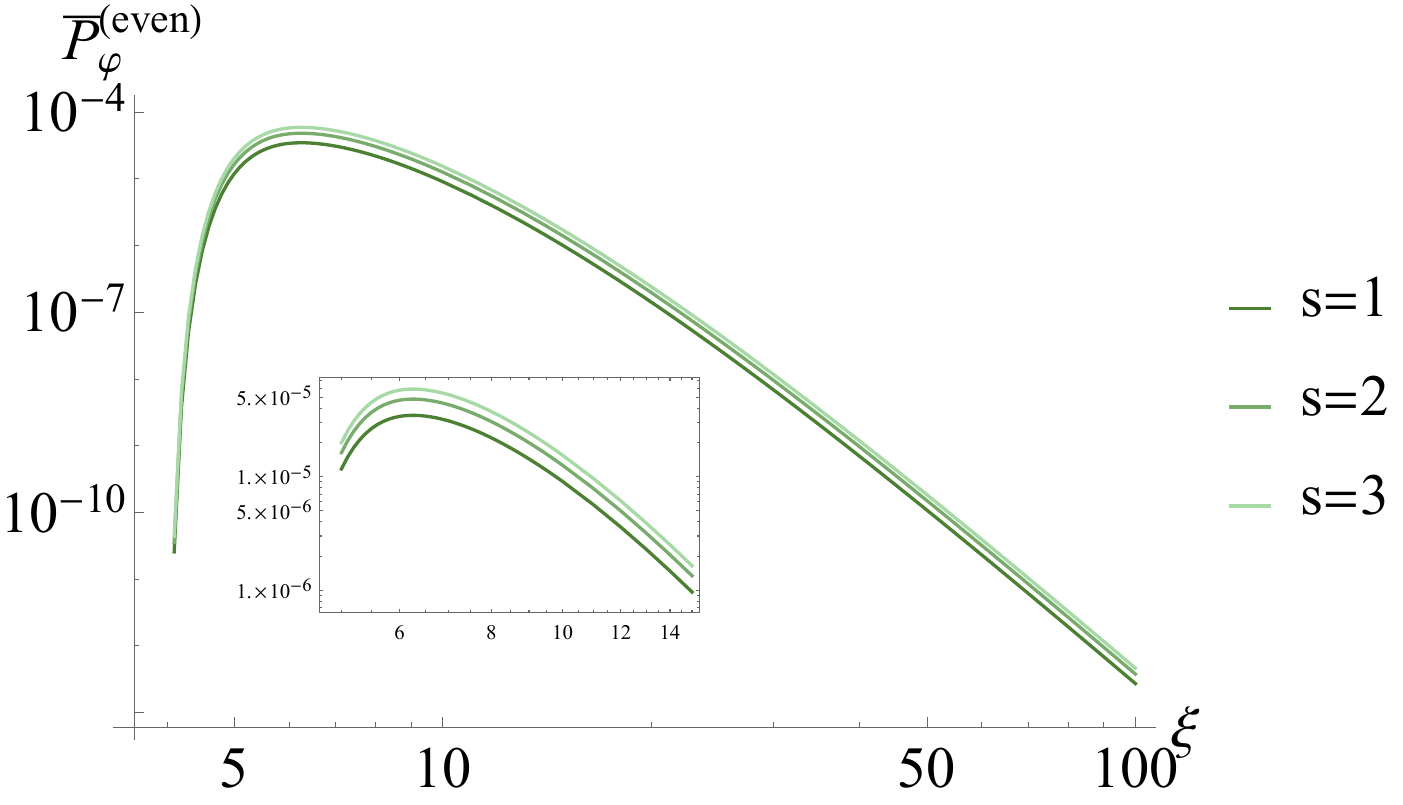}}
\caption{Profiles of the macroscopic observables derived from the energy-momentum-stress tensor: the energy density, the radial, polar and azimuthal pressures, for fixed values of $s=1,2,3$, $\varepsilon_0=1$ and some values of $k$. All profiles are normalized by $\mathcal{N}_{\textrm{gas}}$, and describe asymptotically infinite gas configurations with zero total angular momentum. Left up panel: plot of the energy density for $k=8$. Right up panel: plot of the radial pressure for $k=6$. Left down panel: plot of the polar pressure for $k=6$. Right down panel: plot of the azimuthal pressure for $k=8$.}
\label{Fig:ePrPtPp02}
\end{figure}

Figures~\ref{Fig:ePrPtPp02} show the normalized profiles of observables obtained from the energy-momentum-stress tensor for fixed values of $k$ in an asymptotically infinite non-rotating kinetic gas configuration. From the plots we can notice that they follow a similar behavior as the one in the right panel of figures~\ref{Fig:neven01} when varying the parameter $s$, except for the polar pressure, which at higher values of $s$, the peak of the pressure $P_{\hat{\vartheta}}^{\text{(even)}}$ decreases in amplitude.

\begin{figure}[H]
\centerline{
\includegraphics[scale=0.325]{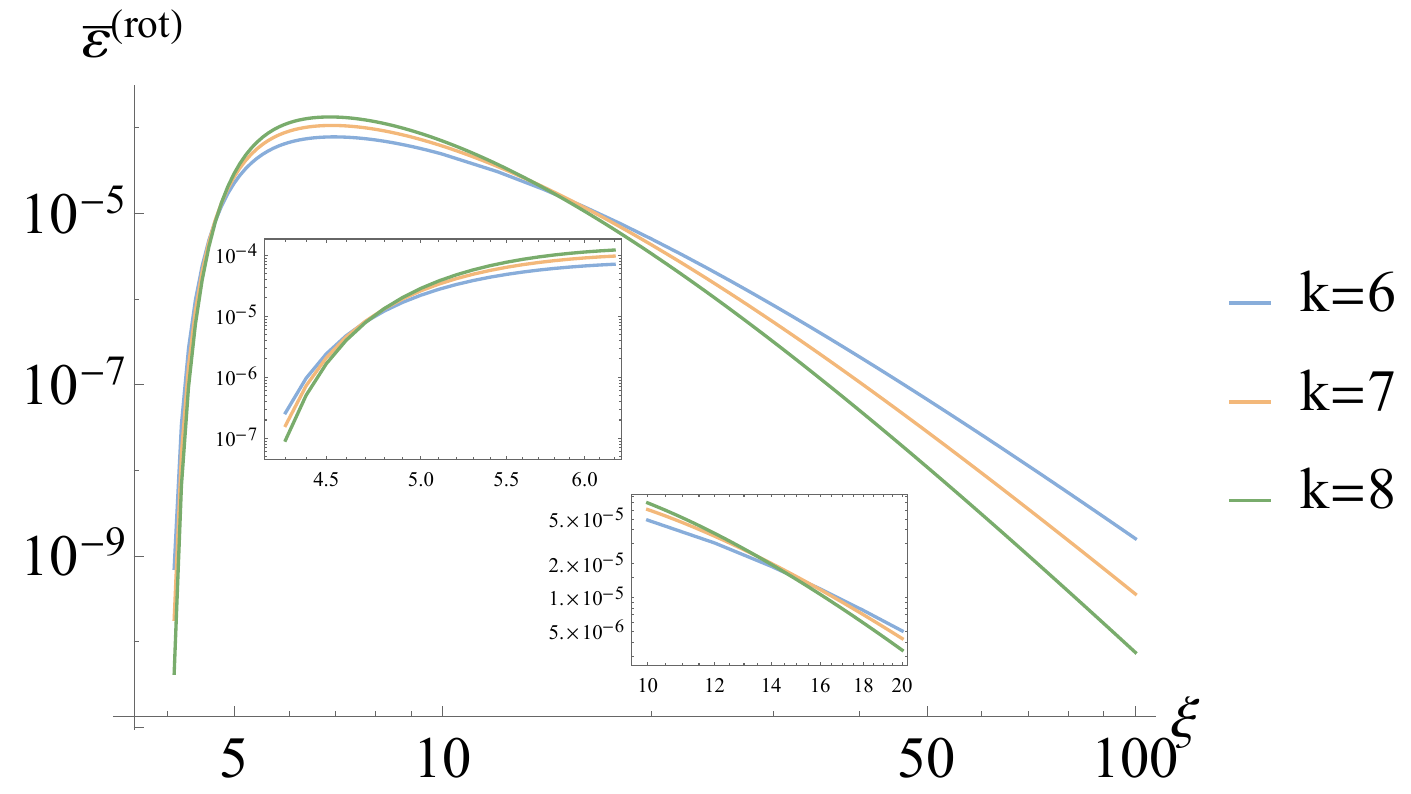}
\includegraphics[scale=0.325]{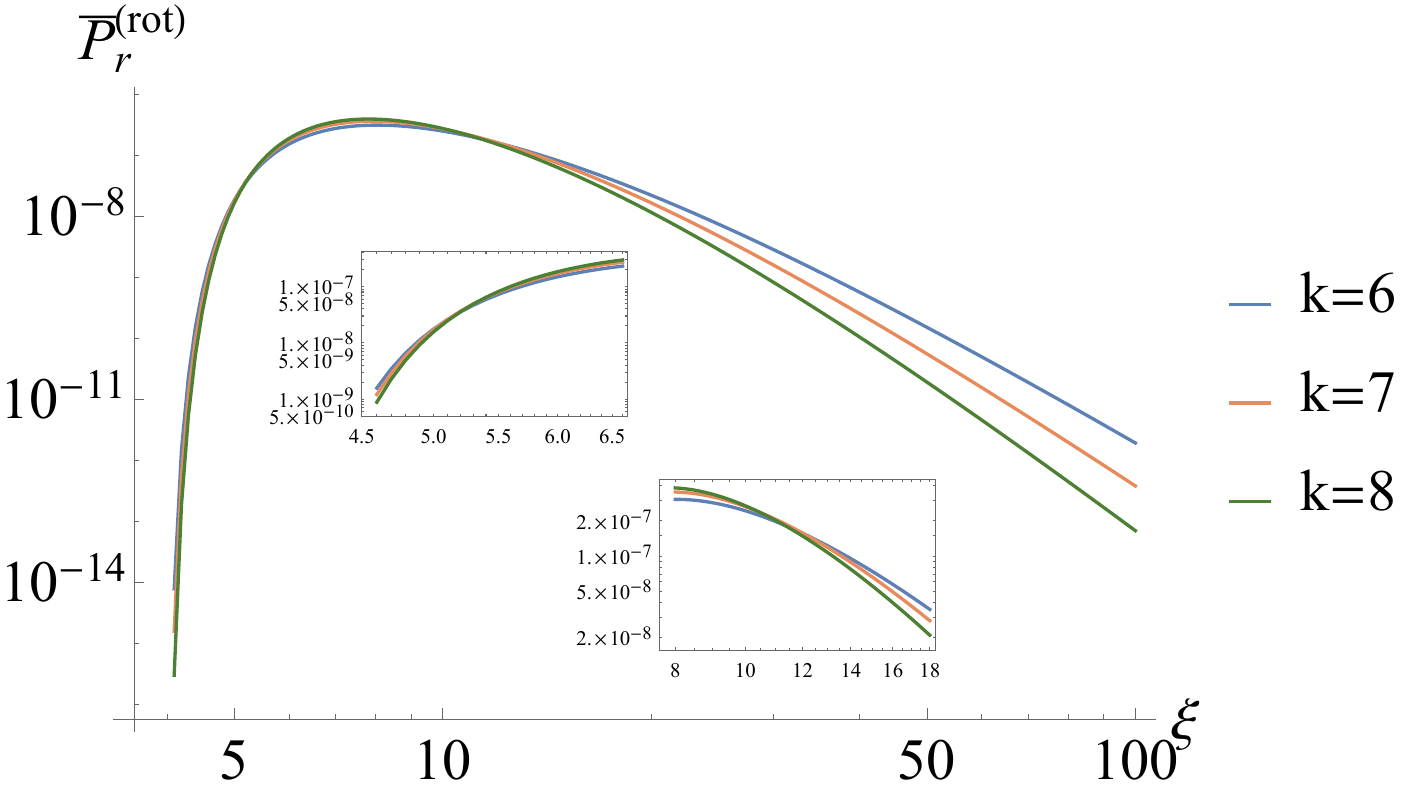}}
\centerline{
\includegraphics[scale=0.325]{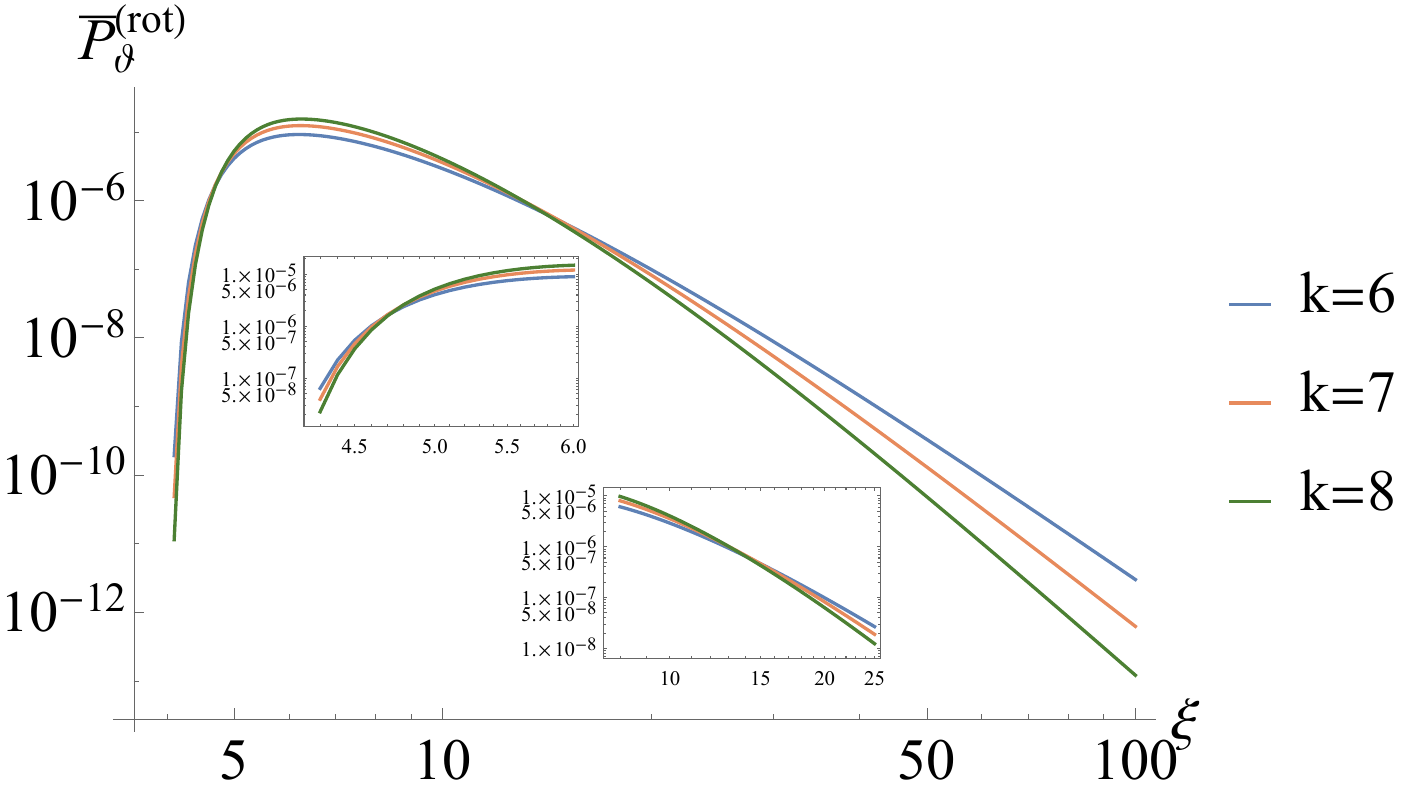}
\includegraphics[scale=0.325]{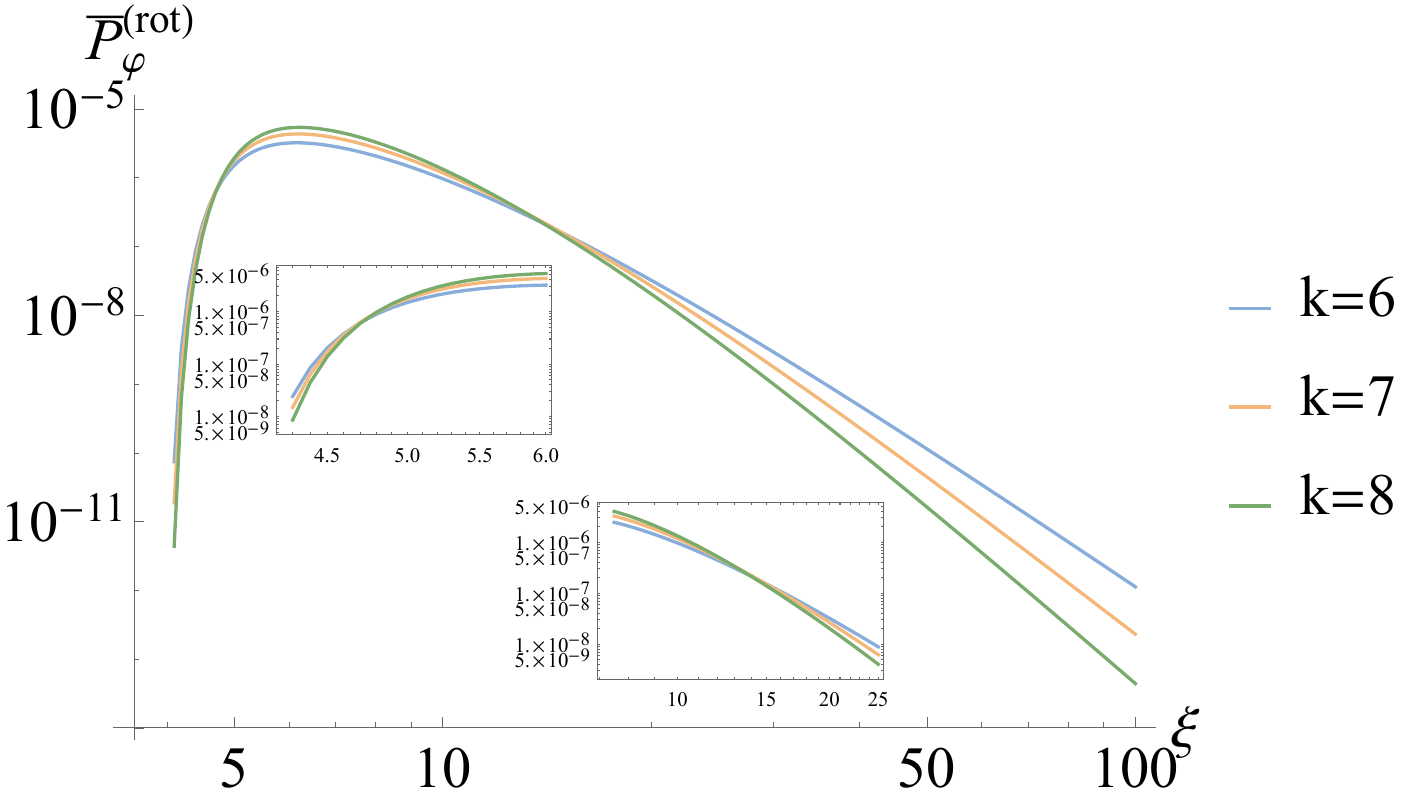}}
\caption{Profiles of the macroscopic observables derived from the energy-momentum-stress tensor: the energy density, the radial, polar and azimuthal pressures, for fixed values of $k=6,7,8$, $\varepsilon_0=1$ and some values of $s$. All profiles are normalized by $\mathcal{N}_{\textrm{gas}}$, and describe asymptotically infinite gas configurations with non-zero total angular momentum. Left up panel: plot of the energy density for $s=2$. Right up panel: plot of the radial pressure for $s=1$. Left down panel: plot of the polar pressure for $s=1$. Right down panel: plot of the azimuthal pressure for $s=2$.}
\label{Fig:ePrPtPp03}
\end{figure}

For the observables derived from the rotating model~(\ref{Eq:ObservablesRote}-\ref{Eq:ObservablesRotPphi}) one can observe that in the energy density and azimuthal pressure terms between the non-rotating gas model and the rotating gas model, there are meaningful differences. The presence of the term $T^{\textrm{(rot)}}_{\hat{0}\hat{3}}$ in the energy-momentum-stress tensor for a rotating kinetic gas results in the expressions for the aforementioned quantities being more complex than in the non-rotating gas model. In fact, if the component $T^{\textrm{(rot)}}_{\hat{0}\hat{3}}$ vanishes in the rotating case, one cannot recover the same expression for the non-rotating case, as the angular expressions for each model are different (see equations~(\ref{Eq:I01}) and~(\ref{Eq:App01})). Despite this, as shown in figures~\ref{Fig:ePrPtPp03}, the profiles of the energy density and the main pressures for the rotating gas model do not differ in behavior from those of the non-rotating gas model when controlling the variation of the parameter $k$: as its value increases, it can be noted that the observables tend to have maximum values in the vicinity of the black hole and that their amplitudes decay faster at distances far from the black hole. 

\begin{figure}[H]
\centerline{
\includegraphics[scale=0.325]{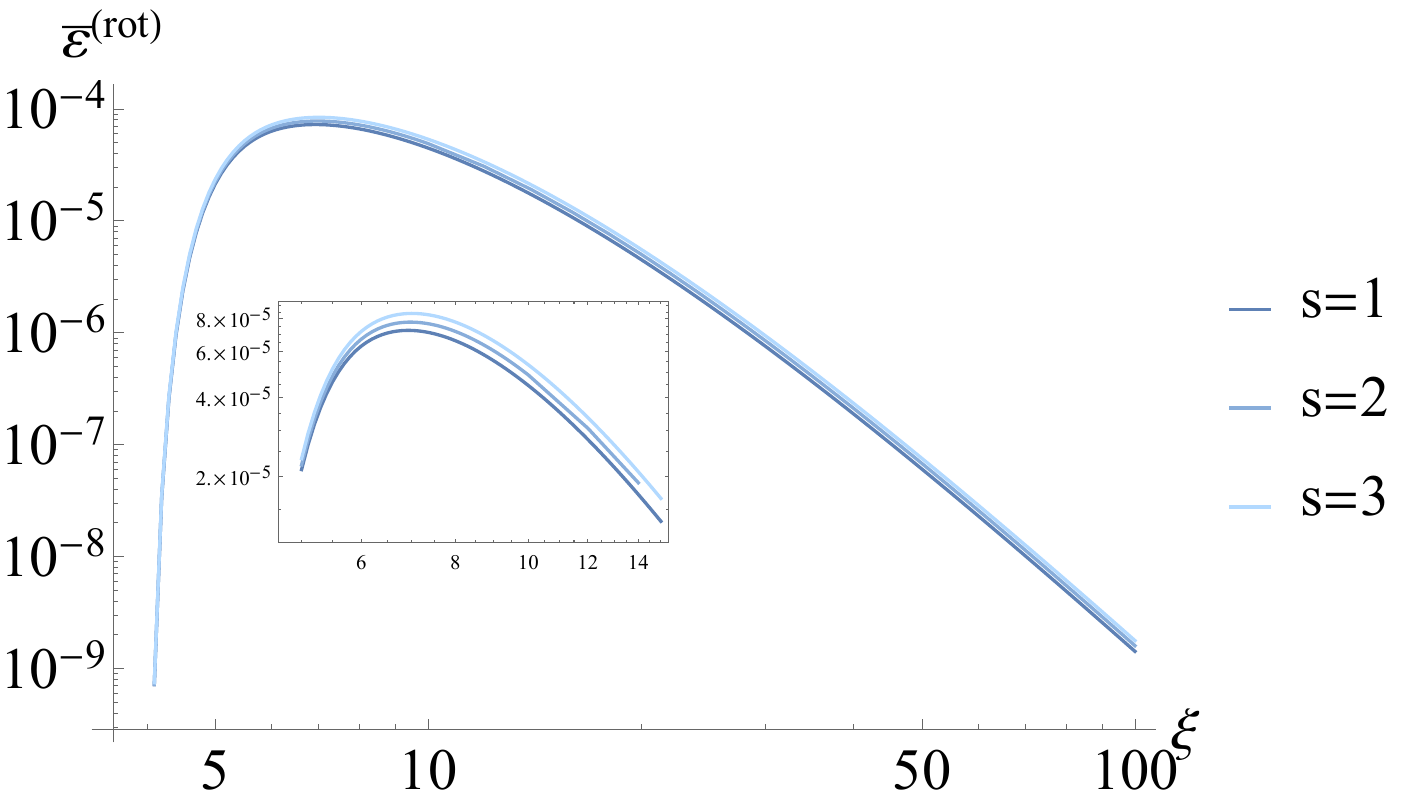}
\includegraphics[scale=0.325]{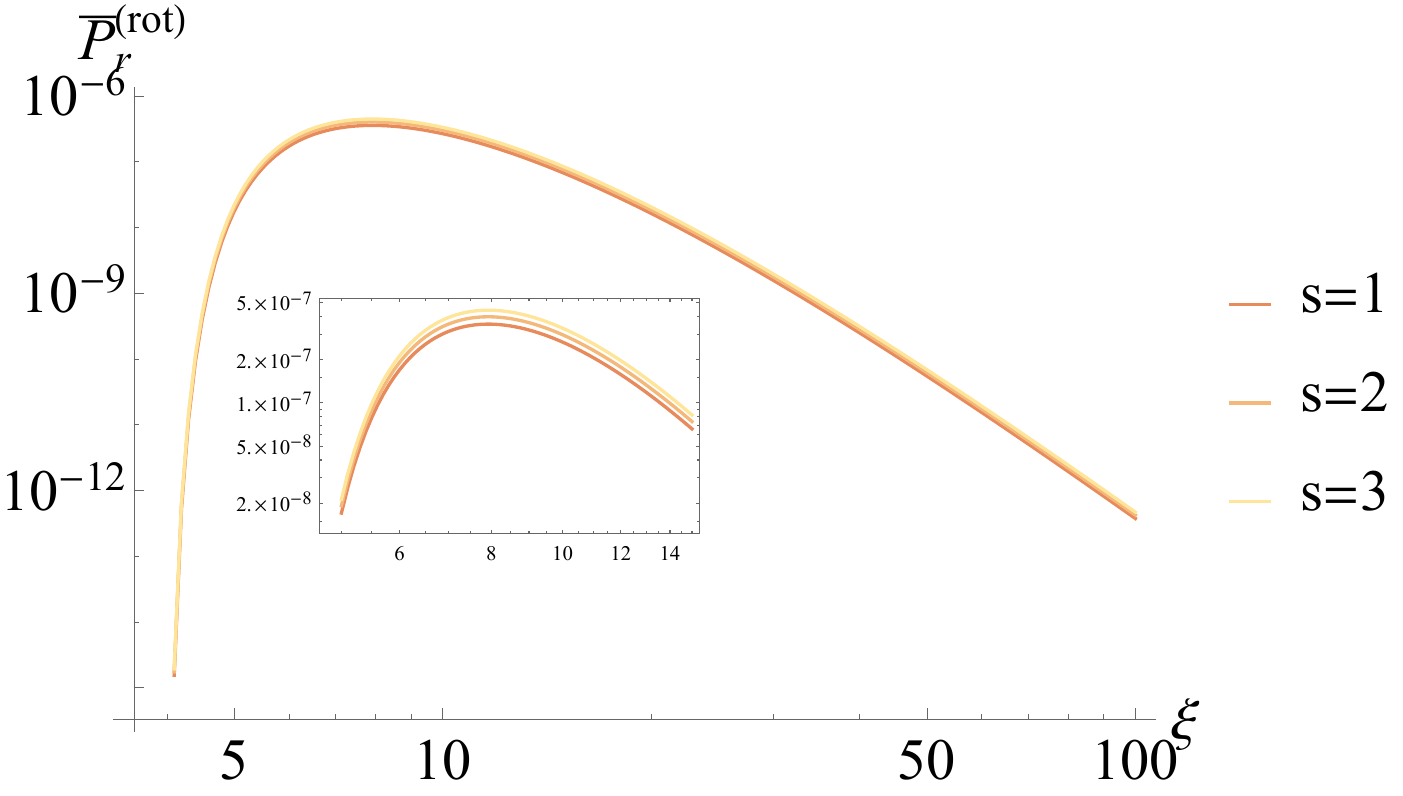}}
\centerline{
\includegraphics[scale=0.325]{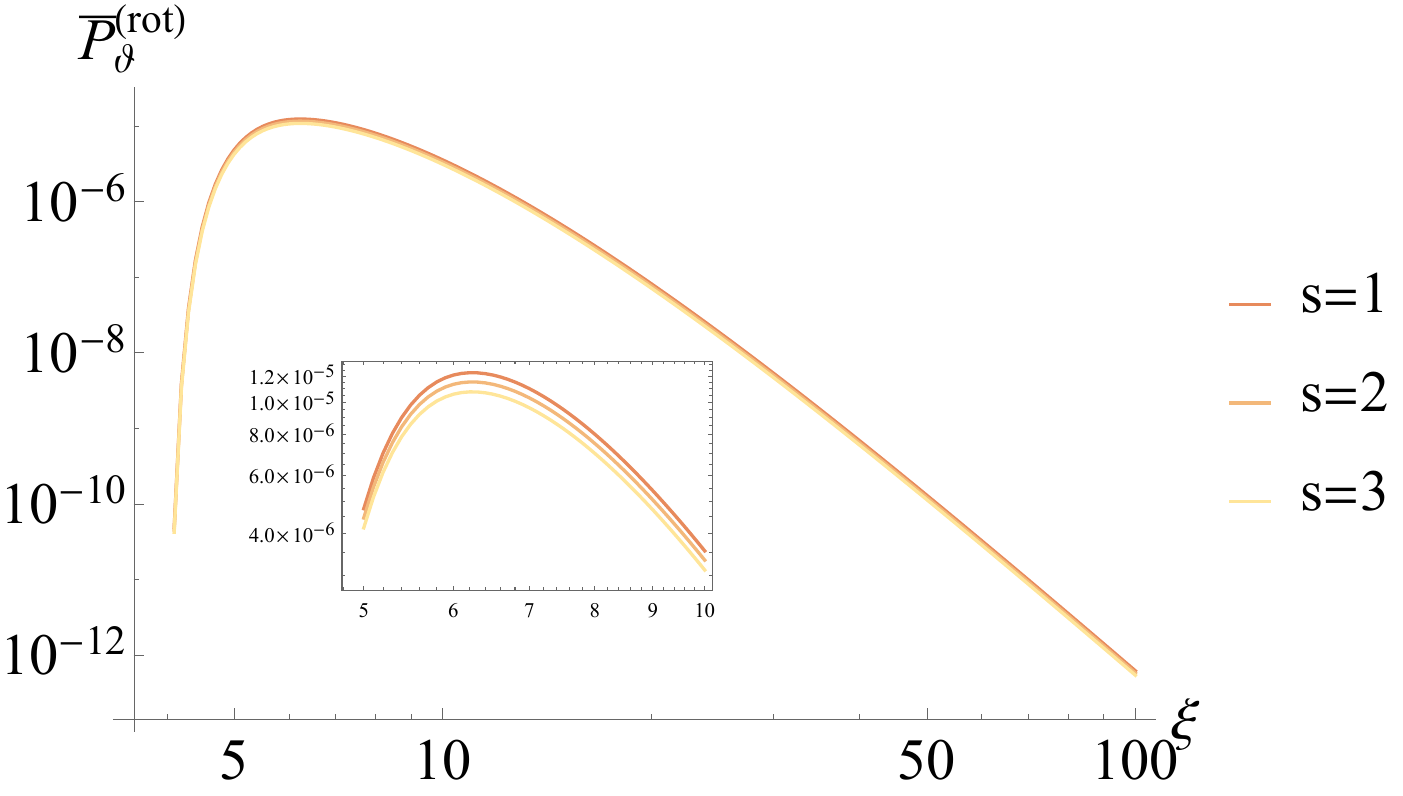}
\includegraphics[scale=0.325]{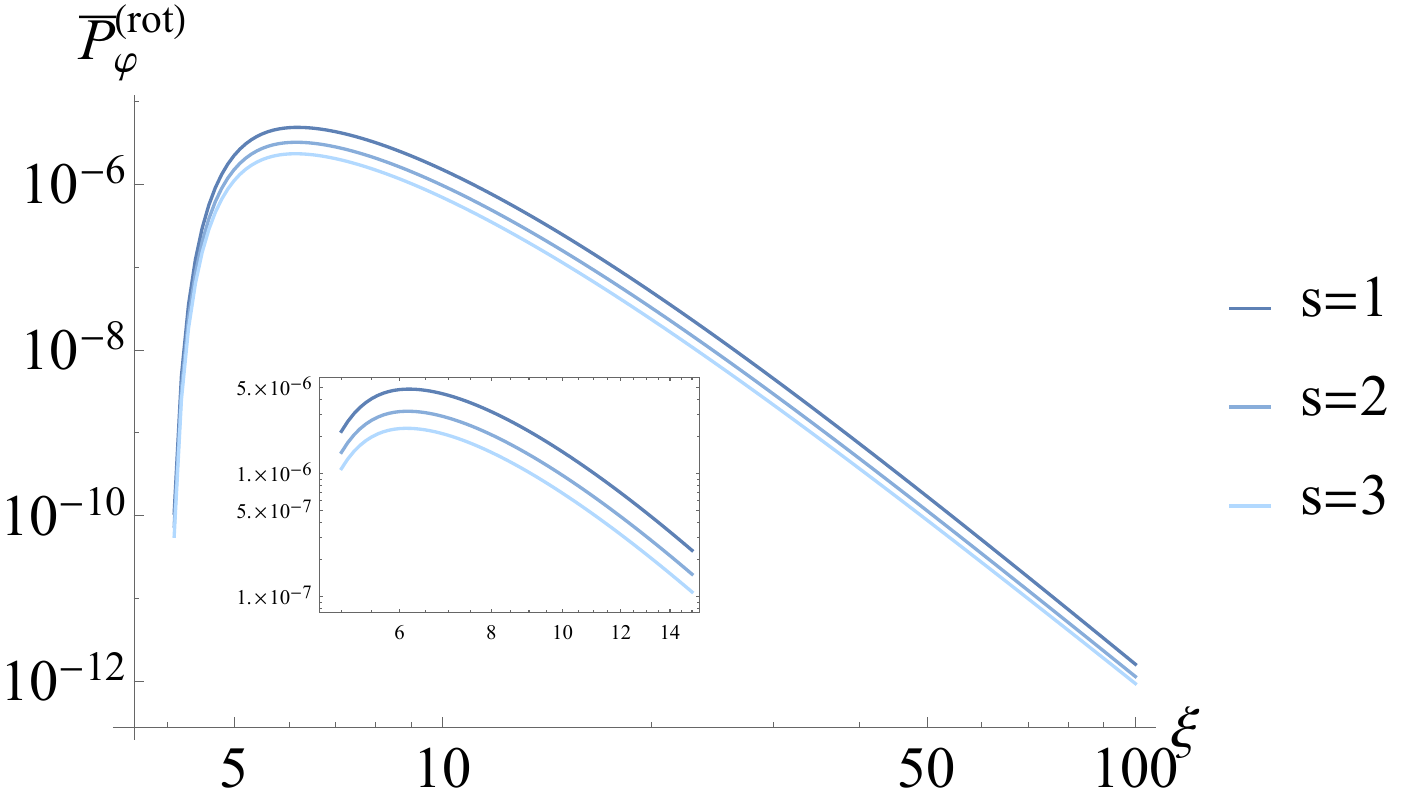}}
\caption{Profiles of the macroscopic observables derived from the energy-momentum-stress tensor: the energy density, the radial, polar and azimuthal pressures, for fixed values of $s=1,2,3$, $\varepsilon_0=1$ and some values of $k$. All profiles are normalized by $\mathcal{N}_{\textrm{gas}}$, and describe asymptotically infinite gas configurations with non-zero total angular momentum. Left up panel: plot of the energy density for $k=6$. Right up panel: plot of the radial pressure for $k=7$. Left down panel: plot of the polar pressure for $k=7$. Right down panel: plot of the azimuthal pressure for $k=6$.}
\label{Fig:ePrPtPp04}
\end{figure}

In figures~\ref{Fig:ePrPtPp04}, when increasing the value of the parameter $s$, the amplitude of each observable increases slightly, except for the polar $P_{\hat{\vartheta}}^{\text{(rot)}}$ and azimuthal $P_{\hat{\varphi}}^{\text{(rot)}}$ pressures, in which the amplitude rather decreases as the value of the parameter $s$ increases. Particularly in the case of azimuthal pressure, it can be noted that the effect of the parameter $s$ on its behavior is more significant than it is for the rest of the profiled observables, which also differs substantially from its analogous quantity in the non-rotating gas model. At first glance, these differences in the behavior of the main pressures are linked to which term is or is not more dominant in the components of the energy-momentum-stress tensor. A more detailed analysis of these consequences will be presented in the article~\cite{dMcG2025} through the anisotropy parameter, which gives us more information about the behavior of the gas configuration.

A similar behavior is displayed for finite gas configurations with a cut-off value of $\varepsilon_0<1$, as the one exhibited in figures~\ref{Fig:ePrPtPp01}-\ref{Fig:ePrPtPp04} for all macroscopic quantities.

\subsection{Brief comparison with a fluid model}
\label{SubSec:Results03}
\begin{figure}
\centering
\begin{subfigure}{.5\textwidth}
  \includegraphics[width=0.8\linewidth]{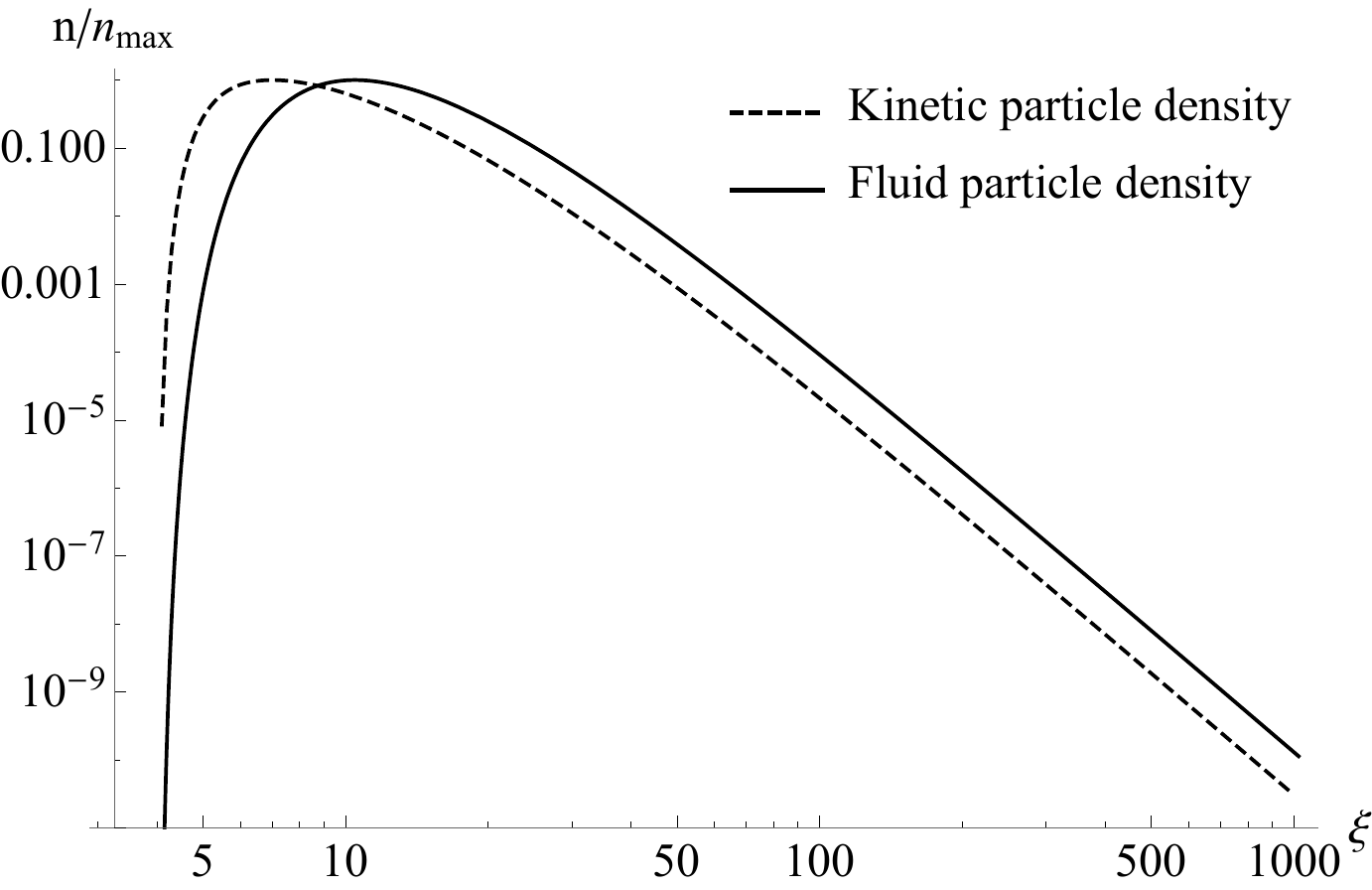}
  \caption{}
  \label{fig:nnmax}
\end{subfigure}%
\begin{subfigure}{.5\textwidth}
  \includegraphics[width=0.8\linewidth]{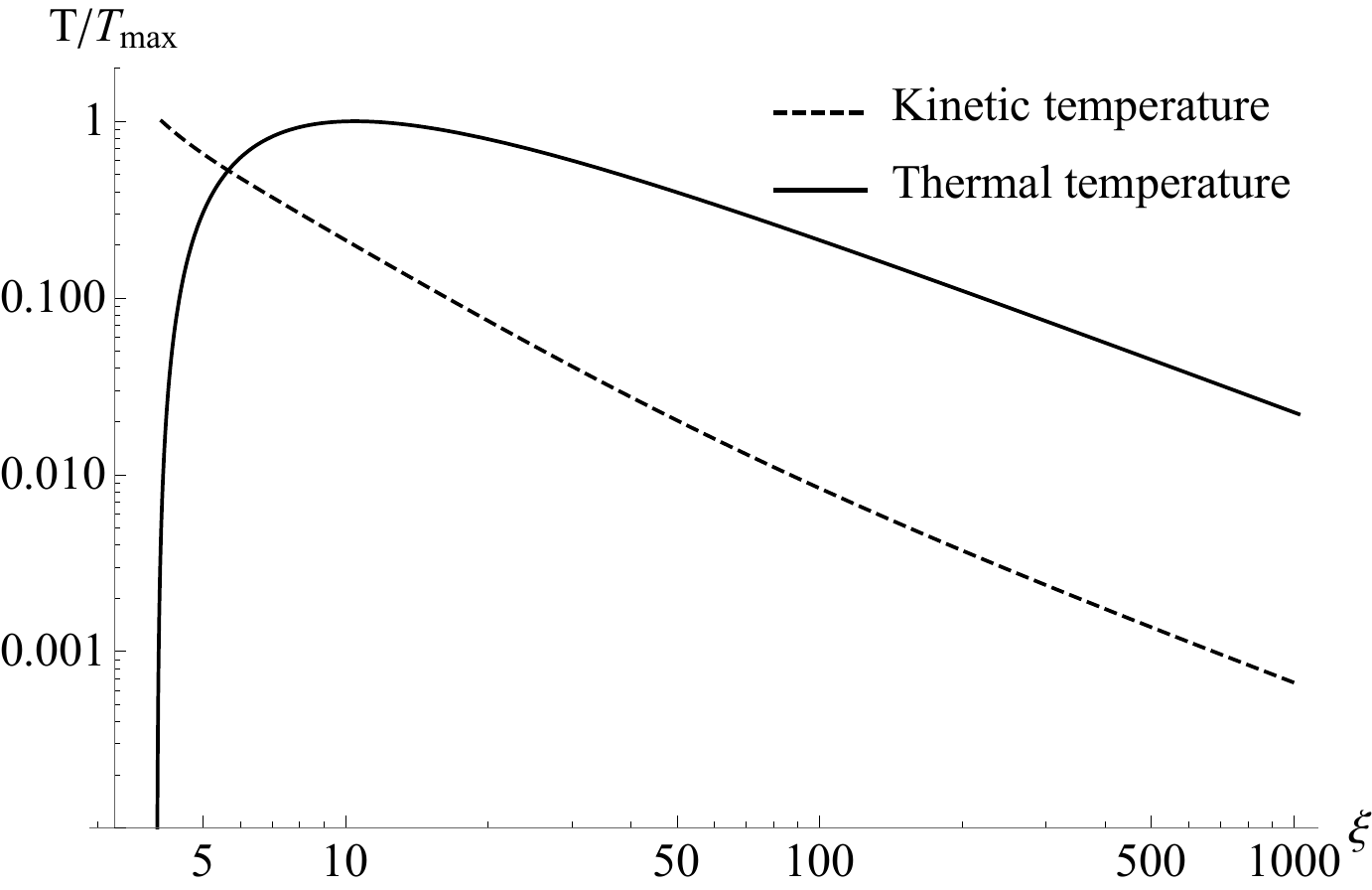}
  \caption{}
  \label{fig:TTmax}
\end{subfigure}
\caption{Normalized profiles between the kinetic and fluid descriptions. The left panel shows the profile of the particle density for adiabatic index $\gamma=1.17$ (continuous line) and the corresponding particle density for the non-rotating kinetic model (dashed). The right panel shows the profile of thermal (continuous line) and kinetic (dashed) temperatures. Both of the kinetic model profiles are plotted with $(k,s)=(6,2)$.}
    \label{fig:compFluidVsKin}
\end{figure}

To characterize the macroscopic behavior of the kinetic model proposed in this work, we compare its predictions with those of a relativistic fluid model, such as the ``polish doughnuts''. For the fluid model, we assume a polytropic equation of state $P = K n^\gamma$ with $K$ a constant and $\gamma$ the adiabatic index subject to $1 < \gamma \leq 2$, which is related to the polytropic index $k$ by $\gamma = 1 + 1/k$. Here, $n$ and $T$ are related to the specific enthalpy $h$, (see for example~\cite{Rezzolla-Book} or Appendix E in~\cite{cGoS2023b} for more details) by:
\begin{equation}
h - 1 
= \frac{\gamma}{\gamma-1}\frac{K}{\bar{m}} n^{\gamma-1}
= \frac{\gamma}{\gamma-1}\frac{k_B}{\bar{m}} T,
\end{equation}
with $\bar{m}$ being the averaged rest mass per particle and both models are obeying the ideal gas equation $P = n k_B T$. Figure \ref{fig:nnmax} shows the profile of the particle density for the kinetic and fluid models, both of them normalized over the maximum of their respective configurations. Both profiles are seen to share the same global qualitative shape: in each case the density increases from large radii up to a maximum and then decreases toward the inner region. This indicates that the fluid model and the kinetic gas configuration exhibit the same morphology. However, relevant systematic differences are present. In particular, the radial positions of the maxima do not coincide: the density maximum in the kinetic model is displaced with respect to that of the fluid model, independent from the choice of parameters $k$ and $s$ used in the kinetic gas model.\\
The comparison of the temperature profiles (Fig. \ref{fig:TTmax}) reveals a qualitatively different behavior. In contrast to the density case, no clear correlation is observed between the temperature distributions of the kinetic and the fluid models. The radial variations of one do not track those of the other, neither in the position of their maxima nor in the overall profile shape.\\
In particular, the radii at which the temperature reaches characteristic values differ significantly between the two models, and regions where one temperature increases do not correspond to analogous regions in the other.\\
Therefore, while in the density there exists at least a global morphological correspondence between both descriptions, in the temperature case the predictions of the fluid and kinetic models are different between each other.\\
The same qualitative pattern is observed in rotating kinetic gas configurations constructed from non-even distribution functions. In such cases, although rotation modifies the radial location of the maxima and the width of the profiles, the comparison with the corresponding fluid model retains the same features: agreement in the morphology of the density of particles profiles but with displaced maxima and discrepancies at large radii, and absence of correlation between the fluid and kinetic temperatures for these models.\\
This suggests that the differences observed between the fluid and kinetic descriptions do not depend on the absence of rotation, but rather constitute a generic property of the collisionless kinetic model considered here.

Based on the results shown so far, the next chapter will state the conclusions derived from this study.


\section{Conclusions}
\label{Sec:Conclusions}

Analytical expressions were obtained to calculate the particle current density and the energy–momentum-stress tensor, which in turn allowed us to profile some macroscopic observables of interest in a stationary relativistic kinetic gas configuration, namely the particle density, energy density, and the principal pressures. Analytical expressions were also obtained to calculate various conserved quantities throughout the configuration, being those the total number of particles, energy, and angular momentum of the gas. A description of a relativistic kinetic gas was constructed from a distribution function that considers the polytropic \emph{ansatz} and the inclination angle of the orbits followed by the particles of the gas. Based on the dependence on the parameters on which the \emph{ansatz} depends, different configurations with finite total mass, energy, and angular momentum are constructed and explored, allowing comparisons between these configurations, both with infinite and finite extensions. In this case, larger values of $k$ cause the physical observables to tend to show maximum values in regions closer to the black hole, but they also cause the configurations to drop more significantly as the particles are farther from the black hole. In turn, depending on the values of $s$, the profiles of the observables of the gas configuration exhibited different behaviors. For example, at greater values of $s$, the configuration tends toward a thin disk. From the profiles in the non-rotating case, the greatest concentration tends to occur around the equatorial plane, while in the rotating case, more significant concentrations appear in the polar region when $s=1$, due to the combined effect of azimuthal motion and Lorentz contraction. These differences in gas morphology become particularly noticeable when comparing the contour plots of the non-rotating model (figure~\ref{Fig:neven03}) with those of the rotating model (figure~\ref{Fig:nrot03}). Despite this, for values other than $s=1$, the gas particles tend to concentrate around the equatorial plane in both models. However, in the rotating gas model, the particle density amplitudes are significantly larger than in the non-rotating gas model. Consequently, it can be noted in the contour plots associated with the rotating gas model that the particles are more dispersed in space compared to those of the non-rotating gas model, where the contours cancel out at smaller radii.

When comparing the gas configuration profiles between the non-rotating and rotating models, the most significant variations are evident in the principal gas pressures. In the rotating model, polar pressure tends to predominate in the vicinity of the black hole, while in the non-rotating configuration, azimuthal pressure predominates. The effect of including the azimuthal term in the particle current density~(\ref{Eq:CurrentDensity}) substantially affects how the particles are distributed around the black hole. Similarly, the presence of the off-diagonal component of the energy-momentum-stress tensor~(\ref{Eq:EnergyMomentumStress}) generates changes in the gas pressures between the gas particle orbital inclination models that were introduced through the distribution function proposed in Section~\ref{Sec:Models}.

Asymptotically infinite gas configurations and finite gas configurations were constructed by introducing a cut to the system energy, given by the cut-off parameter $E_0$. These configurations exhibit similar characteristics. Furthermore, bounds on the extension of the finite configuration were analytically obtained, as shown in Section~\ref{Sec:Results}, in equations~(\ref{Eq:rmin}) and~(\ref{Eq:rmax}) which depend on the choice of the parameter $E_0$. This result allows one to construct bounded configurations with finite extension analytically, which can be compared with other finite extension configurations constructed with fluid models, such as the \emph{polish doughnuts}~\cite{Rezzolla-Book,cGoS2023b}.

With the results in this study along with the advances currently under development in~\cite{dMcG2025}, the description of relativistic kinetic gas configurations on bounded trajectories around a Schwarzschild black hole, initiated by~\cite{cGoS2022}, is more developed. The proposed model, based on relativistic kinetic theory, has allowed us to characterize the morphology of these configurations by considering models of the inclination angle of the gas particle orbits and the polytropic \emph{ansatz}.


\acknowledgments


The authors thank Olivier Sarbach, Ana L. García-Perciante, and Felix Salazar for their insightful comments and discussions throughout this work. We also acknowledge Miroslava Servellón for her careful technical review of the manuscript. R.R. gratefully acknowledges the support of the National Autonomous University of Honduras during his 2024 research stay at the Autonomous Metropolitan University–Cuajimalpa Unit in Mexico City.


\appendix


\section{Integrals based on the inclination angle models}
\label{Appx:A}

In the models considered in section~\ref{Sec:Models}, we analyze a stationary and axisymmetric relativistic kinetic gas configuration around a Schwarzschild black hole. Within this context, the particle distribution function is typically factorized into one part depending on the energy $E$ and another on the azimuthal angular momentum $L_z$—or, equivalently, on the inclination angle of the particle orbits. When computing macroscopic observables such as particle number density, current vector, and the components of the energy-momentum-stress tensor, an integral over the orbital inclination angle arises. These integrals define a family of angular functions denoted by, 
\begin{equation}
    \mathcal{I}_{\textrm{i-model}},
\end{equation}
in which the subscript $(\textrm{i-model})$ corresponds to even ($i$) or rotating model ($i/2$), which encode the angular contribution of the azimuthal momentum component in the kinetic gas. The distinction between the \emph{even} and \emph{rot} models lies in the symmetry of the distribution with respect to $L_z$. In the \emph{even} model, the distribution is symmetric under the transformation $L_z \mapsto -L_z$, implying a vanishing total angular momentum (non-rotating gas). In contrast, the \emph{rot} model privileges one direction of $L_z$, yielding a net total angular momentum (rotating gas). As a result, the associated angular functions differ in structure: in the \emph{even} case, they exhibit parity properties that suppress net azimuthal flow, whereas in the \emph{rot} model, they include terms that reflect a non-zero average rotation. In the macroscopic expressions, the functions $\mathcal{I}$ and its variants modulate the angular dependence of the macroscopic observables, which play an important role in shaping the morphology of the relativistic kinetic gas.

The explicit form of the functions $\mathcal{I}$ for non-rotating models are:
\begin{eqnarray}
\label{Eq:I01}
\mathcal{I}_i(\vartheta) &:=& \int\limits_0^{2\pi} G_i^{(\textrm{even})}(\vartheta,\chi) d\chi 
= \int\limits_0^{2\pi} (\sin\vartheta \sin\chi)^{2s} d\chi \nonumber \\
&=& 2\sqrt{\pi}\frac{\Gamma(s+1/2)}{\Gamma(s+1)}\sin^{2s}\vartheta, \\
\mathcal{I}_i^*(\vartheta) &:=& \int\limits_0^{2\pi} \cos^2\chi G_i^{(\textrm{even})}(\vartheta,\chi) d\chi 
= \int\limits_0^{2\pi} \cos^2\chi (\sin\vartheta \sin\chi)^{2s} d\chi \nonumber \\
&=& \sqrt{\pi}\frac{\Gamma(s+1/2)}{\Gamma(s+2)}\sin^{2s}\vartheta \nonumber\\
&=& \frac{1}{2(s+1)}\mathcal{I}_i(\vartheta), \\
\mathcal{I}_i^\dag (\vartheta) &=& \int\limits_0^{2\pi} \sin^2\chi G_i^{(\textrm{even})}(\vartheta,\chi) d\chi = \int\limits_0^{2\pi} \sin^2\chi (\sin\vartheta \sin\chi)^{2s} d\chi \nonumber \\
&=& (2s+1) \mathcal{I}_i^*(\vartheta).
\end{eqnarray}
For construction, it can be seen that
\begin{equation}
    \mathcal{I}_i(\vartheta) = \mathcal{I}_i^*(\vartheta) + \mathcal{I}_i^\dag (\vartheta).
\end{equation}

For rotating models, one obtains
\begin{eqnarray}
\label{Eq:App01}
\mathcal{I}_{i/2}(\vartheta) &:=& \int\limits_0^{2\pi} G_{i/2}^{(\textrm{rot})}(\vartheta,\chi) d\chi 
= \frac{1+s}{1+2s}\frac{1}{2^s} \int\limits_0^{2\pi} (1+\sin\vartheta\sin\chi)^{s} d\chi \nonumber \\
&=& \frac{1+s}{1+2s}\frac{2\pi}{2^s} {}_{2}F{}_{1}\left(\frac{1-s}{2} , -\frac{s}{2} ; 1 ; \sin^2\vartheta \right), \\
\mathcal{I}_{i/2}^*(\vartheta) &:=& \int\limits_0^{2\pi} \cos^2\chi G_{i/2}^{(\textrm{rot})}(\vartheta,\chi) d\chi 
= \frac{1+s}{1+2s}\frac{1}{2^s} \int\limits_0^{2\pi} \cos^2\chi (1+\sin\vartheta\sin\chi)^{s} d\chi \nonumber \\ 
&=& \frac{1+s}{1+2s}\frac{\pi}{2^s} {}_{2}F{}_{1}\left(\frac{1-s}{2} , -\frac{s}{2} ; 2 ; \sin^2\vartheta \right), \\
\mathcal{I}_{i/2}^{\dag}(\vartheta) &:=& \int\limits_0^{2\pi} \sin^2\chi G_{i/2}^{(\textrm{rot})}(\vartheta,\chi) d\chi 
= \frac{1+s}{1+2s}\frac{1}{2^s} \int\limits_0^{2\pi} \sin^2\chi (1+\sin\vartheta\sin\chi)^{s} d\chi \nonumber \\
&=& \frac{1+s}{1+2s}\frac{\pi}{2^s} {}_{3}F{}_{2}\left(\frac{3}{2} , \frac{1-s}{2} , -\frac{s}{2} ; \frac{1}{2} , 2 ; \sin^2\vartheta \right), \\
\label{Eq:App02}
\mathcal{I}_{i/2}^\ddag(\vartheta) &:=& \int\limits_0^{2\pi} \sin\chi G_{i/2}^{(\textrm{rot})}(\vartheta,\chi) d\chi 
= \frac{1+s}{1+2s}\frac{1}{2^s} \int\limits_0^{2\pi} \sin\chi (1+\sin\vartheta\sin\chi)^{s} d\chi \nonumber \\
&=& \frac{1+s}{1+2s}\frac{\pi s \sin\vartheta}{2^s} {}_{2}F{}_{1}\left(\frac{1-s}{2} , 1-\frac{s}{2} ; 2 ; \sin^2\vartheta \right),
\end{eqnarray}
where ${}_{p}F{}_{q}(a_1, \ldots, a_p ; b_1, \ldots, b_q ; z)$ is the generalized hypergeometric function. As in the non-rotating model, it can be seen that
\begin{equation}
    \mathcal{I}_{i/2}(\vartheta) = \mathcal{I}_{i/2}^*(\vartheta) + \mathcal{I}_{i/2}^\dag (\vartheta).
\end{equation}

\begin{figure}[h]
    \centering
    \includegraphics[scale=0.25]{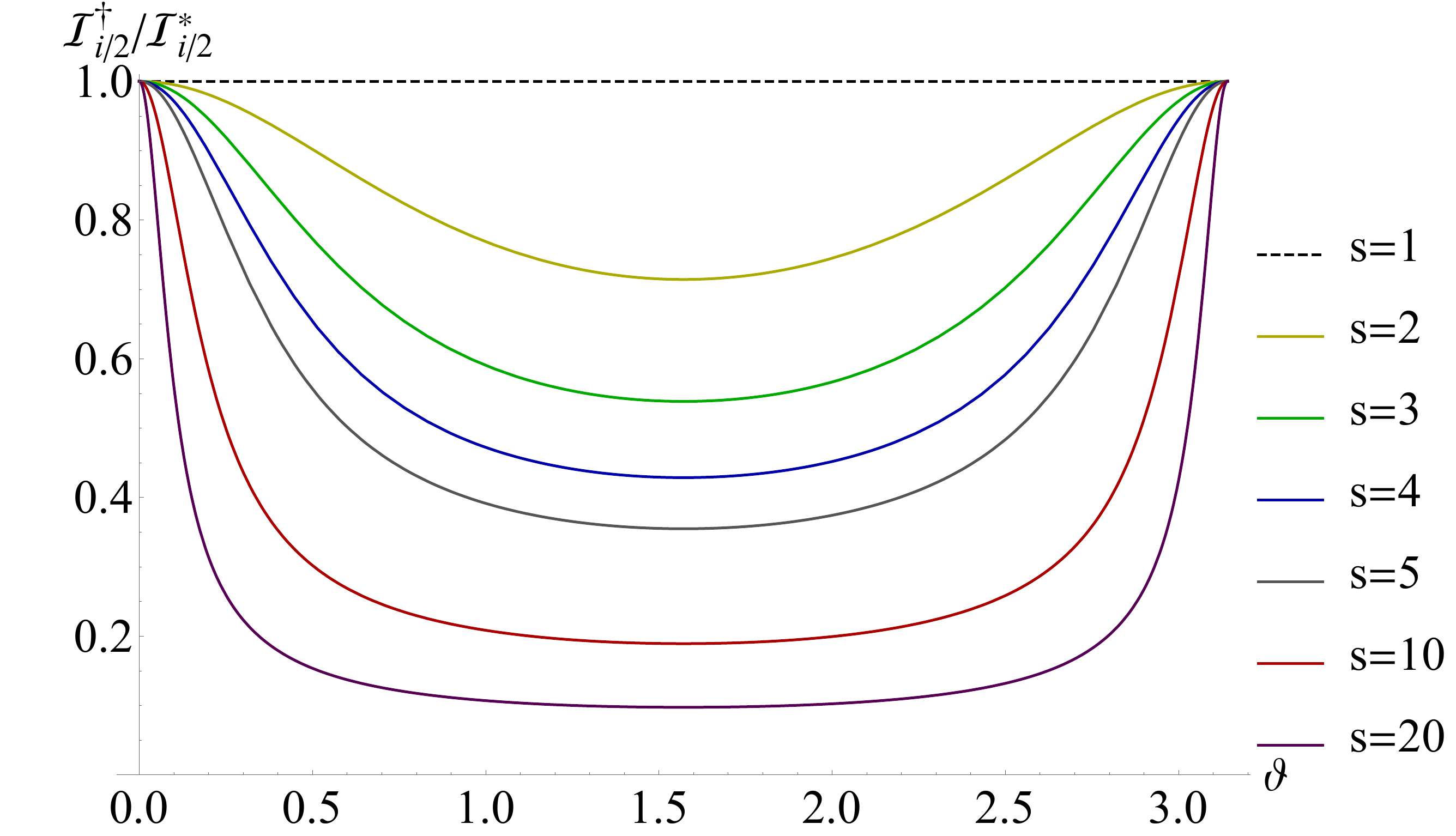}
    \caption{$\mathcal{I}_{i/2}^\dag/\mathcal{I}_{i/2}^*$ ratio vs $\vartheta$ for some values of parameter $s$. As we have seen, the ratio depends on the $\vartheta$-variable, which do not occur in the even models.}
    \label{Fig:IRatios01}
\end{figure}

\begin{figure}[h]
    \centering
    \includegraphics[scale=0.25]{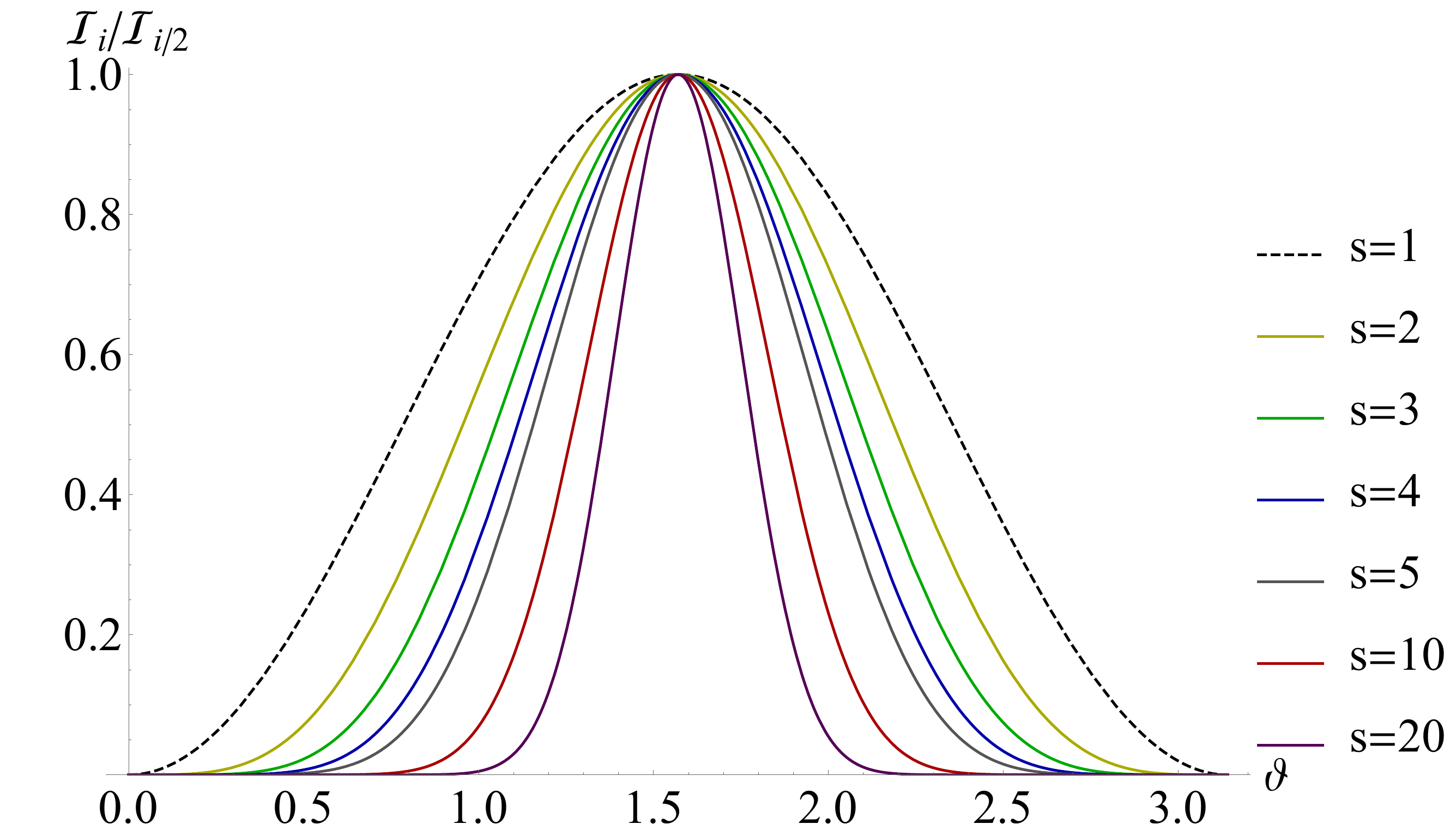}
    \caption{$\mathcal{I}_i/\mathcal{I}_{i/2}$ ratio vs $\vartheta$ for some values of parameter $s$. As we have seen, the ratio depends on the $\vartheta$-variable, which do not occur in the even models.}
    \label{Fig:IRatios02}
\end{figure}

\begin{figure}[h]
    \centering
    \includegraphics[scale=0.25]{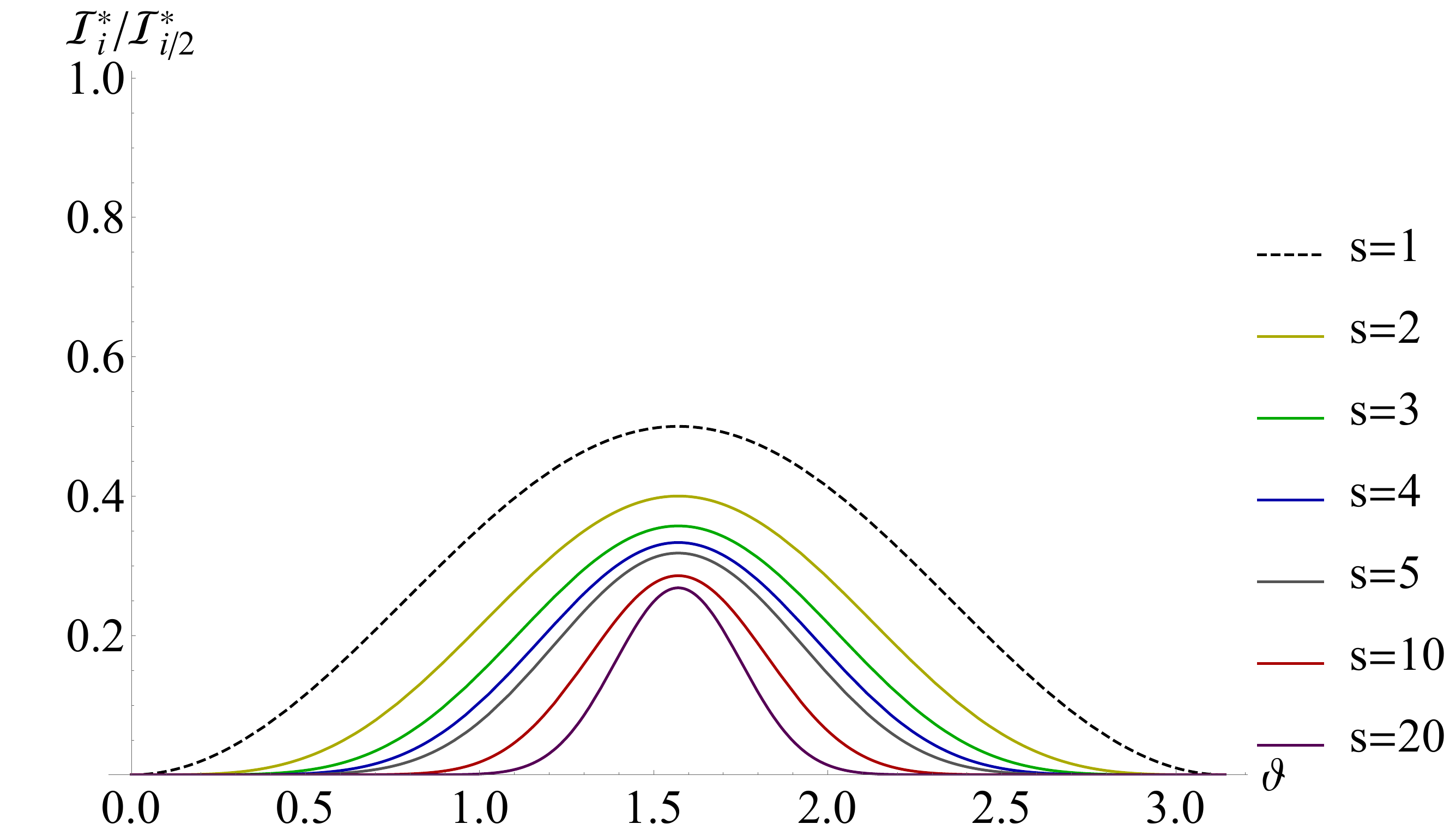}
    \caption{$\mathcal{I}_i^*/\mathcal{I}_{i/2}^*$ ratio vs $\vartheta$ for some values of parameter $s$. As we have seen, the ratio depends on the $\vartheta$-variable, which do not occur in the even models.}
    \label{Fig:IRatios03}
\end{figure}

\begin{figure}[h]
    \centering
    \includegraphics[scale=0.25]{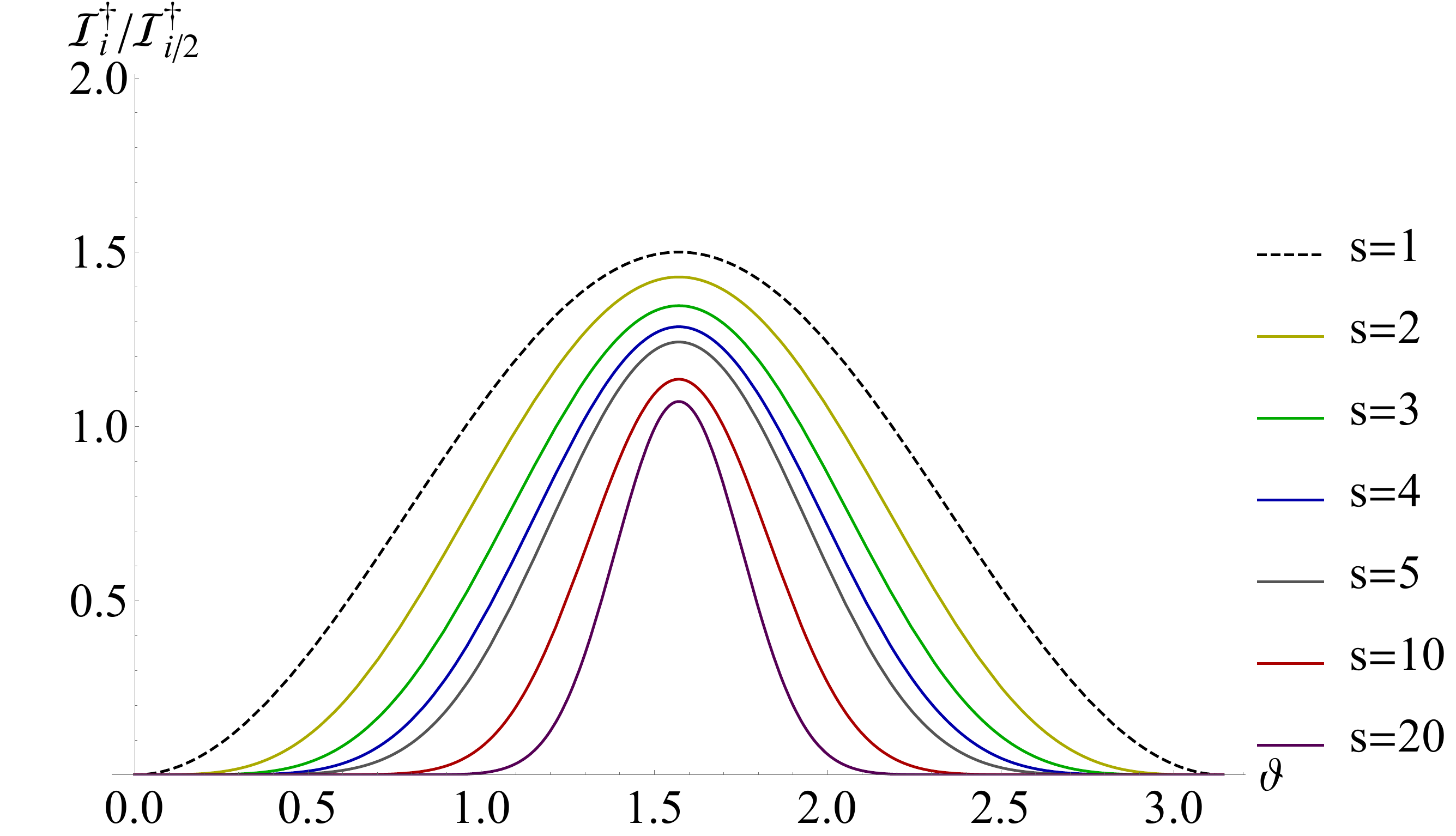}
    \caption{$\mathcal{I}_i^\dag /\mathcal{I}_{i/2}^\dag$ ratio vs $\vartheta$ for some values of parameter $s$. As we have seen, the ratio depends on the $\vartheta$-variable, which do not occur in the even models.}
    \label{Fig:IRatios04}
\end{figure}

\begin{figure}[h]
    \centering
    \includegraphics[scale=0.25]{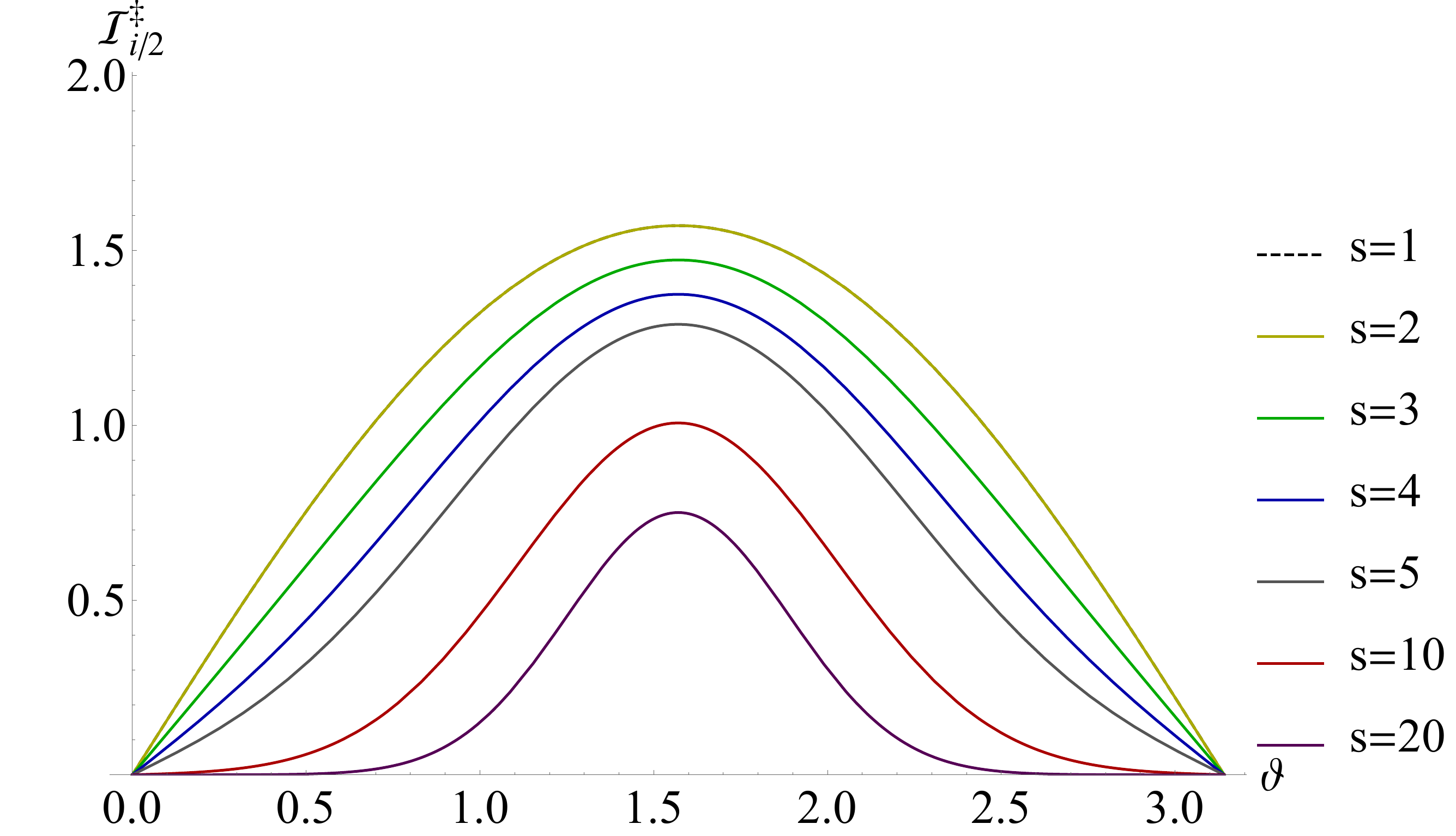}
    \caption{$\mathcal{I}_i^\ddag$ vs $\vartheta$ for some values of parameter $s$. As we have seen, the ratio depends on the $\vartheta$-variable, which do not occur in the even models.}
    \label{Fig:IRatios05}
\end{figure}

In the even model, the relation for the integrals $\mathcal{I}_i^*/\mathcal{I}_i^\dag$ is a simple factor $1/(2s+1)$; however, in the rotating model, the relation is more complex and depends on the $\vartheta$-variable. For sake of illustration, the figure~\ref{Fig:IRatios01} shows the ratio $\mathcal{I}_{i/2}^\dag/\mathcal{I}_{i/2}^*$ vs $\vartheta$-variable for some values of $s$.
In a same way, the figures~\ref{Fig:IRatios02},~\ref{Fig:IRatios03},~\ref{Fig:IRatios04} and~\ref{Fig:IRatios05} show the behavior of the ratio of the different integrals from non-rotating and rotating models.

\section{Integrals over the angular momentum $L$ variable}
\label{Appx:B}

After performing the integrations over the angle variable $\chi$, the remaining integrals are over the energy and angular momentum. At this point, the integrals over angular momentum $L$ variable can also be solved analytically, and the following results are found to be useful throughout Section~\ref{Sec:Models}
\begin{eqnarray}
\int\limits_{L_{\textrm{c}}(E)}^{L_{\textrm{max}}(E,r)} L \sqrt{E^2 - V_{m,L}(r)} dL &=& \frac{\sqrt{N(r)}}{3r} L_{\textrm{max}}^3(E,r) \sqrt{(1 - b^2(E,r))^3} \\
\int\limits_{L_{\textrm{c}}(E)}^{L_{\textrm{max}}(E,r)} \frac{L}{\sqrt{E^2 - V_{m,L}(r)}} dL &=& \frac{r}{\sqrt{N(r)}} L_{\textrm{max}}(E,r) \sqrt{1 - b^2(E,r)} \\
\int\limits_{L_{\textrm{c}}(E)}^{L_{\textrm{max}}(E,r)} \frac{L^2}{\sqrt{E^2 - V_{m,L}(r)}} dL &=& \frac{r}{2\sqrt{N(r)}} L_{\textrm{max}}^2(E,r) \left(\frac{\pi}{2} + b(E,r) \sqrt{1 - b^2(E,r)} - \arctan\frac{b(E,r)}{\sqrt{1 - b^2(E,r)}} \right) \\
\int\limits_{L_{\textrm{c}}(E)}^{L_{\textrm{max}}(E,r)} \frac{L^3}{\sqrt{E^2 - V_{m,L}(r)}} dL &=& \frac{2r}{3\sqrt{N(r)}} L_{\textrm{max}}^3(E,r) \left(1+\frac{b^2(E,r)}{2}\right)\sqrt{1 - b^2(E,r)}
\end{eqnarray}
where $b$ and $L_{\textrm{max}}$ are defined in equation~(\ref{Eq:b}) and~(\ref{Eq:MaximumAngularMomentum}) respectively.


\bibliographystyle{unsrt}
\bibliography{referencias.bib}

\end{document}